\documentclass[a4paper,11pt]{article}
\pdfoutput=1 % if your are submitting a pdflatex (i.e. if you have
\usepackage{jcappub} % for details on the use of the package, please
\usepackage[T1]{fontenc} % if needed

\usepackage{graphicx}
\usepackage{float}
\usepackage{hyperref}
\usepackage{xcolor, xstring}
\usepackage{amssymb, amsmath, mathtools}
\usepackage{listings}
\usepackage{lscape}
\usepackage{lipsum}
\usepackage{multicol}
\usepackage{physics}
\usepackage{natbib}
\usepackage{txfonts}
\usepackage[small]{caption}
\usepackage{xspace}
\usepackage{comment}

\newcommand{\COBEF}{{\it COBE/FIRAS}\xspace}

\newcommand{\Mpc}{{\rm Mpc}}

\newcommand{\expf}[1]{{{\rm e}^{#1}}}

\newcommand{\nS}{n_{\rm S}}

\newcommand{\Planck}{{\it Planck}\xspace}

\newcommand{\vbeta}{{\boldsymbol{\beta}}}
\newcommand{\vbh}{{\boldsymbol{\hat{\beta}}}}

\newcommand{\gh}{{\hat{\gamma}}}

\newcommand{\vgh}{{\hat{\boldsymbol\gamma}}}

\newcommand{\xg}{x}

\newcommand{\id}{{\,\rm d}}

\newcommand{\beq}{\begin{equation}}   %

\newcommand{\eeq}{\end{equation}}   %

\newcommand{\beqa}{\begin{eqnarray}}   %

\newcommand{\eeqa}{\end{eqnarray}}   %

\newcommand{\bealf}[1]{\begin{align} #1 \end{align}}

\newcommand{\beal}{\begin{align}}
\newcommand{\enal}{\end{align}}

\newcommand{\bspl}{\begin{split}}

\newcommand{\espl}{\end{split}}

\newcommand{\bsub}{\begin{subequations}}

\newcommand{\esub}{\end{subequations}}

\newcommand{\bmulti}{\begin{multline}}   %

\newcommand{\beqm}{\begin{mathletters}}   %

\newcommand{\eeqm}{\end{mathletters}}   %

\newcommand{\Ne}{N_{\rm e}}

\newcommand{\Tg}{T_{\gamma}}

\newcommand{\sigT}{\sigma_{\rm T}}

\newcommand{\nPl}{n_{\rm Pl}}

\newcommand{\vek} [1]{\mbox{\boldmath${#1}$\unboldmath}}

\newcommand{\pot}[2]{#1 \times 10^{#2}}

\newcommand{\Thz}{\theta_{z}}

\newcommand{\fNL}{f_{\rm NL}}

\usepackage{bbm}
\def\i{\mathbbm{i}}

\title{CMB Spectral Distortion Anisotropies from Acoustic Damping with primordial non-Gaussianity}

\author[a]{Jens Chluba}

\author[b]{, Atsuhisa Ota}

\author[c]{and Nicola Bartolo}

\affiliation[a]{Jodrell Bank Centre for Astrophysics, School of Physics and Astronomy, The University of Manchester, Oxford Road, Manchester, M13 9PL, U.K.}
\affiliation[b]{Department of Physics and Chongqing Key Laboratory for Strongly Coupled Physics, Chongqing University, Chongqing 401331, People's Republic of China}

\affiliation[c]{Dipartimento di Fisica e Astronomia "Galileo Galilei"
Università degli Studi di Padova
Via Marzolo 8, I-35131 Padova, Italy}

\emailAdd{Jens.Chluba@Manchester.ac.uk}
\emailAdd{aota@cqu.edu.cn}

\date{Aug 2026}
\begin{document}

\abstract{In this paper, we evaluate the precise distortion source and transfer functions caused by mixing of blackbodies of different temperatures using the recently developed frequency hierarchy (FH) treatment of {\tt CosmoTherm}. With this we are able to evaluate the effects of primordial non-Gaussianity (PNG) on the $\mu T$, $\mu E$, $y T$ and $y E$ cross-power spectra including the coupled spectro-spatial evolution and important photon-transport effects retained by the FH treatment. 
For local-type PNG, we compare our results with those from previous works, illustrating new aspects that were previously not captured. We then demonstrate how the $\mu T$ and $\mu E$ signals change in the presence of enhanced curvature perturbations at small scales.
For a nearly scale-invariant primordial spectrum, our results agree broadly with previous estimates on large angular scales but exhibit additional small-scale damping and modified $y$-distortion correlations arising from the scale dependence of the heating source and the distinct transport of distortion perturbations. Tight-coupling and monopole-source approximations accurately reproduce the $\mu$-distortion spectra, while the $y$-distortion signals remain more sensitive to the detailed source evolution. For enhanced small-scale power, anisotropies generated by propagation of the distorted average spectrum can become comparable to those from anisotropic dissipation. Their different angular dependences and parameter scalings principally allow the small-scale power amplitude and primordial non-Gaussianity to be constrained separately.
The results presented here thus pave the path for studying PNG in new regimes using existing and upcoming high precision CMB anisotropy data to measure primordial distortion correlations.

}

\maketitle

\section{Introduction}
%---------------------------------------
Measurements of cosmic microwave background (CMB) temperature and polarisation anisotropies have placed stringent constraints on the statistics of primordial fluctuations \citep[e.g.,][]{WMAP7yrCosmo, Planck2015params, Planck2013ng, Planck2019PNG}. Nevertheless, photon diffusion erases primary temperature perturbations on scales much smaller than those directly accessible through the conventional CMB anisotropy power spectra. However, the energy carried by these perturbations is not completely lost: the dissipation of acoustic modes mixes blackbody spectra of different temperatures and thereby generates small departures from a perfect blackbody spectrum \citep{Sunyaev1970diss, Daly1991, Hu1994, Chluba2012}. At sufficiently early times, Compton scattering and photon-production processes partially thermalise the released energy, producing predominantly $\mu$-type distortions, while energy release closer to recombination gives rise mainly to $y$-type and residual distortions from the intermediate stages \citep{Sunyaev1970mu, Illarionov1975, Burigana1991, Hu1993, Chluba2011therm, Khatri2012mix, Chluba2013Green, Chluba2013PCA}. CMB spectral distortions therefore retain information about primordial perturbations on scales that are otherwise inaccessible to direct observation \citep[e.g.,][]{Chluba2012inflaton, Clesse2014, Cabass2016Discovery}.

Most studies of acoustic damping have focused on the spatially-averaged CMB spectrum. For approximately Gaussian and nearly scale-invariant adiabatic perturbations, the dissipation of acoustic modes produces an average distortion of order $\bar{\mu}\simeq \pot{2}{-8}$ \citep{Chluba2012, Chluba2016}. 
The precise distortion depends on the amplitude and scale dependence of the primordial power spectrum and can consequently be used to constrain small-scale adiabatic, isocurvature, vector and tensor perturbations \citep{Chluba2012inflaton, Chluba2013iso, Chluba2013PCA, Ota2014, Chluba2015}. However, acoustic dissipation is intrinsically {\it inhomogeneous}. Its efficiency varies spatially because the short-wavelength perturbations responsible for the heating can be modulated by longer-wavelength modes. The resulting spectral-distortion anisotropies provide information not only about the small-scale power spectrum but also about correlations between perturbations on widely separated scales.

This makes distortion anisotropies particularly sensitive to primordial non-Gaussianity \citep{Pajer2012, Ganc2012, Biagetti2013, Emami2015muT, Ota2016, Chluba2017muT, Bartolo2016trispec, Shiraishi2016, Ravenni2017, Ota2017, Orlando2022muB}.
For local-type non-Gaussianity, a long-wavelength curvature perturbation modulates the power carried by short-wavelength modes. The local acoustic-heating rate is consequently correlated with the large-scale CMB temperature and polarisation anisotropies, generating $\mu T$, $\mu E$, $y T$ and $y E$ cross-correlations. These observables probe squeezed configurations of the primordial bispectrum in which the short modes can lie far beyond the scales accessible through the primary CMB anisotropies. Distortion anisotropies therefore offer a distinctive test of scale-dependent non-Gaussianity and of (departures from) the standard single-field inflationary consistency relations \citep[e.g.,][]{Emami2015muT, Dimastrogiovanni2016}. Existing CMB data have already been used to constrain anisotropic $\mu$-distortion signals \citep{Rotti2022muT, Bianchini2022}, and the continued improvement of multifrequency CMB measurements \citep{Remazeilles2018muT, Zegeye2023S4, Zegeye2024} correspondingly motivates more complete theoretical predictions.

Previous calculations established the principal scaling of the distortion–temperature/polarization correlations and clarified several aspects of their subsequent evolution. These treatments necessarily relied on simplifying assumptions. The acoustic-damping source was commonly represented through approximate heating window functions, often evaluated in the squeezed limit and with the two dissipating modes taken to be exactly back-to-back. The resulting distortion was then propagated using approximate transfer functions or by assigning a fixed $\mu$- or $y$-type spectral shape according to the epoch of energy release. Such approximations capture the broad physical origin of the signal, but they do not generally follow the continuous conversion between spectral shapes, the scale dependence of the anisotropic source, or the full transport of distortion perturbations through recombination and reionisation. They can also miss geometric contributions arising from the relative orientation of the two short-wavelength modes entering the quadratic acoustic-dissipation source.

A complete calculation is challenging because spectral-distortion anisotropies depend simultaneously on frequency, position and photon direction. Directly discretising the photon spectrum at every angular multipole leads to a large and numerically stiff Boltzmann system, while a description in terms of only temperature, $\mu$ and $y$ amplitudes does not remain closed under Comptonization and thermalization. The {\it frequency hierarchy} (FH) approach was developed to bridge these descriptions \citep{chluba_spectro-spatial_2023-I, chluba_spectro-spatial_2023-II, kite_spectro-spatial_2023-III, Chluba2026}. It expands the spectral dependence in a compact set of frequency-space basis functions derived from the thermalization problem \citep{chluba_spectro-spatial_2023-I} and promotes the corresponding spectral coefficients to angular multipoles \citep{chluba_spectro-spatial_2023-II}. The evolution of the photon field can then be written as a generalised Boltzmann hierarchy in which spectral evolution and spatial transport are treated within one system \citep{kite_spectro-spatial_2023-III}.

This formulation has several advantages for the present problem. Energy released by blackbody mixing is initially injected with a $y$-type spectral dependence, but its final form depends continuously on the injection redshift. Compton scattering redistributes the injected energy in frequency space, converting the initial $y$-type perturbation towards a $\mu$-type shape, while double Compton scattering and bremsstrahlung erase distortions at very early times. At the same time, free streaming, Thomson scattering and photon diffusion redistribute the distortion spatially and generate its angular hierarchy. The FH approach follows all of these effects self-consistently, without assigning the distortion to a predetermined spectral era or applying a separate distortion visibility function after the spatial evolution has been calculated. It also provides frequency-dependent transfer functions that can be projected onto experimentally relevant spectral components after the full evolution has been completed \citep{chluba_spectro-spatial_2023-II}.

The method is especially valuable for distortion anisotropies because their transport differs qualitatively from that of ordinary temperature perturbations. In the tightly coupled photon–baryon fluid, temperature monopole and dipole perturbations exchange energy and undergo acoustic oscillations. A distortion perturbation does not possess the same restoring mechanism. Its multipoles are instead strongly damped by Thomson scattering and photon diffusion, with a damping scale that differs from that of the primary temperature anisotropies \citep{Pajer2012b, Chluba2017muT, Chluba2026TC}. The resulting scale dependence affects the shape of the observable cross-power spectra, particularly at high multipoles. The effect cannot be captured fully by attaching a standard temperature transfer function to a local heating-rate modulation, as we demonstrate here.

In this paper, we use the FH to obtain a detailed treatment of spectral-distortion anisotropies sourced by the damping of primordial acoustic perturbations. We begin by deriving the relevant second-order blackbody-mixing source from the Liouville and Thomson collision terms, including velocity-dependent scattering contributions. The result can be expressed compactly in terms of gauge-independent photon temperature variables. We then project the source onto its angular multipoles and transform it to Fourier space. This calculation shows explicitly that the local heating source depends on the full triangle formed by the observable mode and the two dissipating modes. In particular, the angle between the two short-wavelength modes modulates the heating rate as we show here. This angular dependence is absent when the source is evaluated only in the exactly squeezed, back-to-back configuration used in earlier treatments.

Rather than evolving the full second-order photon distribution directly, we exploit the structure of the observable correlators to simplify the source and transfer calculations. The quadratic acoustic-damping source is first correlated with the large-scale curvature perturbation, reducing the required second-order information to a set of effective, scale- and redshift-dependent anisotropic heating rates. These source functions are constructed from linear temperature and velocity transfer functions while retaining the primordial bispectrum and the complete mode-coupling geometry. They can then be inserted into the linear FH as external spectral-distortion sources. This separation makes the calculation computationally tractable while preserving all contributions relevant to the distortion–temperature and distortion–polarisation cross-spectra.

Using this framework as implemented in {\tt CosmoTherm} \citep{Chluba2011therm}, we compute the $\mu T$, $\mu E$, $yT$ and $yE$ power spectra for local-type primordial non-Gaussianity (PNG), including the continuous spectral evolution, Thomson transport, photon diffusion, recombination and reionisation. This provides predictions for these correlations in which the acoustic-damping source and the subsequent spectro-spatial evolution are treated within a common framework. We compare the results with previous approximate calculations and identify the origin of the visible differences. The refined treatment reproduces the broad large-scale behaviour found previously, but modifies the scale dependence through the exact mode-coupling geometry, the $k$-dependence of the heating source and the additional damping of distortion perturbations. These effects are particularly relevant for the $y$-type correlations and at small angular scales.

We also assess which elements of the full calculation are essential for accurate predictions. By comparing the numerical transfer-function database with tight-coupling expressions, we determine when analytic approximations to the heating rate are sufficient. We separately examine the importance of anisotropic source multipoles beyond the local monopole and the consequences of approximating the heating rate as independent of the observable wavenumber. These tests provide a controlled route towards faster calculations for more general primordial bispectra. Finally, we consider scenarios with enhanced small-scale curvature power. Such an enhancement increases the mean distortion, but it also changes the redshift and scale dependence of the anisotropic source. The FH naturally includes both the direct modulation of acoustic dissipation and the secondary distortion anisotropies generated as the modified average spectrum propagates through the perturbed Universe.

The paper is organised as follows. In section~\ref{sec:mixing}, we derive the blackbody-mixing source and its angular multipoles, including the full Fourier-space mode-coupling geometry. In section~\ref{sec:av_heating}, we discuss the average acoustic-heating rate and the additional contribution generated by local-type non-Gaussianity. Section~\ref{sec:computation_heating} describes the numerical construction of the effective anisotropic heating rates and useful tight-coupling approximations. In section~\ref{sec:models}, we incorporate these sources into the FH and compute the resulting distortion cross-power spectra, comparing different physical and numerical approximations. There we also illustrate that in principle different scenarios can be distinguished before concluding in section~\ref{sec:conclusions}.

As a note in preparation of Sect.~\ref{sec:model-II}, where we consider an simple model with enhanced small scale power for illustration, here we also compute the propagation response of the homogeneous average distortion through the linearly perturbed medium. When the average distortion is itself generated by acoustic damping, these terms {\it formally} correspond to a subset of third-order contributions in the expansion in primordial perturbations. Here, we do not attempt a complete calculation at this order, also because formally the primordial third order terms have different transfer functions and statistical properties. Our results should therefore be regarded as an estimate of this particular propagation contribution rather than as the full (non-)Gaussian correction. However, we expect our calculations to already capture the leading order effects.

\newpage

\section{Main distortion sources from photon mixing}
\label{sec:mixing}
%---------------------------
Solving the full problem of photon mixing requires a second-order treatment in the cosmological perturbations \citep[e.g.,][]{Bartolo2006, Bartolo2007, Pettinari2013}. Here we are only interested in terms that generate {\it distortions} from the damping of first order temperature perturbations. Due to the mixing of blackbody spectra \citep{Chluba2004, Stebbins2007}, these are initially always related to $y$-distortions~\citep{Pitrou2009, Chluba2012, Khatri2012short2x2}, but then evolve under thermalization processes and photon transport to generate distortion anisotropies \citep{Chluba2017muT, chluba_spectro-spatial_2023-II, kite_spectro-spatial_2023-III}.

\paragraph{Distortion anisotropies from the average.}
%---------------------------
We shall assume that no extra source of distortions other than the mixing of blackbodies with different temperatures is present. This means that on average, we expect a distortion that to leading order is sourced by perturbation correlators of the type\footnote{We will symbolically use $\bar{\mu}$ and $\delta \mu$ to denote the average and fluctuating distortions having $\mu$-type distortions in mind even if in this context the distortion shape can be more general.} $\bar{\mu}\leftrightarrow \langle \mathcal{R}(\vek{k}_1)\mathcal{R}(\vek{k}_2)\rangle$, where $\mathcal{R}(\vek{k})$ denotes the primordial curvature perturbations in Fourier space. For standard inflation scenarios, this is expected at a level $\bar{\mu}\simeq \pot{2}{-8}$~\citep{Chluba2012}. In the presence of local-type non-Gaussian perturbations, also correlators of the type $\bar{\mu}\leftrightarrow \fNL^2 \langle \mathcal{R}(\vek{k}_1)\mathcal{R}(\vek{k}_2)\mathcal{R}(\vek{k}_3)\mathcal{R}(\vek{k}_4)\rangle$ will contribute to the average, which for large value of $\fNL$ can become noticeable and even comparable to the two-point Gaussian contribution if $\fNL\gtrsim \pot{2}{4}$ (see section~\ref{sec:av_heating}). Due to effects of standard perturbations \citep{Chluba2012, Chluba2017muT, chluba_spectro-spatial_2023-II}, the average distortions then source distortion anisotropies at a level $\delta \mu(\vek{k})\leftrightarrow \mathcal{R}(\vek{k}) \times \bar{\mu}$, which can correlate with the CMB temperature and polarization perturbations. The corresponding terms have order $\langle \delta \mu(\vek{k}) \delta T(\vek{K})\rangle \leftrightarrow \langle \mathcal{R}(\vek{k})\mathcal{R}(\vek{K})\rangle \times \bar{\mu}$.
If the primordial perturbations sourcing $\bar{\mu}$ have as sufficiently high amplitude, this could cause noticeable distortion anisotropies.

\paragraph{Distortion anisotropies from anisotropic dissipation.}
%---------------------------
In addition to the average contribution and related transfer effects through the perturbed Universe, the photon mixing process directly sources distortion anisotropies as a function of $\vek{k}$ at various eras through anisotropic dissipation. As we will show in detail below, these are related to convolution integrals involving the primordial perturbations like $\delta \mu(\vek{k})\propto \int \id^3 k_1\,\mathcal{R}(\vek{k}_1)\mathcal{R}(\vek{k}-\vek{k}_1)$, which cause anisotropic heating. The temperature correlators then take the form $\langle \delta \mu(\vek{k})\,\delta T(\vek{K})\rangle \leftrightarrow \langle \mathcal{R}(\vek{K})\,\mathcal{R}(\vek{k}_1)\mathcal{R}(\vek{k}-\vek{k}_1)\rangle $, a fact that can be used to evaluate the required distortion sources, as we show below. For Gaussian fluctuations, this term vanishes and only the higher order correlators of the form $\langle \delta \mu(\vek{k})\delta \mu(\vek{k}')\rangle \leftrightarrow \langle \mathcal{R}(\vek{k}_2)\mathcal{R}(\vek{k}'-\vek{k}_2)\,\mathcal{R}(\vek{k}_1)\mathcal{R}(\vek{k}-\vek{k}_1)\rangle $ matter. Here, we evaluate the anisotropic source terms $\propto \langle \mathcal{R}(\vek{K})\,\mathcal{R}(\vek{k}_1)\mathcal{R}(\vek{k}-\vek{k}_1)\rangle $ and compute related transfer functions to obtain the correct predictions for the cross power spectra in the presence of PNG, placing previous estimates \citep{Pajer2012, Ganc2012, Emami2015muT, Chluba2017muT, Ravenni2017} on a solid footing.

\subsection{Collecting the main distortion source terms from photon mixing}
%---------------------------
At first order in the standard perturbation variables there are no distortion sources unless some process releases energy to the average CMB spectrum \citep[e.g.,][]{Chluba2012, kite_spectro-spatial_2023-III}. To describe the distortion transfer effects, an extended Boltzmann hierarchy at first order in perturbation theory is sufficient \citep[e.g.,][]{chluba_spectro-spatial_2023-II}, so the task at hand is to include all second order source terms from photon temperature mixing.

Denoting the collision term at second order in perturbation theory by $\mathcal{C}^{(2)}[n]$ the Boltzmann equation for the photon occupation number, $n(\eta, \vek{r}, \vgh)$, reads \citep[cf.][]{Bartolo2007, Pitrou2009, Nitta2009}
%-----------
\beal
\label{eq:Boltzmann_second_order_general}
&\partial_\eta n^{(2)} + \vgh\cdot \nabla n^{(2)} 
- \xg\partial_{\xg} n^{(0)}\left[\partial_\eta\Phi^{(2)}+ \vgh\cdot \nabla \Psi^{(2)}  \right]
- \xg\partial_{\xg} n^{(1)}\left[\partial_\eta\Phi^{(1)}+ \vgh\cdot \nabla \Psi^{(1)}\right]
-\left(\Phi^{(1)}-\Psi^{(1)}\right) \vgh\cdot \nabla n^{(1)} 
\nonumber\\[2mm]
&\;
+
\left[\partial_{x_i} \left(\Phi^{(1)}-\Psi^{(1)}\right)
-\gh_i\gh_j \,\partial_{x_j} \left(\Phi^{(1)}-\Psi^{(1)}\right) 
 \right] \partial_{\gh_i} n^{(1)}
+ \xg\partial_{\xg} n^{(0)}
\left(\Phi^{(1)}-\Psi^{(1)}\right)\vgh\cdot \nabla \Psi^{(1)}
= \mathcal{C}^{(2)}[n]
\end{align}
%-----------
where $\Phi$ and $\Psi$ are the Newtonian potential perturbations\footnote{Here we have $\Psi=\Phi^{\rm Bartolo}$ and $\Phi=-\Psi^{\rm Bartolo}$.} and $\vgh$ denotes the direction in which the photon is traveling, and $\vek{r}$ denotes the spatial coordinate. 
In the above expression the Hubble term was absorbed using the frequency variable $\xg\propto \nu/(1+z)$.\footnote{Commonly, one uses the average CMB temperature to write $\xg=h\nu/k_{\rm B} \Tg$ where $\Tg \propto (1+z)$, but {\it any} reference temperature with a scaling $\propto (1+z)$ is enough \citep{Chluba2011therm}.} Also, we directly made use of $\partial_\eta \nPl(\xg)=0$, $\nabla \nPl=0$, and neglected vector and tensor perturbations. 

All terms $\propto \xg\partial_{\xg} n^{(0)}$ can be neglected as they only source temperature perturbations at any significant level with the related distortion anisotropies being treated at first order. The terms $\propto \vgh\cdot \nabla n^{(1)}$ and $\propto \partial_{\gh_i} n^{(1)}$ do not change the spectral shape and hence constitute higher order corrections to the evolution of the photon field, which we neglect here.\footnote{
%---------------
If we source average distortions $\simeq\langle \mathcal{R}(\vek{k}_1)\mathcal{R}(\vek{k}_2)\rangle$, we expect distortion terms $n^{(1)}(\vek{k})\leftrightarrow \mathcal{R}(\vek{k})\times \langle \mathcal{R}(\vek{k}_1)\mathcal{R}(\vek{k}_2)\rangle$. These sources then give rise to $n^{(2)}(\vek{k})\leftrightarrow \mathcal{R}(\vek{k}')\mathcal{R}(\vek{k}-\vek{k}')\times \langle \mathcal{R}(\vek{k}_1)\mathcal{R}(\vek{k}_2)\rangle$, which yield temperature-distortion correlators $\langle \delta \mu(\vek{k})\delta T(\vek{k}')\rangle \leftrightarrow \langle\mathcal{R}(\vek{k}')\mathcal{R}(\vek{k}'')\mathcal{R}(\vek{k}-\vek{k}'')\rangle \times \langle \mathcal{R}(\vek{k}_1)\mathcal{R}(\vek{k}_2)\rangle$ that we do not consider here.}
%-------
This means that we only have two terms on the left hand side of Eq.~\eqref{eq:Boltzmann_second_order_general} with contributions to distortions. 

To obtain the relevant distortion source terms, we only consider temperature anisotropies for the first order field. Inserting
the Ansatz $n^{(1)}=G \Theta^{(1)}$ and $n^{(2)}=\Delta n^{(2)}+ G [\Theta^{(2)}+(\Theta^{(1)})^2]+\frac{1}{2} Y (\Theta^{(1)})^2$, where $\Delta n^{(2)}$ is a distortion, we then find:
%-----------
\bsub
\beal
\label{eq:Boltzmann_second_order_general_dist_source}
\partial_\eta n^{(2)} + \vgh\cdot \nabla n^{(2)} &\quad\longrightarrow\quad
\Theta^{(1)}\,\left[\partial_\eta \Theta^{(1)} + \vgh\cdot \nabla \Theta^{(1)} \right] Y
\\
- \xg\partial_{\xg} n^{(1)}\left[\partial_\eta\Phi^{(1)}+ \vgh\cdot \nabla \Psi^{(1)}\right]
&\quad\longrightarrow\quad  
\Theta^{(1)}\,\left[\partial_\eta\Phi^{(1)}+ \vgh\cdot \nabla \Psi^{(1)}\right]Y,
\end{align}
\esub
%-----------
where we used $- \xg\partial_{\xg} G=3 G + Y$ and then dropped the temperature term. Put together this gives the photon mixing distortion source \citep{Chluba2012}
%-----------
\beal
\label{eq:Boltzmann_second_order_general_dist_source_LHS}
\mathcal{S}^{\rm mix}_{\rm L}&=-\frac{\Theta^{(1)}}{\tau'}\!\left[\partial_\eta \Theta^{(1)} + \vgh\cdot \nabla \Theta^{(1)} + \partial_\eta\Phi^{(1)}+ \vgh\cdot \nabla \Psi^{(1)}\right]
%\nonumber \\
%&
\equiv\Theta^{(1)}\!\left[\Theta^{(1)}-\Theta^{(1)}_0-\frac{1}{10}\Theta^{(1)}_2-\vbeta^{(1)}\!\cdot \vgh \right]
\end{align}
%-----------
from the left hand side ($\leftrightarrow$ the Liouville terms) of the second order Boltzmann equation. In the last step we used that $\partial_\eta \Theta^{(1)} + \vgh\cdot \nabla \Theta^{(1)} + \partial_\eta\Phi^{(1)}+ \vgh\cdot \nabla \Psi^{(1)}\equiv \mathcal{C}^{(1)}[n]=\tau'\left[\Theta^{(1)}_0+\frac{1}{10}\Theta^{(1)}_2-\Theta^{(1)}+\vbeta^{(1)}\!\cdot \vgh \right]$. We also defined 
$\Theta^{(1)}_\ell\equiv \Theta^{(1)}_\ell(\eta, \vgh, \vek{r})=\sum_m \Theta^{(1)}_{\ell m}(\eta, \vek{r})\,Y_{\ell m}(\vgh)$ for the first order temperature fluctuations (see Appendix~\ref{app:Theta_conventions}) and the baryon speed $\beta^{(1)}\equiv|\vbeta^{(1)}|=\beta^{(1)}(\eta, \vek{r})$ and $\vbeta^{(1)}=\beta^{(1)}\,\vbh$. 

We next obtain all terms from the second order collision term, $\mathcal{C}^{(2)}[n]$. From Eq.~(C16b) of \citep{Chluba2012}, we directly have the Thomson terms
%-----------
\beal
\label{eq:Thomson}
\mathcal{S}^{\rm mix}_{\rm T}&=\frac{1}{2}\left\{ \left[(\Theta^{(1)})^2\right]_0+\frac{1}{10}\left[(\Theta^{(1)})^2\right]_2-(\Theta^{(1)})^2\right\}
\end{align}
%-----------
from resting electrons. Here $[\ldots]_\ell=\sum_m [\ldots]_{\ell m}\,Y_{\ell m}(\vgh)$, where $[\ldots]_{\ell m}$ is the spherical harmonic coefficient of the relevant quantity.

Using Appendix~C of \citep{Chluba2012} and Appendix~E of \citep{ChlubaBO25}, for the velocity-dependent distortion source terms from blackbody mixing we find
%-----------
\beal
\label{eq:velocity}
\mathcal{S}^{\rm mix}_{\beta}&=
\beta^{(1)}\left[\frac{\beta^{(1)}}{3}-C^0_1 \,\Theta^{(1)}_{10} \, Y_{00}(\vgh)
\right]
+\frac{\beta^{(1)}}{10}
\sum_{m=-1}^1 
C^m_2\left[\Theta^{(1)}_{2m}
Y_{1m}(\vgh)
-
\Theta^{(1)}_{1m}
Y_{2m}(\vgh)
\right]
\\ \nonumber 
&\qquad -\frac{\beta^{(1)}}{10}
\sum_{m=-2}^2 
C^m_3\left[\Theta^{(1)}_{3m}
Y_{2m}(\vgh)
-
\Theta^{(1)}_{2m}
Y_{3m}(\vgh)
\right]
+\beta^{(1)} \Theta^{(1)}_{0}\,P_{1}(\vgh\cdot\vbh)
+\frac{11}{30}\,(\beta^{(1)})^2\,P_{2}(\vgh\cdot\vbh)
,
\end{align}
%-----------
with $C^m_\ell=\sqrt{(l^2-m^2)/(4\ell^2-1)}$ and where $\vbh^{(1)}$ is assumed to be parallel to the $z$-axis.
To obtain a coordinate independent version, we refer to Appendix~\ref{app:coordinate_indep}, which yields
%-----------
\beal
\label{eq:velocity_coordinate_independent_final}
\mathcal{S}^{\rm mix}_{\beta}&=
\frac{\vbeta^{(1)}\!\cdot\vbeta^{(1)}}{3}
-\left[\vbeta^{(1)}\!\cdot \vgh\,\Theta^{(1)}(\vgh)\right]_0
-\frac{1}{10}\left[\vbeta^{(1)}\!\cdot \vgh\,\Theta^{(1)}(\vgh)\right]_2
\nonumber \\ 
&\qquad +\vbeta^{(1)}\!\cdot \vgh \left[ \Theta^{(1)}_{0}(\vgh)+\frac{1}{10}\,\Theta^{(1)}_{2}(\vgh)\right]
+\frac{11}{30}\,\vbeta^{(1)}\!\cdot\vbeta^{(1)}\, P_{2}(\vgh\cdot\vbh).
\end{align}
%-----------
The sum of the three contributions is what we need in our computation of distortion sources. The final source term then has to be evaluated for the various multipoles and also transformed to Fourier space to perform the computation. To obtain the relevant transfer functions, we compute the relevant correlators with the observable in mind. This simplifies the general computation significantly. 

\subsubsection{Simplified expression for the real-space source term from photon mixing}
%---------------------------
Before proceeding, we give the expression for the sum of all three terms. For this we introduce the variable $V^{(1)}=\vbeta^{(1)}\!\cdot \vgh$. This has the benefit that $V$ now behaves the same way as a CMB temperature variable and thus can be expanded into spherical harmonics \citep{Ota2017}. 
With Eq.~\eqref{eq:velocity_coordinate_independent_rederived_V}, we then obtain the distortion source function\footnote{For convenience, we will drop the explicit statement of the order of the transfer functions, knowing that all are computed from first order perturbations.}
%-----------
\beal
\label{eq:dis_source_V}
\mathcal{S}^{\rm mix}(\eta, \vek{r}, \vgh)&=\mathcal{S}^{\rm mix}_{\rm L}+\mathcal{S}^{\rm mix}_{\rm T}+\mathcal{S}^{\rm mix}_{\rm \beta}
\nonumber\\
&=
\Theta\left[\Theta-\Theta_0-\frac{1}{10}\Theta_2\right]+
[V^2]_0
-2\left[V\Theta\right]_0
+\frac{1}{2}\left\{[\Theta^2]_0+\frac{1}{10}[\Theta^2]_2-\Theta^2\right\}
\nonumber \\
&\qquad 
+\left[V\Theta\right]_0
-V\Theta+V\Theta_0
+\frac{1}{10}\Big\{V\Theta_2-\left[V \Theta\right]_2\Big\}
+\frac{11}{20}\,V^2-\frac{11}{20}\,[V^2]_0,
\end{align}
%-----------
where we grouped terms to make cancellations of various multipole moments more obvious. For example, only the first three groups of terms contribute to the photon monopole distortion source.
By introducing the gauge-independent temperature variable $\Theta_{\rm g}=\Theta+\Psi-V$ with $\Theta_{\rm g, \ell}=\Theta_{\ell}+\delta_{\ell 0} \Psi-\delta_{\ell 1} V_1$ and noticing that $V^2=[V^2]_2 + [V^2]_0$, we can further regroup terms as
%-----------
\beal
\label{eq:dis_source_V_B}
\mathcal{S}^{\rm mix}(\eta, \vek{r}, \vgh)
&=
\frac{1}{2}\left\{[(\Theta-V)^2]_0+\frac{1}{10}[(\Theta-V)^2]_2-(\Theta-V)^2\right\}
-(\Theta-V)\left[\Theta_0+\frac{1}{10}\Theta_2-(\Theta-V)\right]
\nonumber \\
&\equiv 
\Theta_{\rm g}\left[\Theta_{\rm g}-\Theta_{\rm g,0}-\frac{1}{10}\Theta_{\rm g,2}\right]
-\frac{1}{2}\left\{\Theta_{\rm g}^2-[\Theta_{\rm g}^2]_0-\frac{1}{10}[\Theta_{\rm g}^2]_2\right\},
\end{align}
%-----------
where the first line directly agrees with Eq.~(6) of \citep{Ota2019}.
This means that all required quantities are determined by the projections $[\Theta_{\rm g}]_\ell=\Theta_{\rm g, \ell}$ and $[\Theta_{\rm g}^2]_\ell$.

We also note that for the distortion anisotropy evolution the anisotropic temperature sources are not crucial as they only cause distortions at higher order in the perturbations \citep{Chluba2012}. In addition, these will source non-Gaussian temperature anisotropies and locally also lead to varying electron temperature corrections. However, the related {\it photon cooling} is negligible since the heat capacity of the electrons/baryons is very small.

\subsection{Monopolar distortion sources}
%------------------------
With the results above, we can evaluate the distortion source for the local monopole \citep[compare also][]{Chluba2012}. Since all the contributions lead to a $y$-distortion source, this can equivalently be thought of as heating of the electrons. Averaging Eq.~\eqref{eq:dis_source_V} or alternatively Eq.~\eqref{eq:dis_source_V_B} over photon directions we find
%-----------
\beal
\label{eq:local_monopole_dis_source}
\mathcal{S}_0^{\rm mix}(\eta, \vek{r})
&=\int \mathcal{S}^{\rm mix}(\eta, \vek{r}, \vgh)\frac{\id \vgh}{4\pi}
\equiv [\Theta_{\rm g}^2]_0 -\Theta_{\rm g, 0}^2 -\frac{1}{10}[\Theta^2_{\rm g,2}]_0
\nonumber\\
&=
\int 
\left[\Theta^2-\Theta^2_0-\frac{1}{10}\Theta^2_2\right]\frac{\id \vgh}{4\pi}
+
[V^2]_0
-2\left[V\Theta\right]_0
\nonumber\\
&=
\sum_{m=-1}^1\frac{|\Theta_{1m}|^2}{4\pi}
+\frac{9}{10}\sum_{m=-2}^2\frac{|\Theta_{2m}|^2}{4\pi}
+\sum_{\ell\geq3, m} \frac{|\Theta_{\ell m}|^2}{4\pi}
+
\frac{\vbeta\!\cdot\vbeta}{3}
-2\left[\vbeta\cdot \vgh \,\Theta\right]_0,
\end{align}
%-----------
where in the last line we used Eq.~\eqref{eq:V2}.
This source term has a uniform contribution and a fluctuating part, which we now evaluate. 

To make progress, we first go to Fourier space. Since all source terms are real, we can think of $\vbeta\cdot\vbeta\equiv \vbeta\cdot\vbeta^*$ and $2\left[\vbeta\cdot \vgh \,\Theta\right]_0\equiv \beta^*\left[\vbh\cdot \vgh \,\Theta\right]_0+\beta\left[\vbh\cdot \vgh \,\Theta\right]^*_0$ in this step. This then yields the convolution integral:
%-----------
\beal
\label{eq:local_monopole_dis_source_k}
\mathcal{S}^{\rm mix}_0(\eta, \vek{k})
%&=
%\int \frac{\id^3 k'}{(2\pi)^3}\frac{\id^3 k''}{(2\pi)^3} \id^3 x\,
%\expf{\i \vek{x}\cdot (\vek{k}'-\vek{k}''-\vek{k})}
%\,\mathcal{R}(\vek{k}')\,\mathcal{R}^*(\vek{k}'')\,
%\hat{\mathcal{M}}_0(\eta, \vek{k}', \vek{k}'')
%\nonumber\\[-1mm]
%&=
%\int \frac{\id^3 k'}{(2\pi)^3}\,\id^3 k''
%\delta(\vek{k}'-\vek{k}''-\vek{k}) 
%\,\mathcal{R}(\vek{k}')\,\mathcal{R}(-\vek{k}'')\,
%\hat{\mathcal{M}}_0(\eta, \vek{k}', \vek{k}'')
%\nonumber\\
%&=
%\int \frac{\id^3 k'}{(2\pi)^3}
%\,\mathcal{R}(\vek{k}')\,\mathcal{R}(\vek{k}-\vek{k}')\,
%\hat{\mathcal{M}}_0(\eta, \vek{k}', \vek{k}'-\vek{k})
&=
\int \frac{\id^3 k'}{(2\pi)^3}
\,\mathcal{R}(\vek{k}')\,\mathcal{R}^*(\vek{k}'-\vek{k})\,
\hat{\mathcal{M}}_0(\eta, \vek{k}', \vek{k}'-\vek{k})
\\[1mm]
\hat{\mathcal{M}}_0(\eta, \vek{k}_1, \vek{k}_2)
&=\sum_{m=-1}^1\frac{\Theta_{1m}(\vek{k}_1)\,\Theta^*_{1m}(\vek{k}_2)}{4\pi}
+\frac{9}{10}\sum_{m=-2}^2\frac{\Theta_{2m}(\vek{k}_1)\,\Theta^*_{2m}\,(\vek{k}_2)}{4\pi}
+\sum_{\ell\geq3, m} \frac{\Theta_{\ell m}(\vek{k}_1)\,\Theta^*_{\ell m}(\vek{k}_2)}{4\pi}
\nonumber
\\
\nonumber
&\qquad
+\frac{\vbeta(\vek{k}_1)\cdot \vbeta^*(\vek{k}_2)}{3}
-\frac{1}{3}
\sum_{m}
\bigg\{ 
\beta^*(\vek{k}_1)
Y_{1m}(\vbh_1) \, \Theta_{1 m}(\vek{k}_2)
+\beta(\vek{k}_1) Y^*_{1m}(\vbh_1) \, \Theta^*_{1 m}(\vek{k}_2)
\bigg\}.
\end{align}
%-----------
Here, we suppressed $\eta$ in the variables and used the transfer functions $X(\eta, \vek{k})=\hat{X}(\eta, \vek{k})\,\mathcal{R}(\vek{k})$ to relate to the initial scalar perturbations $\mathcal{R}(\vek{k})$ although we dropped the hats in the expression for $\mathcal{M}_0$. Inserting $\Theta_{\ell m}(\eta, \vek{k}_i)
\equiv 4\pi (-\i)^\ell\, \tilde{\Theta}_{\ell}(\eta, k_i)\,Y^*_{\ell m}(\hat{\vek{k}}_i)$ from Eq.~\eqref{eq:Theta_ellm_irrotational} and $\beta(\eta, \vek{k}_i)=-\i \tilde{\beta}(\eta, k_i)$ and $\vbh_i=\hat{\vek{k}}_i$
for irrotational velocity fields then yields
%-----------
\beal
\label{eq:local_monopole_dis_source_k_MM}
\hat{\mathcal{M}}_0(\eta, \vek{k}_1, \vek{k}_2)
&=\left[
3\tilde{\Theta}_1(k_1)\,\tilde{\Theta}_1(k_2)
+\frac{\tilde{\beta}(k_1)\,\tilde{\beta}(k_2)}{3}
-\tilde{\beta}(k_1)\,\tilde{\Theta}_1(k_2)
-\tilde{\beta}(k_2)\,\tilde{\Theta}_1(k_1)\right] P_1(\hat{\vek{k}}_1\cdot \hat{\vek{k}}_2)
\\
\nonumber
&\qquad
+\frac{9}{2}\tilde{\Theta}_2(k_1)\,\tilde{\Theta}_2(k_2)\, P_2(\hat{\vek{k}}_1\cdot \hat{\vek{k}}_2)
+\sum_{\ell\geq 3}(2\ell +1 )\,\tilde{\Theta}_\ell(k_1)\,\tilde{\Theta}_\ell(k_2)\, P_\ell(\hat{\vek{k}}_1\cdot \hat{\vek{k}}_2),
\end{align}
%-----------
with the first group of terms further simplifying to 
%-----------
\beal
3\tilde{\Theta}_1(k_1)\,\tilde{\Theta}_1(k_2)
+\frac{\tilde{\beta}(k_1)\,\tilde{\beta}(k_2)}{3}
-\tilde{\beta}(k_1)\,\tilde{\Theta}_1(k_2)
-\tilde{\beta}(k_2)\,\tilde{\Theta}_1(k_1)=3\tilde{\Theta}_{1,\rm g}(k_1)\,\tilde{\Theta}_{1,\rm g}(k_2)
\end{align}
%-----------
for the gauge-independent dipole term $\tilde{\Theta}_{1,\rm g}(k_i)=\tilde{\Theta}_1(k_i)-\tilde{\beta}(k_i)/3$. This finally gives
%-----------
\beal
\label{eq:local_monopole_dis_source_k_MM_final}
\hat{\mathcal{M}}_0(\eta, \vek{k}_1, \vek{k}_2)
&=3\tilde{\Theta}_{1,\rm g}(k_1)\,\tilde{\Theta}_{1,\rm g}(k_2) \,P_1(\hat{\vek{k}}_1\cdot \hat{\vek{k}}_2)
\\
\nonumber
&\qquad
+\frac{9}{2}\tilde{\Theta}_2(k_1)\,\tilde{\Theta}_2(k_2)\, P_2(\hat{\vek{k}}_1\cdot \hat{\vek{k}}_2)
\\
\nonumber
&\qquad\qquad
+\sum_{\ell\geq3}(2\ell +1 )\,\tilde{\Theta}_\ell(k_1)\,\tilde{\Theta}_\ell(k_2)\, P_\ell(\hat{\vek{k}}_1\cdot \hat{\vek{k}}_2).
\end{align}
%-----------
An important feature is that $\hat{\mathcal{M}}_0(\eta, \vek{k}_1, \vek{k}_2)=\hat{\mathcal{M}}_0(\eta, k_1, k_2, \hat{\vek{k}}_1\cdot \hat{\vek{k}}_2)$, meaning that the angle between the modes modulates the distortion anisotropies, an aspect that was neglected in previous treatments.

\subsubsection{Uniform heating terms}
%----------------------------

\paragraph{Uniform heating term - Gaussian part.}
%----------------------------
To obtain the uniform part of the distortion source we carry out the ensemble average using the correlator $\langle\mathcal{R}(\vek{k}')\mathcal{R}^*(\vek{k}'-\vek{k})\rangle=(2\pi)^3\,\delta(\vek{k})\,P(k')$ for Gaussian fluctuations and then integrate Eq.~\eqref{eq:local_monopole_dis_source_k} over $\id^3 k/(2\pi)^3$ to find:
%-----------
\beal
\label{eq:local_monopole_dis_source_av}
\left<\mathcal{S}^{\rm mix}_0\right>
&=
\int \Delta^{\rm G}(k')\,
\Bigg\{
\frac{9}{2}\,\tilde{\Theta}^2_2(\eta,k')
+3\tilde{\Theta}^2_{1, \rm g}(\eta,k')
+\sum_{\ell\geq 3} (2\ell+1) \,\tilde{\Theta}^2_\ell(\eta,k')
\Bigg\}\,\id \ln k'.
\end{align}
%-----------
with $\Delta^{\rm G}(k)\equiv k^3 P(k)/[2\pi^2]$.
The average distortion source is naturally only a function of time which depends on the temperature perturbations at first order in perturbation theory. 
In the pre-recombination era, only the first term contributes significantly and provides the shear viscosity of the medium. The second term remains small, as tight-coupling prevents baryons and photons from slipping, which suppresses heat conduction ($\propto \tilde{\Theta}^2_{1, \rm g}$). Similarly, higher multipoles are strongly suppressed and only contribute close to the recombination era \citep{Chluba2012}.
We note that this expression does not include all polarization terms \citep[see][for all contributions]{Chluba2015}, but otherwise reproduces previous results \citep[e.g.,][]{Chluba2012, Chluba2013iso, Ota2017}. With a similar method, one can correctly evaluate the contributions from tensor and vector perturbations \citep{Ota2014, Chluba2015}, the latter of which has not been explicitly studied. 

\paragraph{Uniform heating term - non-Gaussian part.}
%----------------------------
We next evaluate the average contributions in the presence of local-type non-Gaussianity.
For this we start with
%-----------
\beal
\mathcal{R}(\vek{x})=\mathcal{R}^{\rm G}(\vek{x})+\frac{3}{5}\,f_{\rm NL}\left\{[\mathcal{R}^{\rm G}(\vek{x})]^2-\left<[\mathcal{R}^{\rm G}(\vek{x})]^2\right>\right\},
\end{align}
%-----------
where $\mathcal{R}^{\rm G}(\vek{x})$ is the Gaussian part of the primordial scalar perturbations. We then obtain \citep[compare Appendix~A of][]{Chluba2017muT}
%-----------
\bsub
\beal
\langle \mathcal{R}_1\,\mathcal{R}_2\rangle
&=(2\pi)^3\,\delta^{(3)}(\vek{k}_1+\vek{k}_2)\left[ 
P(k_1) 
+\frac{18}{25}\,f_{\rm NL}^2 \int \frac{\id^3 q}{(2\pi)^3}P(q)\,P(|\vek{k}_1-\vek{q}|)\right]
\\
\langle \mathcal{R}_1\,\mathcal{R}_2\,\mathcal{R}_3\rangle
&=
(2\pi)^3\,\delta^{(3)}(\vek{k}_1+\vek{k}_2+\vek{k}_3)
\times\frac{6}{5}\,f_{\rm NL}\,\left[P(k_1)\,P(k_2) + P(k_1)\,P(k_3)+P(k_2)\,P(k_3)\right].
%\\
%\langle \mathcal{R}_1\,\mathcal{R}_2\,\mathcal{R}_3\,\mathcal{R}_4\rangle
%&=\langle \mathcal{R}^{\rm G}_1\,\mathcal{R}^{\rm G}_2\,\mathcal{R}^{\rm G}_3\,\mathcal{R}^{\rm G}_4\rangle
%+(2\pi)^3\,\delta^{(3)}(\vek{k}_1+\vek{k}_2+\vek{k}_3+\vek{k}_4)
%\\ \nonumber
%&\!\!\!\!\times\frac{36}{25}\,f^2_{\rm NL}\,\left\{P(k_1)\,P(k_2)[P(k_{13})+P(k_{14})] + 5\, {\rm permutations} \right\}.
\end{align}
\esub
%-----------
Here we introduced the bispectrum parametrization
%------------
\bealf{
\label{eq:B_def}
B(k, k_1,k_2)=\frac{6}{5}\,f_{\rm NL}[P(k_1)P(k_2)+P(k_1)P(k)+P(k_2)P(k)]
}
%------------
as commonly used for local-type primordial non-Gaussianity.
In the two-point correlation function for the source term, we then have the non-Gaussian source of average distortions
%-----------
\beal
\label{eq:local_monopole_dis_source_NG}
\left<\mathcal{S}^{\rm mix, NL}_0\right>
&=
\frac{18}{25}\,f_{\rm NL}^2 \int \Delta^{\rm NL}(k')
\,\Bigg\{
\frac{9}{2}\,\tilde{\Theta}^2_2(\eta,k')
+3\tilde{\Theta}^2_{1, \rm g}(\eta,k')
+\sum_{\ell\geq 3} (2\ell+1) \,\tilde{\Theta}^2_\ell(\eta,k')
\Bigg\}\,\id \ln k'
\nonumber\\
\Delta^{\rm NL}(k')&=
\frac{{k'}^3}{2\pi^2}
\int \frac{\id^3 q}{(2\pi)^3}
\frac{\id \hat{\vek{k}}'}{4\pi} P(q)\,P(|\vek{k}'-\vek{q}|)
\equiv 
\frac{{k'}^3}{2\pi^2}
\int \Delta^{\rm G}(q)\,\frac{\id q}{q}
\int_{|q-k'|}^{q+k'}
\frac{\tilde{k} \id \tilde{k}}{2 k' q} P(\tilde{k})
\nonumber\\
&=\frac{{k'}^2}{2}\int \Delta^{\rm G}(q)\,\frac{\!\id q}{q^2}
\int_{|q-k'|}^{q+k'}
\Delta^{\rm G}(\tilde{k})\,
\frac{\!\id \tilde{k}}{\tilde{k}^2} %
\end{align}
%-----------
where we set $\tilde{k}(k', q, \chi\equiv \hat{\vek{k}}'\cdot\hat{\vek{q}})=|\vek{k}'-\vek{q}|\equiv ({k'}^2+q^2-2 q k' \chi)^{1/2}$ and then traded $\chi$ for $\tilde{k}$. We note that this expression does not converge without introducing scale cuts in the Gaussian fields. 
We furthermore mention that primordial trispectra are omitted here but could also contribute at the level $\propto \fNL^2$ \citep[e.g.][]{Bartolo2016trispec, Shiraishi2016}.
Instead of Eq.~\eqref{eq:local_monopole_dis_source_NG} we can alternatively write
%-----------
\beal
\label{eq:Delta_NLlocal_monopole_dis_source_NG}
\Delta^{\rm NL}(k')&=
\frac{{k'}^3}{2\pi^2}\!
\int \frac{\id^3 q}{(2\pi)^3}
\frac{\id \hat{\vek{k}}'}{4\pi} P(q)\,P(|\vek{k}'-\vek{q}|)
%\nonumber \\&
\equiv 
\frac{{k'}^3}{2\pi^2}\!
\int \! \frac{\!\id^3 q}{(2\pi)^3}
\frac{\!\id^3 q'}{(2\pi)^3}
\frac{\!\id \hat{\vek{k}}'}{4\pi} 
\,(2\pi)^3\delta^{(3)}(\vek{q}'+\vek{q}-\vek{k}')
P(q)\,P(q')
\nonumber \\
&\equiv 
\frac{{k'}^3}{2\pi^2}
\int \id^3 r \int \frac{\id^3 q}{(2\pi)^3}
\frac{\id^3 q'}{(2\pi)^3}
\frac{\id \hat{\vek{k}}'}{4\pi} 
\,\expf{\i\vek{r}\cdot\left(\vek{q}'+\vek{q}-\vek{k}'\right)}
P(q)\,P(q')
\nonumber \\
&\equiv 
\frac{{k'}^3}{2\pi^2}
\int \id^3 r j_0(r k') 
\int 
j_0(r q)\,j_0(r q')\,
\Delta^{\rm G}(q)\,\Delta^{\rm G}(q')\id \ln q
\id \ln q'
\nonumber \\
&\equiv 
\frac{2{k'}^3}{\pi}
\int r^2\id r j_0(r k') 
\left[\int 
j_0(r q)\,
\Delta^{\rm G}(q)\id \ln q\right]^2.
\end{align}
%-----------
Assuming a simple power-law for $\Delta^{\rm G}(k)$ does not yield a finite result for $\Delta^{\rm NL}(k')$ without introducing cut-offs. For a single mode, $\Delta^{\rm G}_\delta(k)=A_\delta \delta(\ln  k-\ln k_0)$ one finds:
%-----------
\beal
\label{eq:Delta_NL_delta}
\Delta^{\rm NL}_\delta(k')
&=
\frac{2{k'}^3}{\pi} A_\delta^2\int r^2\id r j_0(r k') 
\left[ 
j_0(r k_0)\right]^2
=\frac{{k'}^3}{2} A_\delta^2 \,\mathcal{I}_0(k',k_0,k_0)
=\frac{A_\delta^2}{2}\,\frac{{k'}^2}{k_0^2},
\end{align}
%-----------
for $0\leq k'\leq 2k_0$ and where the function $\mathcal{I}_\ell(k',k_1,k_2)$ is discussed in Appendix~\ref{app:Triple_j_int}. 

\subsubsection{Distortion anisotropy source from the local monopole}
%----------------------------
To understand how to compute the required transfer functions in the presence of source terms, we recapped the arguments for temperature variables in Appendix~\ref{app:evol_eq}. This showed that even for second order sources, one can obtain the required solutions from the standard first order perturbation transfer functions, given the correct $k$, $\ell$ and $m$ dependence of the source term and related transfer function solution [see Eq.~\eqref{eq:evol_I_F_kz_ellm_final} and \eqref{eq:correlation_kz_final_B}]. 
%
%The argument can be transferred to distortion transfer functions, as explained in Appendix~\ref{app:distortion_sources} for the $\mu$-parameter. %However, the real source term structure directly depends on the related observable, which allows one to simplify the problem significantly.

To compute the relevant transfer function sources for distortion anisotropies, we have to consider the correlators $\langle \mathcal{R}(\vek{K})\,S^{\rm mix}_0(\vek{k})\rangle$, where $\vek{K}$ is related to the large-scale temperature mode. If we only consider Gaussian fluctuations, we furthermore need $\langle S^{\rm mix}_0(\vek{k})\,S^{\rm mix}_0(\vek{k}')\rangle$ to be able to pre-compute the relevant transfer functions (see Appendix~\ref{app:distortion_sources} for details). These statements rely on the fact that whatever the precise transfer function in each case is, one can always independently compute them at first order in perturbation theory, using the relevant first order temperature transfer functions for the source term as for the temperature case. This is because even if there are second order corrections to the temperature terms, these do not source distortion anisotropies at the same order. Also, even if the distortion anisotropy source has a more general structure, once we consider the observational correlator, only those sources exciting modes that then correlate will have to be computed.

To obtain the non-Gaussian contributions from the photon mixing source for $\ell=0$, we can use
%-----------
\bealf{
\langle \mathcal{R}^*(\vek{K})\,S^{\rm mix}_0(\vek{k})\rangle
&\leftrightarrow 
\langle \mathcal{R}^*(\vek{K})\mathcal{R}(\vek{k}')\mathcal{R}^*(\vek{k}'-\vek{k})\rangle=(2\pi)^3 \delta(\vek{K}-\vek{k})
\times B(K, k', |\vek{k}'-\vek{k}|),
}
%-----------
with an example bispectrum given in Eq.~\eqref{eq:B_def}.
As explained in Appendix~\ref{app:distortion_sources}, we can then compute the final distortion multipoles in the frame with $\vek{k}\parallel z$ using the source term
%-----------
\beal
\label{eq:source_monopole_kz}
s_{00}(\eta, k)=\int 
\frac{k_1^2\!\id k_1}{2\pi^2}\,
\int_{|k_1-k|}^{k_1+k}\,\frac{k^2_2 \id k_2}{2 \pi^2}
\,\frac{\pi^2 B(k, k_1, k_2)}{k k_1 k_2 \, P(k)}\,
\hat{\mathcal{M}}_0\left(\eta, k_1, k_2,  \frac{k_1^2+k_2^2-k^2}{2 k_1 k_2}\right),
\end{align}
%-----------
together with the evolution equation for first order perturbations to obtain the solutions to the sourced distortion anisotropies. The final result is then obtained by using this result in the expression for the cross power spectra [see Eq.~\eqref{eq:correlation_kz_final_B_source}].

\subsection{Higher multipoles of the distortion source}
\label{sec:source_ellm}
%----------------------------
For general $\ell, m$ we find the distortion sources from Eq.~\eqref{eq:dis_source_V_B} as
%-----------
\beal
\label{eq:dis_source_V_B_ellm}
\mathcal{S}^{\rm mix}_{\ell m}(\eta, \vek{r})
&=
[\Theta_{\rm g}^2]_{\ell m}-\Theta_{{\rm g}, \ell m}\Theta_{\rm g,0}
-\frac{1}{10}[\Theta_{\rm g}\Theta_{\rm g,2}]_{\ell m}
-\frac{1}{2}\left\{[\Theta_{\rm g}^2]_{\ell m}-\delta_{\ell 0}[\Theta_{\rm g}^2]_{00}-\frac{\delta_{\ell 2}}{10}[\Theta_{\rm g}^2]_{2m}\right\}
\\ \nonumber
&=
\sum_{\ell',m'}\,\sum_{\ell'',m''}
(-1)^{m}\mathcal{G}^{\ell, \ell',\ell''}_{-m,m',m''}
\left(\frac{1+\delta_{\ell 0}}{2}+\frac{\delta_{\ell 2}}{20}-\delta_{\ell'' 0}-\frac{\delta_{\ell'' 2}}{10}\right)
\,\Theta_{{\rm g},\ell' m'}(\eta, \vek{r})
\,\Theta_{{\rm g},\ell''m''}(\eta, \vek{r})
\end{align}
%-----------
where we used the Gaunt integrals, $\mathcal{G}^{\ell, \ell',\ell''}_{m,m',m''}$ [see Eq.~\eqref{eq:Gaunt}]. Going to Fourier space, we then have
%-----------
\beal
\label{eq:local_monopole_dis_source_k_Ylm}
\mathcal{S}^{\rm mix}_{\ell m}(\eta, \vek{k})
&=
\int \frac{\id^3 k'}{(2\pi)^3}
\,\mathcal{R}(\vek{k}')\,\mathcal{R}^*(\vek{k}'-\vek{k})\,
\hat{\mathcal{M}}_{\ell m}(\eta, \vek{k}', \vek{k}'-\vek{k})
\\[1mm]\nonumber
\hat{\mathcal{M}}_{\ell m}(\eta, \vek{k}_1, \vek{k}_2)
&=
\sum_{\ell',m'} \sum_{\ell'',m''}
(-1)^{m+m''}\mathcal{G}^{\ell, \ell',\ell''}_{-m,m',-m''}
\left(\frac{1+\delta_{\ell 0}}{2}+\frac{\delta_{\ell 2}}{20}-\delta_{\ell'' 0}-\frac{\delta_{\ell'' 2}}{10}\right)
\nonumber\\
&\qquad \qquad \qquad \qquad \qquad \qquad \qquad \qquad 
\times
\,\hat{\Theta}_{{\rm g},\ell' m'}(\eta, \vek{k}_1)
\,\hat{\Theta}^*_{{\rm g},\ell''m''}(\eta, \vek{k}_2)
\end{align}
%-----------
and $\hat{\Theta}_{{\rm g},\ell m}(\eta, \vek{k}_i)
\equiv 4\pi (-\i)^\ell\, \tilde{\Theta}_{{\rm g},\ell}(\eta, k_i)\,Y^*_{\ell m}(\hat{\vek{k}}_i)$. Assuming $\ell=0$, we can sum over $\ell''$ and $m''$, which yields $\ell''\rightarrow \ell'$ and $m''\rightarrow m'$ and $(-1)^{m'}\mathcal{G}^{0, \ell',\ell'}_{0,m',-m'}=1/\sqrt{4\pi}$. With this we then recover the result given above after multiplying by $Y_{00}(\vgh)=1/\sqrt{4\pi}$ and using the spherical harmonic addition theorem.

To obtain the required source term for general $\ell$ and $m$ we compute the correlator $\langle \mathcal{R}^*(\vek{K})\,S^{\rm mix}_{\ell m}(\vek{k})\rangle$. Following the same steps as for $\ell=m=0$ (see Appendix~\ref{app:distortion_sources}) then yields the intermediate result 
%-----------
\beal
\Sigma_{\ell m}(\eta', \vek{k}) 
&=
\int 
\frac{k_1^2\!\id k_1}{2\pi^2}\,
\int_{-1}^1\frac{\id \hat{\vek{k}}_1}{4\pi}\,
B(k, k_1,|\vek{k}_1-\vek{k}|)\,
\hat{\mathcal{M}}_{\ell m} \left(\eta', \vek{k}_1, \vek{k}_1-\vek{k}\right).
\end{align}
%-----------
where $B(k, k_1, k_2)$ is the bispectrum as before.  Considering the system where $\vek{k}\parallel z$, we know that $\phi_1$ of $\vek{k}_1$ is identical to $\phi_2$ of $\vek{k}_2=\vek{k}_1-\vek{k}$. We can therefore carry out the integral
%-----------
\bealf{
&\int \id \phi_1 \hat{\Theta}_{{\rm g},\ell' m'}(\eta, \vek{k}_1)
\,\hat{\Theta}^*_{{\rm g},\ell''m''}(\eta, \vek{k}_2)
\nonumber\\
&\qquad =
2\pi \, \delta_{m' m''}\,
(4\pi)^2 (-\i)^{\ell'-\ell''}\, \tilde{\Theta}_{{\rm g},\ell'}(\eta, k_1)\, \tilde{\Theta}_{{\rm g},\ell''}(\eta, k_2)\,Y^*_{\ell' m'}(\hat{\vek{k}}_1)
\,
Y_{\ell'' m'}(\hat{\vek{k}}_2).
}
%-----------
In the Gaunt integral, $\mathcal{G}^{\ell, \ell',\ell''}_{-m,m',-m''}$, $m''=m'$ then ensures $\delta_{m0}$. For convenience, we therefore introduce
%-----------
\bealf{
\hat{\mathcal{M}}_{\ell} \left(\eta, k_1, k_2, \hat{\vek{k}}\cdot \hat{\vek{k}}_1,\hat{\vek{k}}\cdot \hat{\vek{k}}_2\right) 
&=\int \frac{\id \phi_1}{2\pi}
\,\frac{\hat{\mathcal{M}}_{\ell 0} \left(\eta, \vek{k}_1, \vek{k}_1-\vek{k}\right)}{(-\i)^{\ell} \sqrt{4\pi\,(2\ell+1)}}
\nonumber\\
&=
\sum_{\ell',\ell''} 
(-\i)^{\ell'-\ell''-\ell}\left(\frac{1+\delta_{\ell 0}}{2}+\frac{\delta_{\ell 2}}{20}-\delta_{\ell'' 0}-\frac{\delta_{\ell'' 2}}{10}\right) 
\nonumber\\
&\qquad \times 
\tilde{\Theta}_{{\rm g},\ell'}(\eta, k_1)\, \tilde{\Theta}_{{\rm g},\ell''}(\eta, k_2)\,
\mathcal{P}_{\ell\ell'\ell''}\left(\hat{\vek{k}}\cdot \hat{\vek{k}}_1,\hat{\vek{k}}\cdot \hat{\vek{k}}_2\right)
\nonumber\\
\mathcal{P}_{\ell\ell'\ell''}\left(\hat{\vek{k}}\cdot \hat{\vek{k}}_1,\hat{\vek{k}}\cdot \hat{\vek{k}}_2\right)&=
\sum_{m'}
\frac{\sqrt{4\pi} \,(-1)^{m'}\mathcal{G}^{\ell, \ell',\ell''}_{0,m',-m'}}{\sqrt{(2\ell+1)}}
\, 4\pi \, Y^*_{\ell' m'}(\hat{\vek{k}}_1)
\,
Y_{\ell'' m'}(\hat{\vek{k}}_2),
}
%-----------
where we note that $\mathcal{P}_{\ell\ell'\ell''}$ does not explicitly depend on the angles $\phi_1$ and $\phi_2$. We also have
the identity $\mathcal{P}_{\ell \ell'\,\ell''}\left(\hat{\vek{k}}\cdot \hat{\vek{k}}_1,\hat{\vek{k}}\cdot \hat{\vek{k}}_2\right)=\mathcal{P}_{\ell \ell''\,\ell'}\left(\hat{\vek{k}}\cdot \hat{\vek{k}}_2,\hat{\vek{k}}\cdot \hat{\vek{k}}_1\right)$.
The required distortion source terms then are
%------------
\beal
\label{eq:source_ell_m_kz_gen}
s_{\ell m}(\eta, k)=\delta_{m0} \int 
\frac{k_1^2\!\id k_1}{2\pi^2}\,
\int_{|k_1-k|}^{k_1+k}\,\frac{k^2_2 \id k_2}{2 \pi^2}
\,\frac{\pi^2 B(k, k_1, k_2)}{k k_1 k_2 \, P(k)}\,
\hat{\mathcal{M}}_{\ell}\left(\eta, k_1, k_2, \hat{\vek{k}}\cdot \hat{\vek{k}}_1,\hat{\vek{k}}\cdot \hat{\vek{k}}_2\right),
\end{align}
%-----------
with $\hat{\vek{k}}\cdot \hat{\vek{k}}_1\equiv \chi_1=(k^2+k_1^2-k_2^2)/[2k k_1]$ and $\hat{\vek{k}}\cdot \hat{\vek{k}}_2\equiv \chi_2 =(k_1\chi_1-k)/k_2=(k_1^2-k_2^2-k^2)/[2k k_2]$. 

\subsubsection{Source for $\ell=0$.}
%-----------
From Eq.~\eqref{eq:source_ell_m_kz_gen} and \eqref{eq:Gaunts} we directly recover Eq.~\eqref{eq:source_monopole_kz} for $\ell=0$. Using
%-----------
\bealf{
\mathcal{P}_{0\ell'\ell''}\left(\chi_1,\chi_2\right)&=
\sum_{m'}
\sqrt{4\pi} \,(-1)^{m'}\mathcal{G}^{0, \ell',\ell''}_{0,m',-m'}
\,4\pi \, Y^*_{\ell' m'}(\hat{\vek{k}}_1)
\,
Y_{\ell'' m'}(\hat{\vek{k}}_2)
\nonumber\\ 
& =
\delta_{\ell'\ell''}\,(2\ell'+1)\,
\sum_{m'} \frac{4\pi \, Y^*_{\ell' m'}(\hat{\vek{k}}_1)
\,
Y_{\ell' m'}(\hat{\vek{k}}_2)}{(2\ell'+1)}
=\delta_{\ell'\ell''}\,
(2\ell'+1)\,P_{\ell'}(\chi_{12}),
}
%-----------
then yields $\hat{\mathcal{M}}_{0}\left(\eta, k_1, k_2, \chi_1,\chi_2\right)$ where $\chi_{12}=\hat{\vek{k}}_1\cdot \hat{\vek{k}}_2=\chi_1\chi_2+\cos(\varphi_1-\varphi_2) \sqrt{1-\chi_1^2}\sqrt{1-\chi_2^2}$.

\subsubsection{Source for $\ell=1$.}
%-----------
For $\ell=1$, we have
%-----------
\bealf{
\nonumber
\mathcal{P}_{1\ell'\ell''}\left(\chi_1,\chi_2\right)&=
\delta_{\ell'+1,\ell''}\,\sum_{m'} C^{m'}_{\ell'+1}
4\pi \, Y^*_{\ell' m'}(\hat{\vek{k}}_1)
\,
Y_{\ell'+1 m'}(\hat{\vek{k}}_2)
+ \delta_{\ell'-1,\ell''}\,\sum_{m'} C^{m'}_{\ell'}
4\pi \, Y^*_{\ell' m'}(\hat{\vek{k}}_1)
\,
Y_{\ell'-1 m'}(\hat{\vek{k}}_2).
}
%-----------
Using the identity
%-----------
\bealf{
\label{eq:Ylm_identity}
C^{m'}_{\ell'+1}\,Y_{\ell'+1\, m'}(\hat{\vek{k}}_i)
&=P_1(\chi_i)\,Y_{\ell'\, m'}(\hat{\vek{k}}_i)-C^{m'}_{\ell'}\,Y_{\ell'-1\, m'}(\hat{\vek{k}}_i)
}
%-----------
with $C^m_\ell=\sqrt{(\ell^2-m^2)/(4\ell^2-1)}$ then gives
%-----------
\bealf{
\mathcal{P}_{1\ell'\ell''}\left(\chi_1,\chi_2\right)&=
\delta_{\ell'+1,\ell''}\sum_{m'} 
4\pi \, \left\{   P_1(\chi_2) Y^*_{\ell' m'}(\hat{\vek{k}}_1)
\,
Y_{\ell' m'}(\hat{\vek{k}}_2)
-C^{m'}_{\ell'}
Y^*_{\ell' m'}(\hat{\vek{k}}_1)
\,
Y_{\ell'-1 m'}(\hat{\vek{k}}_2)\right\}
\nonumber\\
&\!\!\!\!\!\!\!\!\!\!\!\!
+ \delta_{\ell'-1,\ell''}
\sum_{m'} 
4\pi \left\{ P_1(\chi_1) \, Y^*_{\ell'-1 m'}(\hat{\vek{k}}_1)
\,
Y_{\ell'-1 m'}(\hat{\vek{k}}_2)
-
C^{m'}_{\ell'-1} Y^*_{\ell'-2 m'}(\hat{\vek{k}}_1)
\,
Y_{\ell'-1 m'}(\hat{\vek{k}}_2)\right\}
\nonumber\\
&\!\!\!\!\!\!\!\!\!\!\!\!\!\!\!\!\!\!\!\!\!\!\!\!=
\delta_{\ell'+1,\ell''}\left\{ (2\ell' +1) P_1(\chi_2) P_{\ell'}(\chi_{12})
-\sum_{m'} 
C^{m'}_{\ell'}
4\pi \, Y^*_{\ell' m'}(\hat{\vek{k}}_1)
\,
Y_{\ell'-1 m'}(\hat{\vek{k}}_2)\right\}
\nonumber\\ 
&\!\!\!\!\!\!\!\!\!\!\!\!\!\!\!\!\!\!\!\!\!\!\!\!\qquad
+ \delta_{\ell'-1,\ell''}
\left\{ (2\ell'-1) P_1(\chi_1) P_{\ell'-1}(\chi_{12})
-\sum_{m'} 
C^{m'}_{\ell'-1}
4\pi \, Y^*_{\ell'-2 m'}(\hat{\vek{k}}_1)
\,
Y_{\ell'-1 m'}(\hat{\vek{k}}_2)\right\}
\\ \nonumber
&\!\!\!\!\!\!\!\!\!\!\!\!\!\!\!\!\!\!\!\!\!\!\!\!=
\delta_{\ell'+1,\ell''}\left\{P_1(\chi_2) P'_{\ell'+1}(\chi_{12})
-P_1(\chi_1) P'_{\ell'}(\chi_{12})\right\}
+ \delta_{\ell'-1,\ell''}
\left\{ P_1(\chi_1) P'_{\ell'}(\chi_{12})
-P_1(\chi_2) P'_{\ell'-1}(\chi_{12})\right\},
}
%-----------
where the last row was obtained by repeatedly applying Eq.~\eqref{eq:Ylm_identity} and then recognizing that 
$$P'_{\ell+1}(x)\equiv \partial_x P_{\ell+1}(x)=(2\ell+1)P_{\ell}(x)+(2\ell-3)P_{\ell-2}(x)+(2\ell-5)P_{\ell-4}(x)+\ldots$$
This means the non-zero cases 
%-----------
\bsub
\bealf{
\mathcal{P}_{1 0 1}\left(\chi_1,\chi_2\right)
&=P_{1}(\chi_2),
&\mathcal{P}_{1 1 0}\left(\chi_1,\chi_2\right)
&=P_{1}(\chi_1)
\\[2mm]
\mathcal{P}_{1 1 2}\left(\chi_1,\chi_2\right)
&=3 P_{1}(\chi_2) P_{1}(\chi_{12})-P_{1}(\chi_1),
&\mathcal{P}_{1 2 1}\left(\chi_1,\chi_2\right)
&=3 P_{1}(\chi_1) P_{1}(\chi_{12})-P_{1}(\chi_2)
}
\esub
%-----------
involving $\ell \leq 2$ required in our computations.

Putting things together, for the $\ell=1$ source functions this finally gives
%-----------
\bealf{
\hat{\mathcal{M}}_{1} \left(\eta, k_1, k_2, \chi_1,\chi_2\right) 
&=
\sum_{\ell',\ell''} 
(-\i)^{\ell'-\ell''-1}\left(\frac{1}{2}-\delta_{\ell'' 0}-\frac{\delta_{\ell'' 2}}{10}\right) 
\,\tilde{\Theta}_{{\rm g},\ell'}(\eta, k_1)\, \tilde{\Theta}_{{\rm g},\ell''}(\eta, k_2)\,
\mathcal{P}_{1 \ell'\ell''}\left(\chi_1,\chi_2\right)
\nonumber\\
&=
-\sum_{\ell'} 
\left(\frac{1}{2}-\frac{\delta_{\ell' 1}}{10}\right) 
\,\tilde{\Theta}_{{\rm g},\ell'}(\eta, k_1)\, \tilde{\Theta}_{{\rm g},\ell'+1}(\eta, k_2)\,
\mathcal{P}_{1 \ell'\,\ell'+1}\left(\chi_1,\chi_2\right)
\nonumber\\
&\qquad 
+\sum_{\ell'} 
\left(\frac{1}{2}-\delta_{\ell' 1}-\frac{\delta_{\ell' 3}}{10}\right) 
\,\tilde{\Theta}_{{\rm g},\ell'}(\eta, k_1)\, \tilde{\Theta}_{{\rm g},\ell'-1}(\eta, k_2)\,
\mathcal{P}_{1 \ell'\,\ell'-1}\left(\chi_1,\chi_2\right)
\nonumber\\
&=
-\sum_{\ell'} 
\left(\frac{1}{2}-\frac{\delta_{\ell' 1}}{10}\right) 
\,\tilde{\Theta}_{{\rm g},\ell'}(\eta, k_1)\, \tilde{\Theta}_{{\rm g},\ell'+1}(\eta, k_2)\,
\mathcal{P}_{1 \ell'\,\ell'+1}\left(\chi_1,\chi_2\right)
\nonumber\\
&\qquad 
+\sum_{\ell'} 
\left(\frac{1}{2}-\delta_{\ell' 0}-\frac{\delta_{\ell' 3}}{10}\right) 
\,\tilde{\Theta}_{{\rm g},\ell'+1}(\eta, k_1)\, \tilde{\Theta}_{{\rm g},\ell'}(\eta, k_2)\,
\mathcal{P}_{1 \ell'+1\,\ell'}\left(\chi_1,\chi_2\right)
}
%-----------
where we introduced $\chi_i=\hat{\vek{k}}\cdot \hat{\vek{k}}_i$. 
The first few terms up to $\ell=2$ are then
%-----------
\bealf{
\hat{\mathcal{M}}_{1} \left(\eta, k_1, k_2, \chi_1,\chi_2\right) 
&\approx 
-\frac{1}{2} 
\,\tilde{\Theta}_{{\rm g},0}(\eta, k_1)\, \tilde{\Theta}_{{\rm g},1}(\eta, k_2)\,\chi_2
-\frac{2}{5}
\,\tilde{\Theta}_{{\rm g},1}(\eta, k_1)\, \tilde{\Theta}_{{\rm g},2}(\eta, k_2)\,\left[3\chi_2 \chi_{12}-\chi_1\right]
\\ \nonumber
&\qquad - \frac{1}{2} 
\, \tilde{\Theta}_{{\rm g},0}(\eta, k_2)\,\tilde{\Theta}_{{\rm g},1}(\eta, k_1)\,\chi_1
+\frac{1}{2}
\,\tilde{\Theta}_{{\rm g},1}(\eta, k_2)\, \tilde{\Theta}_{{\rm g},2}(\eta, k_1)\,\left[3\chi_1 \chi_{12}-\chi_2\right],
}
%-----------
where we used $\chi_{12}=\hat{\vek{k}}_1\cdot \hat{\vek{k}}_2$. 

\subsubsection{Source for $\ell=2$ and $\ell=3$.}
%-----------
For $\ell=2$ and $\ell=3$, we only give the final result including first order transfer functions up to $\ell=2$. These yield
%-----------
\bsub
\bealf{
\hat{\mathcal{M}}_{2} \left(\eta, k_1, k_2, \chi_1,\chi_2\right) 
&\approx 
\frac{9}{20} 
\left[\tilde{\Theta}_{{\rm g},0}(\eta, k_1)\, \tilde{\Theta}_{{\rm g},2}(\eta, k_2)\,P_2(\chi_2)-\tilde{\Theta}_{{\rm g},0}(\eta, k_2)\, \tilde{\Theta}_{{\rm g},2}(\eta, k_1)\,P_2(\chi_1)\right]
\nonumber\\
&\qquad -\frac{33}{100}\tilde{\Theta}_{{\rm g},1}(\eta, k_1)\, \tilde{\Theta}_{{\rm g},1}(\eta, k_2)\left[3\chi_1 \chi_{2}-\chi_{12}\right]
\nonumber\\
&\qquad\qquad -\frac{9}{28}\tilde{\Theta}_{{\rm g},2}(\eta, k_1)\, \tilde{\Theta}_{{\rm g},2}(\eta, k_2)\left[3\chi_1 \chi_{2}\chi_{12}-1\right]
\\[2mm]
\hat{\mathcal{M}}_{3} \left(\eta, k_1, k_2, \chi_1,\chi_2\right) 
&\approx 
\frac{9}{35} 
\tilde{\Theta}_{{\rm g},1}(\eta, k_1)\, \tilde{\Theta}_{{\rm g},2}(\eta, k_2)\,\left[5\chi_1\chi_2^2-2\chi_2\chi_{12}-\chi_1\right]
\nonumber\\
&\qquad -\frac{9}{28} 
\tilde{\Theta}_{{\rm g},1}(\eta, k_2)\, \tilde{\Theta}_{{\rm g},2}(\eta, k_1)\,\left[5\chi_2\chi_1^2-2\chi_1\chi_{12}-\chi_2\right]
}
\esub
%-----------
For $\ell>3$, no sources appear when we only include first order transfer functions with $\ell \leq 2$. We note that the geometric terms relating to the angles between the mode functions clearly alter the results, however, overall these contributions do not affect the final results as strongly.

\subsection{Comparison to previous works}
%-----------
In previous works, the spatially varying heating rates for $\ell=0$ \citep{Chluba2017muT, Ota2017} and $\ell=1, 2$ \citep{Ota2017} have been obtained neglecting the additional modulation from the angle between the dissipating modes. A derivation following these previous treatments is given in Appendix~\ref{app:old_sources}, with the main results given by Eq.~\eqref{eq:local_dipole_dis_source_I}, Eq.~\eqref{eq:local_dipole_dis_source_IIII} and Eq.~\eqref{eq:local_octupole_dis_source}. These expressions are equivalent to those for $\hat{\mathcal{M}}_{\ell}$ when setting 
$k_1=k_2$, $\chi_1=1$ and $\chi_2=-1$, which also implies $\chi_{12}=-1$. However, for detailed computations the full angular dependence has to be taken into account as otherwise the $k$-dependence is incorrectly modeled. This could be especially important for general primordial bispectra.

\section{Computations of the effective average heating rates}
\label{sec:av_heating}
%-----------
The dissipation of small scale acoustic modes on the average CMB spectrum can be modeled as an effective heating rate \citep{Chluba2012}. This is because the local $y$-distortion from the mixing of blackbody spectra directly causes the electrons to up-scatter and thus automatically transfer the energy back to the photon field. The effective average heating rate then takes the form
%-----------
\beal
\label{eq:Q_dot}
\frac{\id\langle\mathcal{Q}\rangle}{\id \ln z}
=4\left<\mathcal{S}^{\rm mix}_0\right>\,\frac{\id \tau}{\id \ln z},
\end{align}
%-----------
where $\left<\mathcal{S}^{\rm mix}_0\right>$ is given by Eq.~\eqref{eq:local_monopole_dis_source_av}, $\tau$ is the Thomson optical depth, and the factor of $4$ is from the normalization of the $y$ spectrum. For adiabatic Gaussian fluctuations, the average distortion signals have been considered in \citep{Chluba2012} and \citep{Chluba2012inflaton} for various models of the small-scale power spectrum. Importantly, the scale-dependence of the small scale power spectrum is directly related to the time-dependence of the effective energy injection rate, with quasi-scale invariant small-scale power spectra yielding $\frac{\id\langle\mathcal{Q}\rangle}{\id \ln z}\approx {\rm const}$ in the pre-recombination era \citep{Chluba2011therm,
Chluba2012}, causing $\bar{\mu}\approx \pot{2}{-8}$ \citep{Chluba2012, Chluba2016}.

In the presence of non-Gaussian fluctuations, another contribution to the average distortion can arise according to Eq.~\eqref{eq:local_monopole_dis_source_NG}. To evaluate this contribution, we have to compute the effective power spectrum, $\Delta^{\rm NL}(k)$, which at large scales exhibits a scaling, $\Delta^{\rm NL}(k)\propto k^3$, as the integral becomes independent of $k\ll 1$. 
As already pointed out, we have to assume a truncation in scales to evaluate this term. 
Let us assume that the relevant small-scale power spectrum is given by 
$\Delta^{\rm G}(k)=A_{\rm p} (k/k_{\rm p})^{\nS-1}$ at $k_{\rm p}\leq k \leq k_{\rm c}$ and zero otherwise. We can then evaluate the inner integral in Eq.~\eqref{eq:local_monopole_dis_source_NG} as
%-----------
\beal
\label{eq:inner_int}
\mathcal{I}^{\rm NL}(k, q)
&=\int_{|q-k|}^{q+k}
\Delta^{\rm G}(\tilde{k})\,
\frac{\!\id \tilde{k}}{\tilde{k}^2} 
=\frac{A_{\rm p}\left\{
\min\left[k_{\rm c},q+k\right]^{\nS-2}-\min\left[k_{\rm c},\max\left[k_{\rm p},|q-k|\right]\right]^{\nS-2}
\right\}}{(\nS-2)\,k_{\rm p}^{\nS-1}} 
\end{align}
%-----------
To make progress, it is best to transform to variables $\xi=k/k_{\rm p}$, $\xi_{\rm c}=k_{\rm c}/k_{\rm p}$ and $\xi_1=q/k_{\rm p}$. The final integral then reads
%-----------
\beal
\label{eq:outer_int}
\Delta^{\rm NL}(k)
&=\frac{A^2_{\rm p}\,\xi^2}{2(\nS-2)}\int_{1}^{\xi_{\rm c}}
\xi_1^{\nS-3}  
\left\{
\min\left[\xi_{\rm c},\xi_1+\xi\right]^{\nS-2}-\min\left[\xi_{\rm c},\max\left[1,|\xi_1-\xi|\right]\right]^{\nS-2}
\right\} \id \xi_1.
%\nonumber\\
%&=\frac{A^2_{\rm p}\,\xi^2}{2(\nS-2)}\int_{1-\xi}^{\xi_{\rm c}-\xi}
%(\tilde{\xi}+\xi)^{\nS-3}  
%\left\{
%\min\left[\xi_{\rm c},\tilde{\xi}+2\xi\right]^{\nS-2}-\min\left[\xi_{\rm c},\max\left[1,|\tilde{\xi}|\right]\right]^{\nS-2}
%\right\} \id \tilde{\xi}.
\end{align}
%-----------
For general $\nS$, the solution can be written in terms of the incomplete Euler Beta function. However, given the many conditions, numerical evaluations of the remaining integral is a lot easier. We note that $\Delta^{\rm NL}(k)=0$ at $\xi>2\xi_{\rm c}$. We also find that rather independently of $\nS$ at $\xi\lesssim 1$, we have $\Delta^{\rm NL}(k)\propto A^2_{\rm p}\,\xi^3$ and $\Delta^{\rm NL}(k)\simeq A^2_{\rm p}\,\sqrt{\xi}$ at $1\lesssim \xi\lesssim \xi_{\rm c}$ (in particular for $\nS\simeq 1$). This means that the effect on the monopole is bounded by $\Delta^{\rm NL}(k)\lesssim A^2_{\rm p}\,\sqrt{\xi}$. From Eq.~\eqref{eq:local_monopole_dis_source_NG}, we then estimate the average $\mu$ contribution to be less than\footnote{A factor order unity to a few should be inserted in here, given that the integral is over $A^2_{\rm p} \xi^{1/2}$ rather than $A^2_{\rm p}$.} 
%-----------
\beal
\label{eq:local_monopole_dis_source_NG_estimate}
\left<\mu^{\rm NL}\right>
&\lesssim 
\frac{18}{25}\,f_{\rm NL}^2 A_{\rm p} \left<\mu^{\rm G}\right>\simeq \pot{3.8}{-13}\,f_{\rm NL}^2 \,\left[\frac{A_{\rm p}}{\pot{2.4}{-9}}\right]^2
%\times \left[\frac{\left<\mu^{\rm G}\right>}{\pot{2}{-8}}\right] 
\times \left[\frac{\left<\mu^{\rm FIRAS}\right>}{\pot{9.0}{-5}}\right]^{-1}\,\left<\mu^{\rm FIRAS}\right>.
\end{align}
%-----------
This means that only for $f_{\rm NL}\gtrsim (10^5-10^6)\times\left[\frac{A_{\rm p}}{\pot{2.4}{-9}}\right]^{-1}$ can one expect the average contribution from the dissipation of the non-Gaussian fluctuations to become comparable to the average distortion limit of \COBEF in the considered model. Evidently, the exact model for the small-scale non-Gaussianity will affect this statement, however, for our computations we often assume that the contribution to the average distortions can be neglected. We also note that using Eq.~\eqref{eq:Delta_NL_delta} does not provide a good estimate for the non-Gaussian contribution to the average distortion. In addition, $f_{\rm NL}\gtrsim (10^5-10^6)$ already implies non-perturbative small-scale fluctuations, which we avoid.

\newpage

\section{Computations of the effective anisotropic heating rates}
\label{sec:computation_heating}
%-----------
We are now in the position to compute the anisotropic effective heating rates. This step has to be done before the evaluation of the distortion anisotropies, as otherwise the computation becomes prohibitive. Using the bispectrum parametrization as in Eq.~\eqref{eq:B_def} we  can also write
%------------
\bealf{
\label{eq:B_def_P}
\frac{B(k, k_1,k_2)}{P(k)}=\frac{6}{5}\,f_{\rm NL}P(k_1)\left[1+\frac{P(k_2)}{P(k)}+\frac{P(k_2)}{P(k_1)}\right].
}
%------------
Inserting this into the definition of the source terms, $s_{\ell m}$, in Eq.~\eqref{eq:source_ell_m_kz_gen}, we then obtain
%------------
\beal
\label{eq:source_ell_m_kz_gen_rewrite}
s_{\ell 0}(\eta, k)=\frac{12}{5}\,f_{\rm NL} \int \Delta^{\rm G}(k_1) \id \ln k_1 
\int_{|k_1-k|}^{k_1+k}\,\frac{k_2 \id k_2}{4 k k_1 }
\,
\left[1+\frac{P(k_2)}{P(k)}+\frac{P(k_2)}{P(k_1)}\right]
\hat{\mathcal{M}}_{\ell}\left(\eta, k_1, k_2, \chi_1,\chi_2\right),
\end{align}
%-----------
where $\chi_i=\chi_i(k, k_1, k_2)$. The strategy is to first build a transfer function database for the standard perturbation variables at first order in perturbations. With this database one can then numerically compute the effective heating rates and tabulate these as a function of redshift and $k$-values. 

\subsection{Transfer function database setup} 
\label{sec:transfer_database}
%-----------
To setup the transfer function  database, we use about 6000 points in $k$ in the range $10^{-4}-10^{2}\,\Mpc^{-1}$. The grid is split into 150 log-space sub-intervals for which the rough ending redshift is estimated using the damping scale. This allows us to stop the calculation of the transfer function once Silk damping has erased anisotropies. Within the sub-intervals, we use a linear grid in $k$ with a density that is based on the value of $\Delta k \,\eta$ at the ending redshift, where $\Delta k=k_{\rm up}-k_{\rm low}$ across the interval. This adds resolution at small scales and late times. 

We compute the transfer functions using $\ell_{\rm max}=10$ in the photon and neutrino hierarchy, including polarization effects for the transfer functions. 
Adiabatic initial conditions are set at $z=10^{11}$ and the runs are maximally carried out until $z_{\rm end}=10^2$. We then store all the transfer function outputs for the gauge-independent variables $\Theta_{\rm g, \ell}=\Theta_{\ell}+\delta_{\ell 0} \Psi-\delta_{\ell 1} V_1$ between $z=10^2$ and $\pot{5}{6}$, as we expect no significant distortions from outside this interval. 
Polarization terms are neglected in our source term. This will affect the heating rates around the recombination era by no more than $5-10\%$ \citep{Chluba2012, Chluba2015}.
If the cosmology changes, this process has to be repeated from scratch.

To add the contributions from $k\geq 10^{2}\,\Mpc^{-1}$, we can use analytic approximations assuming tight coupling. We confirmed that this works very well without significant loss of precision. For these, we have \citep{Hu1995CMBanalytic, Hu1996anasmall, Chluba2013iso}
%-----------
\bsub
\beal
\label{eq:Transfer_funcs}
\hat{\Theta}^{(1)}_0
&\approx \frac{1}{(1+R_{\rm b})^{1/4}}\left[A \cos(k r_{\rm s})+B \sin(k r_{\rm s})\right]\,\expf{-k^2/k_{\rm D}^2}
\\
\hat{\Theta}^{(1)}_1
&\approx \frac{c_{\rm s}}{(1+R_{\rm b})^{1/4}}\left[A \sin(k r_{\rm s})- B \cos(k r_{\rm s})\right]\,\expf{-k^2/k_{\rm D}^2}
\\
\hat{\Theta}^{(1)}_2
&\approx \frac{2}{5f_2}\,\frac{k}{\tau'}\,\hat{\Theta}^{(1)}_1 \quad\text{and}\quad \hat{\Theta}^{(1)}_{\ell>2}\approx 0
\\
\frac{1}{k_{\rm D}^2}&=\int \frac{\id \eta}{6(1+R_{\rm b})\,\tau'}
\left[\frac{R_{\rm b}^2}{1+R_{\rm b}}+\frac{4}{5 f_2}\right]
\end{align}
\esub
%-----------
with $f_2=3/4$ when polarization terms are included and $f_2=9/10$ when they are omitted. We furthermore introduced the baryon loading, $R_{\rm b}$, the sound speed $c_{\rm s}=c/\sqrt{3(1+R_{\rm b})}$, the sound horizon, $r_{\rm s}=\int c_{\rm s}\id \eta$ and Thomson scattering optical depth derivative, $\tau'=\Ne \sigT c a$.

For adiabatic initial conditions, one has $A=(1+4R_\nu/15)^{-1}\approx 0.90$ and $B\approx 0$ with the neutrino fraction $R_\nu\approx 0.41$. In the tight coupling regime, only the quadrupole terms contribute to the heating rates. We find that with the above definitions a slightly modified approximation comes very close to the full numerical results for the average heating rate at $z\gtrsim 2000$:
%-----------
\bsub
\beal
\label{eq:Transfer_funcs_ell2}
\hat{\Theta}^{(1), {\rm m}}_2
%&\approx \frac{2}{5f_2}\,\frac{k}{\tau'}\frac{c_{\rm s}}{(1+R_{\rm b})^{0.3}}\frac{0.9976}{1+\frac{4}{15}\,R_\nu}\,\sin(k r_{\rm s})\,\expf{-k^2/k_{\rm D}^2}
&\approx \frac{2}{5f_2}\,\frac{k}{\tau'}\,\hat{\Theta}^{(1)}_1
\frac{0.99755}{(1+R_{\rm b})^{0.05}}.
\end{align}
\esub
%-----------
The same expression can also roughly reproduce the anisotropic heating rates for the monopole terms at $z\gtrsim 2000$. We therefore use this expression for all evaluations at $k>10^2\,\Mpc^{-1}$.

\vspace{-1.5mm}
\subsection{Effective anisotropic heating rate database}
\label{sec:effect_HR_setup}
%-----------
To perform the heating integrals, Eq.~\eqref{eq:source_ell_m_kz_gen_rewrite}, we use Clenshaw-Curtis quadrature rules for the integral over $k_1$ and a Patterson scheme for the one over $k_2$ developed for {\tt CosmoRec} \citep{Chluba2010, Chluba2010b}. The outer integral is performed in log-space over sub-intervals, while the inner is carried out in linear coordinates. For the computation of the spectral distortion anisotropies using the line of sight approach, we need the heating rates for modes with $k=10^{-4}-1\,\Mpc^{-1}$. These are pre-tabulated on a log-spaced grid in $k$ as a function of $z$ and then interpolated. If the scale-dependence of the bispectrum changes, this process has to be repeated from scratch; however, the dependence on the overall normalization $\propto \fNL$ can be captured by simply rescaling.

\vspace{-1.5mm}
\subsubsection{Approximations for the effective heating rates}
%-----------
The sources of distortion anisotropies at large angular scales and $z>10^5$ are caused by dissipation of modes at very small scales, $k>10^2 \, \Mpc^{-1}$ \citep[see Fig.~5 of][]{Chluba2012}.
In this regime, in Eq.~\eqref{eq:source_ell_m_kz_gen_rewrite} we can assume that most of the integral contributions come from $k\ll k_1 \wedge k_2$.  The range of variations of $k_2$ is then very small, $|k_1-k|\leq k_2 \leq k+k_1$, essentially leaving only $k_2 \approx k_1$ and $\chi_{12}\rightarrow 1$. However, at the same time the full range in $\chi_1$ and $\chi_2$ is covered. From Eq.~\eqref{eq:B_def_P}, we then have
%------------
\bealf{
\label{eq:B_def_P_approx}
\frac{B(k, k_1,k_2)}{P(k)}\approx \frac{12}{5}\,f_{\rm NL}\,P(k_1)
}
%------------
when we assume $k_1\approx k_2$ and $k_2\gg k$ (such that $P(k_2)/P(k)\ll 1$). Using this in Eq.~\eqref{eq:source_ell_m_kz_gen_rewrite} we find
%------------
\bsub
\label{eq:source_ell_m_kz_gen_approx}
\bealf{
s^{\rm L}_{\ell 0}(\eta, k)&\approx \frac{12}{5}\,f_{\rm NL} \int 
\id \ln k_1 \,\Delta^{\rm G}(k_1)\,\bar{\mathcal{M}}_{\ell}\left(\eta, k_1\right),
\\
\bar{\mathcal{M}}_{\ell}\left(\eta, k_1\right)&=\int_{|k_1-k|}^{k_1+k}\,\frac{k_2 \id k_2}{2 k k_1}\,
\hat{\mathcal{M}}_{\ell}\left(\eta, k_1, k_1, \chi_1,\chi_2\right)
}
\esub
%-----------
at long wavelength (hence superscript 'L'), where we introduced the angle average source function, $\bar{\mathcal{M}}_{\ell}\left(\eta, k_1\right)$. By evaluating the integral over $k_2$, we find the following approximate results
%------------
\bealf{
\label{eq:M_av_approx}
\hat{\mathcal{M}}_{0}\left(\eta, k_1\right)
&\approx 
3\tilde{\Theta}^2_{1,\rm g}(k_1)
+\frac{9}{2}\tilde{\Theta}^2_2(k_1)
+\sum_{\ell\geq3}(2\ell+1) \tilde{\Theta}^2_\ell(k_1)
-\frac{k^2}{k^2_1}\left[\tilde{\Theta}^2_{1,\rm g}(k_1)
+\frac{9}{2}\tilde{\Theta}^2_2(k_1)\right],
\nonumber\\
\hat{\mathcal{M}}_{1}\left(\eta, k_1\right)
&\approx 
\frac{k}{k_1}\left[\frac{1}{3}\tilde{\Theta}_{0,\rm g}(k_1)\,\tilde{\Theta}_{1,\rm g}(k_1)
+\frac{17}{15}\tilde{\Theta}_{1,\rm g}(k_1)\,\tilde{\Theta}_2(k_1)\right],
\nonumber\\
\hat{\mathcal{M}}_{2}\left(\eta, k_1\right)
&\approx 
\frac{k^2}{k^2_1}\left[\frac{11}{125}\tilde{\Theta}^2_{1,\rm g}(k_1)
+\frac{9}{100}\tilde{\Theta}_{0,\rm g}(k_1)\,\tilde{\Theta}_2(k_1)
+\frac{9}{35}\tilde{\Theta}^2_2(k_1)\right]\, , 
}
%-----------
where we give terms up to second order in $k/k_1 \ll 1$. For $\ell=0$, at leading order we kept all terms for $\ell\geq 3$, although we only include $\ell=1$ and $\ell=2$ for the those at order $k^2/k_1^2$.

%------------
\begin{comment}
%------------
\bealf{
\label{eq:M_av_approx}
\hat{\mathcal{M}}_{0}\left(\eta, k_1\right)
&\approx 
2\tilde{\Theta}^2_{1,\rm g}(k_1)
+\frac{9}{5}\tilde{\Theta}^2_2(k_1)
+\frac{k^2}{k^2_1}\left[\frac{2}{5}\tilde{\Theta}^2_{1,\rm g}(k_1)
+\frac{9}{35}\tilde{\Theta}^2_2(k_1)\right],
\nonumber\\
\hat{\mathcal{M}}_{1}\left(\eta, k_1\right)
&\approx 
-\frac{k}{k_1}\left[\frac{2}{3}\tilde{\Theta}_{0,\rm g}(k_1)\,\tilde{\Theta}_{1,\rm g}(k_1)
+\frac{14}{75}\tilde{\Theta}_{1,\rm g}(k_1)\,\tilde{\Theta}_2(k_1)\right],
\nonumber\\
\hat{\mathcal{M}}_{2}\left(\eta, k_1\right)
&\approx 
\frac{22}{25}\tilde{\Theta}^2_{1,\rm g}(k_1)
+\frac{18}{35}\tilde{\Theta}^2_2(k_1)
-\frac{k^2}{k^2_1}\left[\frac{44}{125}\tilde{\Theta}^2_{1,\rm g}(k_1)
-\frac{9}{50}\tilde{\Theta}_{0,\rm g}(k_1)\,\tilde{\Theta}_2(k_1)
+\frac{36}{245}\tilde{\Theta}^2_2(k_1)\right],
\nonumber\\
\hat{\mathcal{M}}_{2}\left(\eta, k_1\right)
&\approx 
-\frac{12}{25}\,\frac{k}{k_1}\,
\tilde{\Theta}_{1,\rm g}(k_1)\,\tilde{\Theta}_2(k_1).
}
%-----------
\end{comment}
In the tight coupling regime, we have $\tilde{\Theta}_{1,\rm g}(k_1)\approx 0$. This then implies that the only non-vanishing sources are
%------------
\bealf{
\label{eq:M_av_approx_TC}
\hat{\mathcal{M}}_{0}\left(\eta, k_1\right)
&\approx 
\frac{9}{2}\tilde{\Theta}^2_2(k_1)\times \frac{5}{8}\left\{1-\frac{3}{5\xi_1^2}-\frac{3}{10}\frac{(\xi_1^2-1)^2}{\xi_1^3}\log\left[\frac{\xi_1-1}{\xi_1+1}\right] \right\},
\nonumber\\
\hat{\mathcal{M}}_{2}\left(\eta, k_1\right)
&\approx 
\frac{9}{35}\,\frac{k^2}{k^2_1}\,\tilde{\Theta}^2_2(k_1)
\times \frac{15}{32}\left\{1-\frac{2}{3}\xi_1^2+\xi_1^4+\frac{(\xi_1^2-1)^2(\xi_1^2+1)}{2\xi_1}\,\log\left[\frac{\xi_1-1}{\xi_1+1}\right] \right\},
}
%-----------
\begin{comment}
%------------
\bealf{
\label{eq:M_av_approx}
\hat{\mathcal{M}}_{0}\left(\eta, k_1\right)
&\approx 
\frac{9}{5}\tilde{\Theta}^2_2(k_1)\times \frac{25}{8}\xi_1\left\{\xi_1-\frac{3}{5}\xi_1^3-\frac{3}{10}(\xi_1^2-1)^2\,\log\left[\frac{\xi_1+1}{\xi_1-1}\right] \right\},
\nonumber\\
\hat{\mathcal{M}}_{2}\left(\eta, k_1\right)
&\approx 
\frac{18}{35}\tilde{\Theta}^2_2(k_1)
\times \frac{15}{32}\left\{1-\frac{2}{3}\xi_1^2+\xi_1^4+\frac{(\xi_1^2-1)^2(\xi_1^2+1)}{2\xi_1}\,\log\left[\frac{\xi_1+1}{\xi_1-1}\right] \right\},
}
%-----------
\end{comment}
where we gave the full scaling with $\xi_1=k_1/k\geq 1$. In the limit $\xi_1\rightarrow \infty$ the $\xi_1$-dependent functions become close to unity. Since for the average source of distortions we have
%-----------
\beal
\label{eq:local_monopole_dis_source_av_approx}
\left<\mathcal{S}^{\rm mix}_0\right>
&\approx 
\frac{9}{2}\,\int \id \ln k' \Delta^{\rm G}(k')\,
\tilde{\Theta}^2_2(\eta,k')
\end{align}
%-----------
in the tight coupling regime and for sufficiently small $k$, we then find
%------------
\bealf{
\label{eq:source_ell_m_kz_gen_approx_TC}
s^{\rm L}_{00}(\eta, k)&\approx \frac{12}{5}\,f_{\rm NL}
\left<\mathcal{S}^{\rm mix}_0\right>,
\qquad 
s^{\rm L}_{20}(\eta, k)\approx 
\frac{108}{175}\,f_{\rm NL} \,
\int \id \ln k' \Delta^{\rm G}(k')\,\frac{k^2}{{k'}^2} \tilde{\Theta}^2_2(\eta,k') \rightarrow 0
% 
%s^{\rm L}_{20}(\eta, k)\approx \frac{48}{175}\,f_{\rm NL}
%\, 
%\left<\mathcal{S}^{\rm mix}_0\right>=\frac{2}{7}\,s^{\rm L}_{00}(\eta, k).
}
%-----------
The monopolar source term is roughly scale-independent in the limit we took. In contrast, the quadrupolar source is non-zero and scales $\propto k^2$. However, the extra factor $1/{k'}^2$ in the integrand suppresses contributions from small scales $k' \gg k$, making them sub-dominant.

In the computation of the heating rates, we use the angle average source function according to Eq.~\eqref{eq:M_av_approx} for evaluations in the regime $k/k_1 \leq 10^{-3}$, while we evaluate the full integral over $k_2$ unless stated otherwise. This speeds the computation up significantly while introducing hardly any error.

\subsection{Results for the anisotropic heating rates}
\label{sec:heating_rates}
%------------------------------------
We can cast the problem into computations of effective energy injection rates due to the dissipation process. For convenience we therefore introduce the scaled effective heating rates
%-----------
\beal
\label{eq:Q_ell_dot}
\frac{\id\hat{\mathcal{Q}}_{\ell}}{\id \ln z}=\frac{s_{\ell 0}}{12 \fNL/5}\,\frac{\id \tau}{\id \ln z},
\end{align}
%-----------
which for small $k$ (i.e., large scales) mimics the standard average heating effects and can be directly used in our computations below. We note that this does not include the factor of $4$ since in our computations we actually use direct photon source terms, which do not require this extra factor.

%------------------------------------
\begin{figure}
    \centering
\includegraphics[width=0.82\columnwidth]{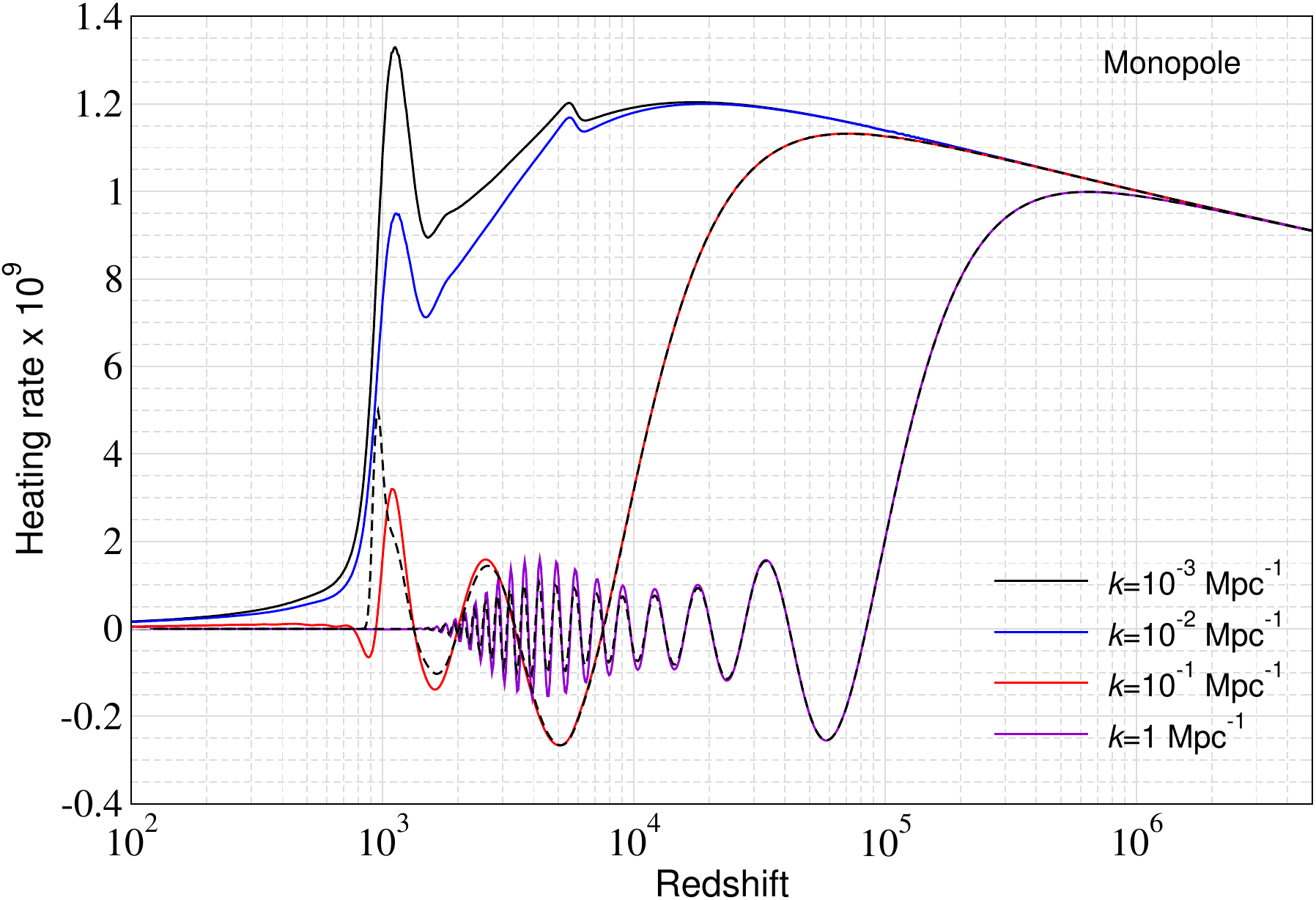}
\\
\caption{Scaled monopole heating rate, $\id\hat{\mathcal{Q}}_{0}/\id \ln z$, for varying values of $k$. At $k\leq 10^{-3}\,\Mpc^{-1}$ the rate does not change significantly, while at smaller scales large variations are visible. The black dashed lines show the results obtained using the tight-coupling approximations for the transfer functions.}
\label{fig:monopole_rate}
\end{figure}
%------------------------------------
In Fig.~\ref{fig:monopole_rate}, we show the effective heating rate for the monopole ($\ell=0$) and varying values of $k$. We computed this source term assuming a Gaussian power spectrum, $\Delta^{\rm G}(k)=A_{\rm S} (k/k_0)^{\nS-1}$, with $A_{\rm S}=\pot{2.4}{-9}$, $\nS=0.96$ and $k_0=0.002\,\Mpc^{-1}$, dividing the result by a factor $\frac{12}{5} \fNL\times 10^{-9}$.
At large scales ($k\geq 10^{-3}\,\Mpc^{-1}$) the effective heating rate becomes $k$-independent and, in terms of the redshift dependence, very similar to the average heating rate for average distortions \citep{Chluba2012}. This is consistent with the discussion around Eq.~\eqref{eq:source_ell_m_kz_gen_approx}. However, at smaller scales, significant scale and time-dependence becomes visible, as already explained in \citep{Chluba2017muT}. Increasing the value of $k$ leads to a suppression of the effective heating around the recombination era ($z\simeq 10^3$). This means that at small scales, most of the heating occurs in the $\mu$-era as is also clear from the fact that modes at these scales damp away at these redshifts. The picture is therefore that at small scales, mostly $\mu$ distortions are sourced, while at large angular scales, both $\mu$ and $y$ distortion anisotropies are generated. Since the overall redshift dependence of the acoustic heating rate is directly linked to the scale-dependence of the power spectrum \citep{Chluba2012inflaton}, the CMB distortion anisotropies thus contain valuable information about the scale-dependence of the Gaussian and non-Gaussian fluctuations.

%------------------------------------
\begin{figure}
    \centering
\includegraphics[width=0.82\columnwidth]{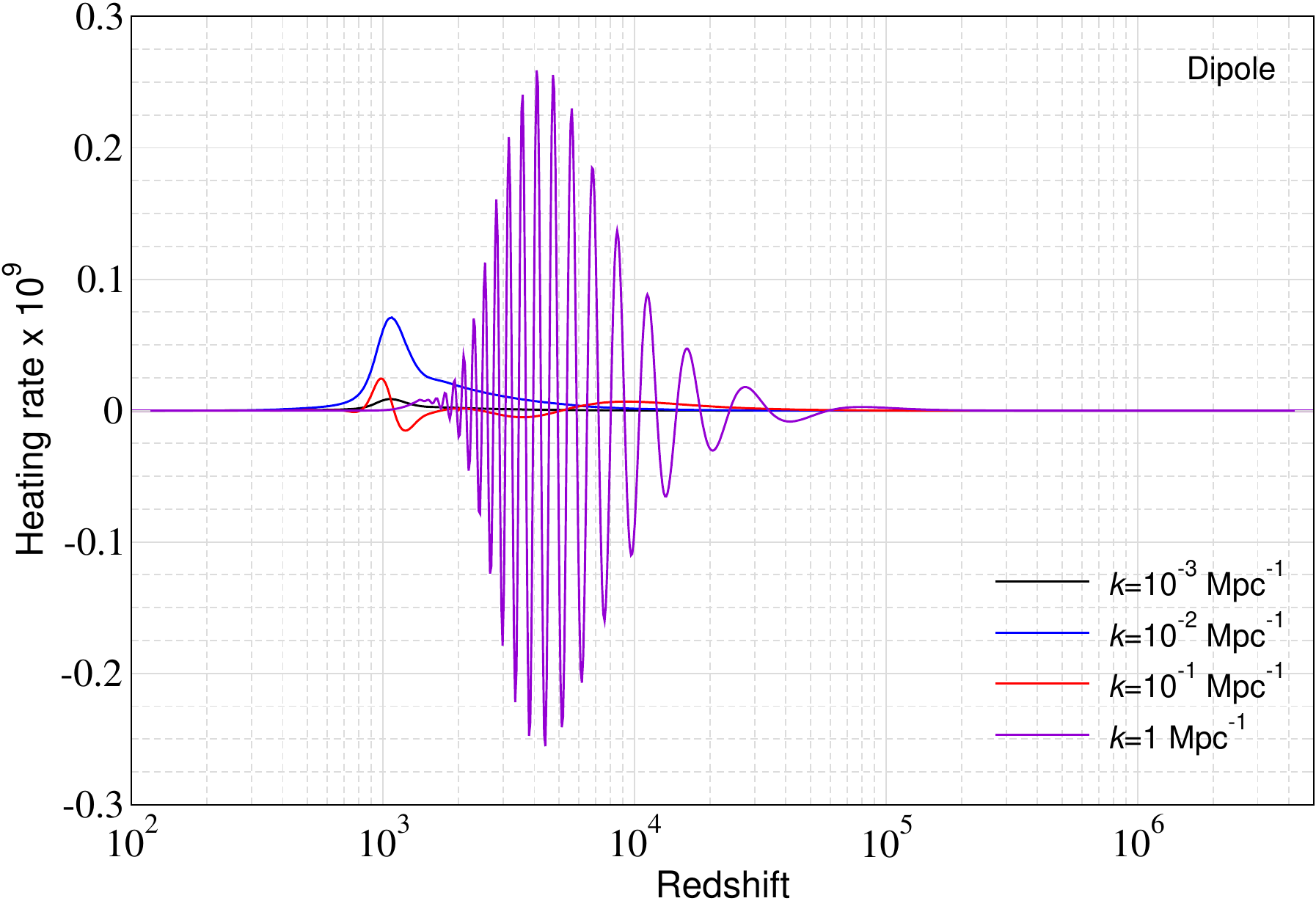}
\\[5mm]
\includegraphics[width=0.82\columnwidth]{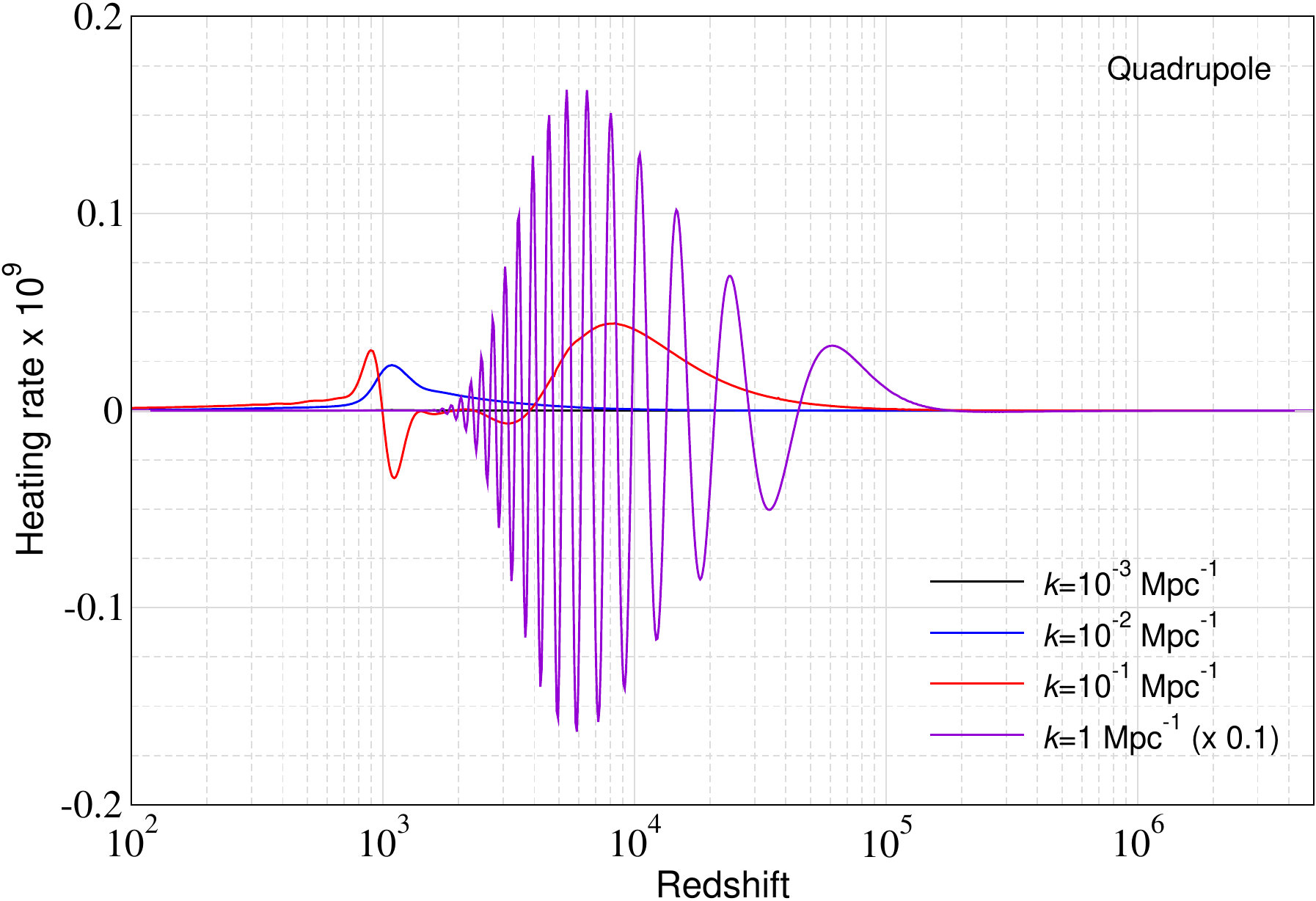}
\\
\caption{Scaled dipolar (top) and quadrupolar (bottom) heating rate, $\id\hat{\mathcal{Q}}_{\ell}/\id \ln z$, for varying values of $k$. At large scales, only small contributions around recombination are noticeable, while at small scales baryon slip starts becoming noticeable. Note that the quadrupole heating rate for $k=1\,\Mpc^{-1}$ was scaled by $1/10$.}
\label{fig:dipole_rate}
\end{figure}
%------------------------------------

As examples, Fig.~\ref{fig:monopole_rate} also illustrates the result for the heating rates of modes with $k=0.1$ and $1\,\Mpc^{-1}$ when applying the analytic tight-coupling solutions for the transfer functions instead of our numerical database. The results agree very well down to $z\simeq 5000-10^4$ in these cases, confirming the robustness of the approximations. Even if not explicitly shown in the figure, this statement is also true for the larger scale modes, suggesting that for $\mu$-distortion anisotropy sourcing one can simplify the computations significantly. We indeed demonstrate this in section~\ref{sec:models}.

In Fig.~\ref{fig:dipole_rate} we show the corresponding effective heating rates for $\ell=1$ and $2$. The contributions are significantly suppressed relative to the monopole contribution, In addition, in the $\mu$-era, when tight-coupling is extremely strong, very small contributions are expected. The biggest contributions for large scale modes come from around the recombination era, while some noticeable contributions in the pre-recombination arise at small scales, $k>0.1\,\Mpc^{-1}$. However, given the rapid oscillations and damping of the source function amplitude towards the recombination period, the net distortion is not expected to be as important. Again, we will illustrate these aspects in section~\ref{sec:models}.

%------------------------------------
\section{Distortion cross power spectra in the presence of PNG}
\label{sec:models}
%------------------------------------
Given the effective heating rates and procedures to compute them for various models, we are now in the position to compute the distortion anisotropies using the FH framework of {\tt CosmoTherm} \citep{chluba_spectro-spatial_2023-I, kite_spectro-spatial_2023-III, Chluba2026}. For this we clarify a few aspects. To cause distortion anisotropies, two primary routes are possible \citep{Chluba2012,kite_spectro-spatial_2023-III, Chluba2026TC}: i) the average distortions source new (Gaussian) distortion anisotropies due to the propagation through the perturbed medium
and ii) anisotropies can be directly created by a local modulation of the distortion source functions.
As we will see below, for a standard (close to scale-independent) power spectrum with local-type PNG (see section~\ref{sec:model-I}) the distortion anisotropies are dominated by ii), while for curvature power spectra with enhanced small-scale power (see section~\ref{sec:model-II}) both i) and ii) become relevant. This leads to a rich phenomenology of distortion anisotropy signals that can be used to distinguish various scenarios. Additional corrections to this picture may arise from higher order effects too \citep[e.g.,][]{Ota2017arXiv-v3, Ota2019}, but are omitted here.

%------------------------------------
\subsection{Distortion signals for standard power spectrum with PNG}
\label{sec:model-I}
%------------------------------------
As mentioned in section~\ref{sec:av_heating}, for a standard (Gaussian) small-scale power spectrum we expect an average distortion at the level $\bar{\mu}\simeq \pot{2}{-8}$. This means that distortion anisotropies from the propagation through the perturbed medium will be at the level $\simeq [10^{-4}-10^{-5}]\,\bar{\mu} \simeq 10^{-12}-10^{-13}$, which is about a factor of $f\simeq 5000$ smaller than what we are currently sensitive to with \Planck through $\mu T$ and $\mu E$ cross correlations \citep{Rotti2022muT}. Therefore, unless the small-scale power spectrum departs from the standard scale-invariant case and has an amplitude around $k\simeq 10^{3}\,\Mpc^{-1}$ (i.e., most relevant to $\mu$-distortions) that is enhanced by a factor $\gtrsim 10^3$ we can neglect this contribution.

In the presence of local type non-Gaussianity, according to Eq.~\eqref{eq:local_monopole_dis_source_NG_estimate} we have an additional average distortion contribution $\lesssim \frac{18}{25}\fNL^2 A_p \, \bar{\mu}$, which means that only for 
%---------------------------
\bealf{
\fNL\gtrsim \sqrt{\frac{25}{18 A_p}}=\pot{2.4}{4}\left[\frac{A_p}{\pot{2.4}{-9}}\right]^{-1/2}
}
%---------------------------
can the non-Gaussian contribution to the average distortion become comparable to the Gaussian contribution. Thus again unless we have 1) an enhanced small-scale power spectrum amplitude or 2) a value of $\fNL>10^5-10^6$ as explained in section~\ref{sec:av_heating}, this remains out of reach of current\footnote{We will use $
\mu T/E$ as shorthand for $\mu T$ and $\mu E$ and similarly for $yT/E$.} $\mu T/E$ constraints.\footnote{We remark that these statements depend on the specific (primordial non-Gaussianity) model.} 
For our discussion below, we can therefore neglect the sourcing of distortion anisotropies caused by distortions to the average spectrum, although we will return to cases with enhanced small scale power later in section~\ref{sec:model-II}.

With these statements in mind, we therefore have to solve a FH system of the form
%------------------------------------------
%---------------------------
\bsub
\label{eq:evol_1_final_hierarchy}
\bealf{
\label{eq:evol_1_final_hierarchy_bg}
\frac{\partial \vek{y}^{(0)}_0}{\partial \eta}
&\approx 0,
\\[1mm]
%-----------------
\frac{\partial \vek{y}^{(1)}_0}{\partial \eta}
&=-k\,\vek{y}^{(1)}_1\!-\!
\frac{\partial \Phi^{(1)}}{\partial \eta}
\vek{e}_G+{\vek{S}'^{(1)}_0}
+\tau' \Thz \Bigg\{M_{\rm T} \vek{y}^{(1)}_0+\left[(\delta^{(1)}_{\rm b} +\Theta^{(1)}_0+ \Psi^{(1)})\, M_{\rm T} +\Theta^{(1)}_0 M_{\rm BK}\right]\vek{y}^{(0)}_0 \Bigg\},
\nonumber \\
%-------------------
\frac{\partial \vek{y}^{(1)}_1}{\partial \eta}
&=k \,
\left(\frac{1}{3}
\vek{y}^{(1)}_{0}-\frac{2}{3}
\vek{y}^{(1)}_{2}\right)
+\frac{k}{3}\Psi^{(1)} \,\vek{e}_G
-\tau'\left[
\vek{y}^{(1)}_1-\frac{\beta^{(1)}}{3}\vek{e}_G\right] + {\vek{S}'^{(1)}_1}
\nonumber \\
%------------------------
\label{eq:evol_1_final_hierarchy_pt}
\frac{\partial \vek{y}^{(1)}_2}{\partial \eta}
&=
k \,
\left(\frac{2}{5}\vek{y}^{(1)}_1-\frac{3}{5}
\vek{y}^{(1)}_3
\right)
-\frac{9}{10}\,\tau'\,\vek{y}^{(1)}_2 +{\vek{S}'^{(1)}_2},
\\ \nonumber
%-------------------------
\frac{\partial \vek{y}^{(1)}_{\ell\geq 3}}{\partial \eta}
&=
k \,
\left(\frac{\ell}{2\ell+1}
\vek{y}^{(1)}_{\ell-1}-
\frac{\ell+1}{2\ell+1}
\vek{y}^{(1)}_{\ell+1}\right)
-\tau'\vek{y}^{(1)}_\ell + {\vek{S}'^{(1)}_\ell},
}
\esub
%---------------------------
where the effective $y$-distortion source term is 
%-----------
\beal
\label{eq:S_ell_dot}
\vek{S}'^{(1)}_\ell=\vek{e}_Y\,\tau'\,s_{\ell 0}.
\end{align}
%-----------
Detailed definitions for all quantities in these equations can be found in \citep{Chluba2026}, but in a few words, $\vek{y}_\ell$ is the distortion vector that describes the CMB spectrum and its distortion at a multipole $\ell$, $\tau'=\id \tau/\id \eta$ is the conformal time derivative of the Thomson scattering optical depths, and the vectors $\vek{e}_G$ and $\vek{e}_Y$ are the temperature and $y$-distortion unit vectors in the spectral basis. We can also recognize the standard Liouville and Thomson scattering terms \citep{Ma1995, BaumannBook}. The remaining terms describe how the spectrum Comptonizes (conversion of $y\rightarrow \mu$) and thermalizes (conversion of $\mu\rightarrow \Theta$) \citep{chluba_spectro-spatial_2023-I, chluba_spectro-spatial_2023-II}.

One important feature of the system in Eq.~\eqref{eq:evol_1_final_hierarchy} is that for the related distortion transfer function, {\it no oscillating waves} are being created \citep{Chluba2017muT, Chluba2026TC}. This is because even in tight-coupling there is no back and forth transfer of energy between the monopole and dipole parts of the distortion variable, which is in stark contrast to the temperature variables, and consistent with \citep{Pajer2012b}. This means that at early times, the sourced modes are overdamped (and thus frozen) until $k$ exceeds the damping scale. As argued in \citep{Chluba2017muT, Chluba2026TC}, this means that the distortions variables behave as 
%-----------
\bsub
\label{eq:y_ell}
\bealf{
\vek{y}_0(\eta_i, \eta, k)&\approx \vek{y}_0(\eta_i, k)\,\expf{-\frac{15}{8} \frac{k^2}{k^2_{\rm D}(\eta_i, \eta)}}
\\
\vek{y}_1(\eta_i, \eta, k)&\approx\frac{k}{3\tau'}\,\vek{y}_0(\eta_i, \eta, k)
}
\esub
%-----------
between $\eta_i$ and $\eta$, where $k^{-2}_{\rm D}(\eta_i, \eta)=k^{-2}_{\rm D}(\eta)-k^{-2}_{\rm D}(\eta_i)$ is directly related to the standard photon damping scale, $k_{\rm D}(\eta)$, and higher order anisotropies are suppressed by additional powers of $\tau'$. Since $\sqrt{8/15}\,k_{\rm D}<k_{\rm D}$, this implies that we expect extra damping of distortion anisotropies.

%------------------------------------
\begin{figure}
\centering
\includegraphics[width=0.85\columnwidth]{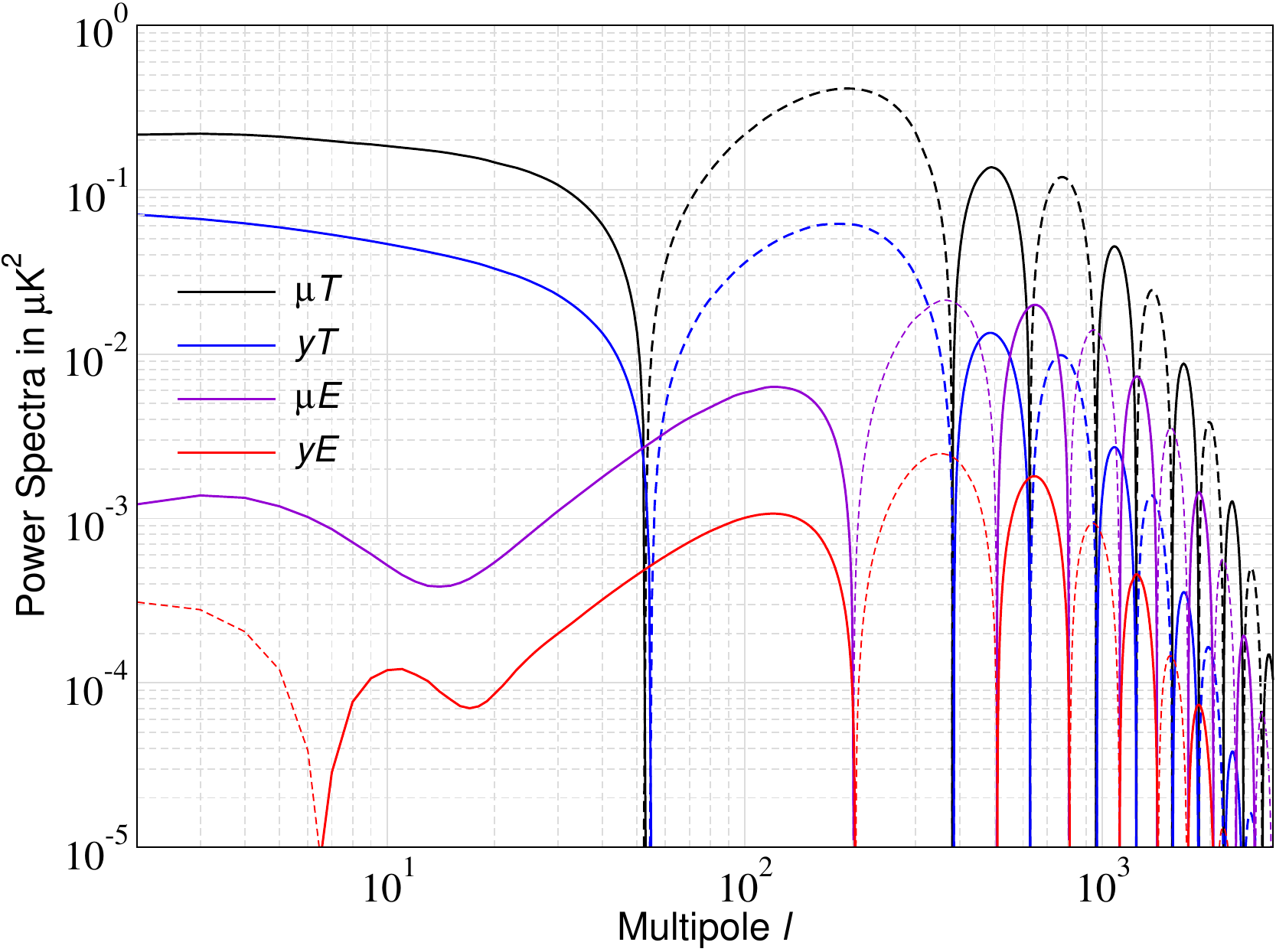}
\\
\caption{Distortion cross power spectra for a quasi-scale invariant small-scale power spectrum with $\fNL=10^3$. The curves were computed using {\tt CosmoTherm} and include the anisotropic source function up to $\ell_{\rm max}=2$. Negative branches are shown as dashed lines.}
\label{fig:PS_all}
\end{figure}
%------------------------------------
With these comments in mind, in figure~\ref{fig:PS_all} we present the results for the distortion cross power spectra for a quasi-scale invariant Gaussian power spectrum with $\fNL=10^3$. We see that the $\mu T$ signal reaches a level that is about four orders of magnitude below the $TT$-power spectrum. Much like the $TE$ power spectrum, we see an alternating pattern between regions of correlation and anti-correlation. The damping envelope of the $\mu T$ power spectrum is well-approximated by $C^{\mu T}_\ell \approx \pot{5}{-4}\,\fNL\expf{-[\ell/630]^{1.35}}\mu{\rm K}^2$, whereas for the standard temperature power spectrum we find $C^{TT}_\ell \approx 8000\,\expf{-[\ell/830]^{1.35}}\mu{\rm K}^2$. The change in the damping scale is indeed roughly consistent with $\sqrt{15/8}\approx 1.36\simeq 830/630$, confirming the analytic expectations. 

The $y T$ power spectrum has a very similar $\ell$-dependence as the $\mu T$ signals, but is about a factor of $\simeq 7$ lower in amplitude around $\ell \simeq 200$. Although at large scales and for a quasi scale invariant Gaussian power spectrum about the same energy is released during the $y$ and $\mu$ eras (e.g., see figure~\ref{fig:monopole_rate}), the differences in the energy normalizations of the $y$ and $\mu$ distortion spectra mostly explain this suppression \citep{chluba_spectro-spatial_2023-I}. 
We also notice that the $y T$ signal damps more rapidly at small scales, as already anticipated in \citep{Chluba2017muT}. This is because in addition to the damping from photon diffusion, at small scales the source functions drop significantly in the $y$-era (see figure~\ref{fig:monopole_rate}). 

The $\mu E$ and $y E$ cross power spectra are further suppressed relative to the $\mu T$ and $y T$ power spectra, with $\pi/2$ phase shift in the oscillatory structure just like for the $TE$ signals. At large scales, we can observe the reionization bump, which for the $y E$ power spectrum has the opposite correlation in comparison to the $\mu E$ at $\ell <7$. However, a detection of this feature is currently very futuristic and will also suffer from cosmic variance uncertainties.

%------------------------------------
\begin{figure}
\centering
\includegraphics[width=0.82\columnwidth]{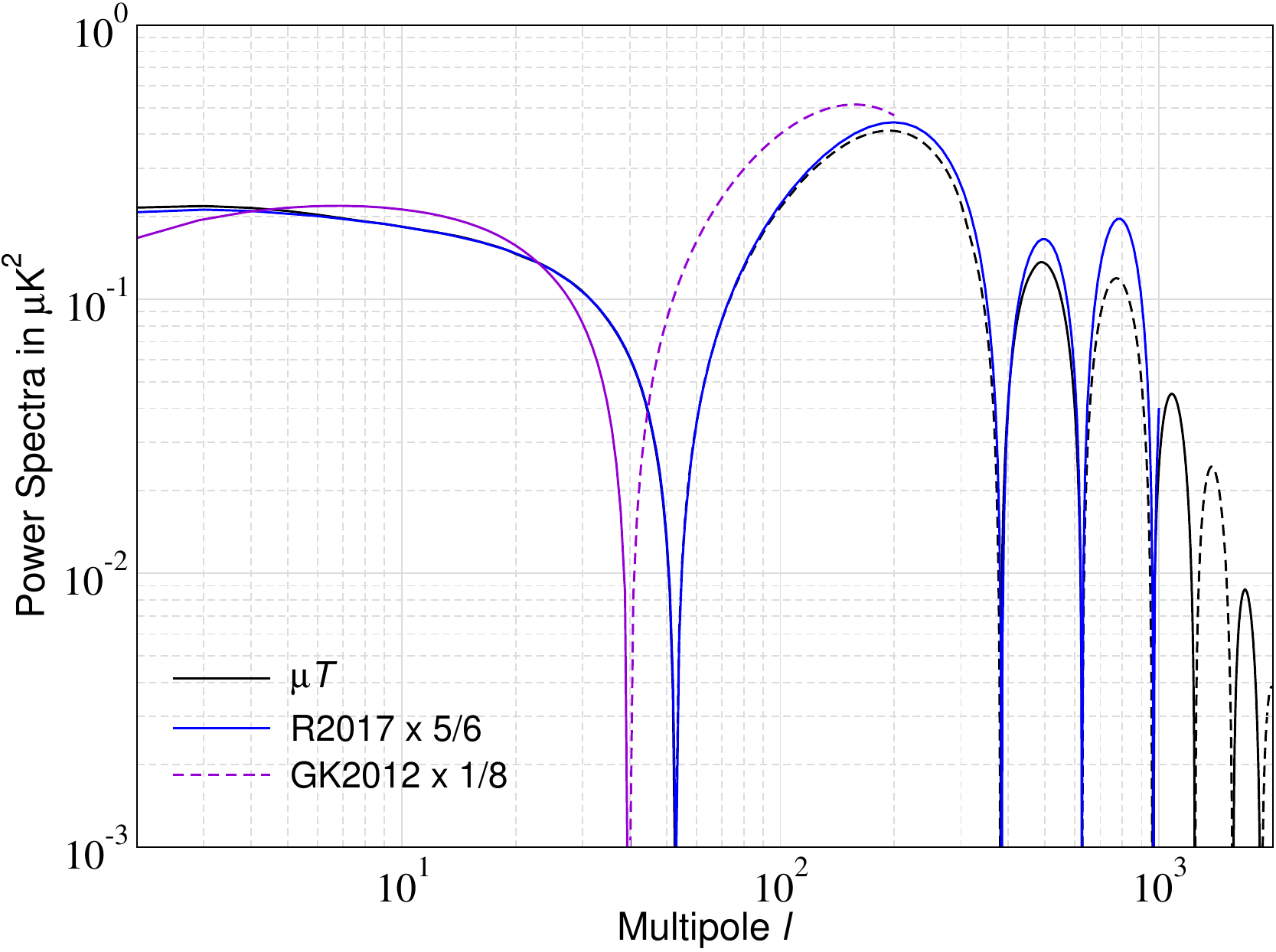}
\\[3mm]
\includegraphics[width=0.82\columnwidth]{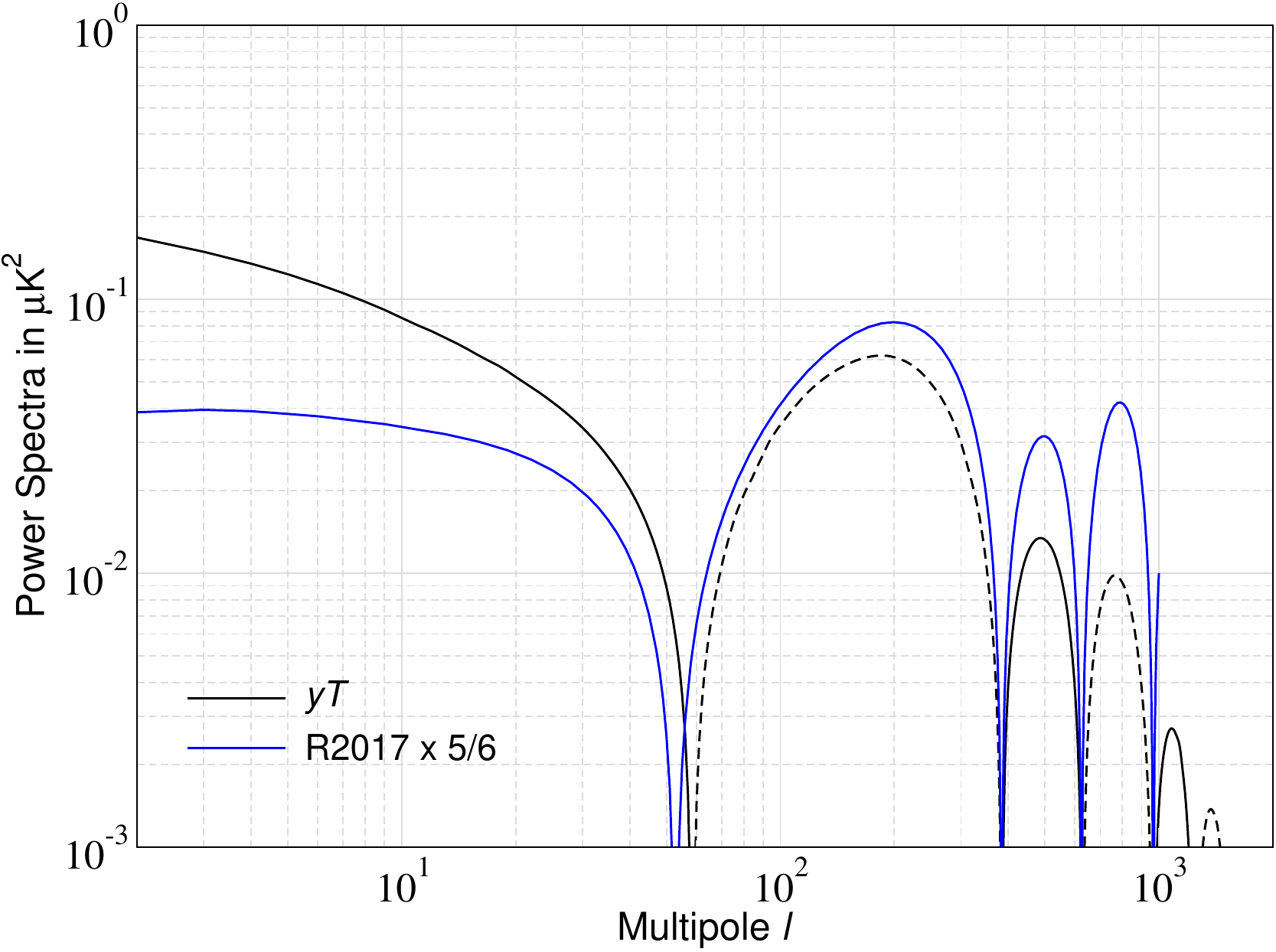}
\\
\caption{Comparison of the $\mu T$ (top) and $y T$ (bottom) cross power spectra for a quasi-scale-invariant Gaussian power spectrum and $\fNL=10^3$ with the results of \citep{Ravenni2017} and \citep{Ganc2012}. The results were obtained using the FH module of {\tt CosmoTherm}. The average $\mu$-distortion is $\bar{\mu}\simeq \pot{2}{-8}$ and leads to negligible additional anisotropies in comparison to the direct sourcing terms. Negative branches are shown as dashed lines for our computations.}
\label{fig:PS_compare}
\end{figure}
%------------------------------------

%------------------------------------
\subsubsection{Comparison to previous calculations}
%------------------------------------
In figure~\ref{fig:PS_compare} we compare our results to those of previous computations. 
Overall we conclude that the broad features of the detailed results obtained using {\tt CosmoTherm} are in good agreement with previous approximate treatments.
The estimate for the $\mu T$ signal given by \citep{Ganc2012} at $\ell \leq 200$ shows clear departures in terms of the $\ell$-dependence, while the difference in amplitude stems from differences in the modeling of the Gaussian part of the average distortion \citep[see][for discussion]{Chluba2017muT}. 
The numerical result for the $\mu T$ signal given by \citep{Ravenni2017} matches our refined calculation very well at $\ell \lesssim 400$, but does not capture the extra damping at small scales. Similarly, we can see that \citep{Ravenni2017} underestimated the $y T$ power spectra at large angular scales ($\ell \lesssim 50$), while extra damping present in the detailed solution causes an overestimation at small angular scales. 

%------------------------------------
\begin{figure}
\centering
\includegraphics[width=0.82\columnwidth]{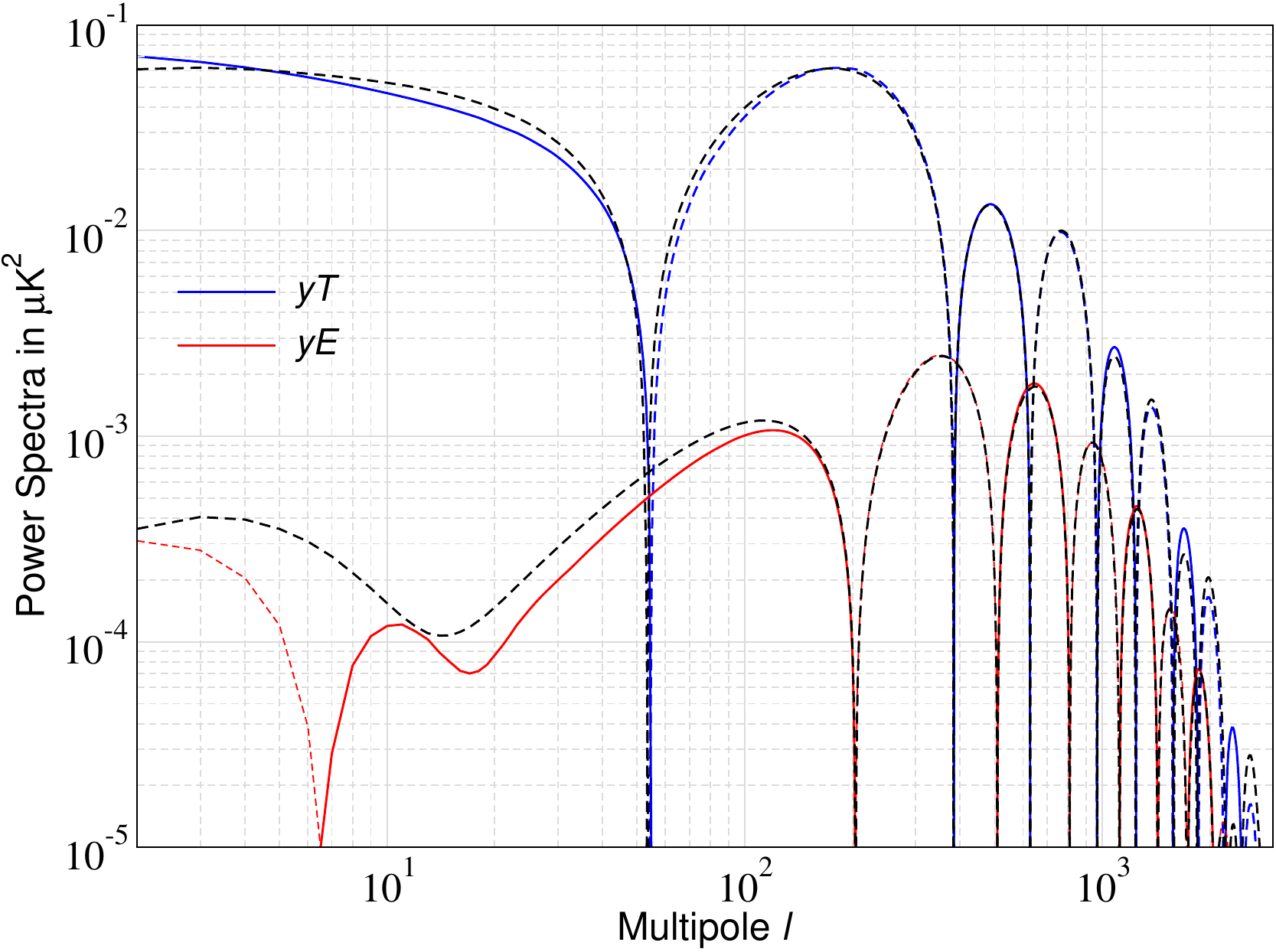}
\\
\caption{Comparison of the detailed prediction for the $y T$ and $y E$ power spectra with those obtained using the tight coupling solutions in the calculation of the heating rates (dashed black lines).}
\label{fig:PS_compare_TC}
\end{figure}
%------------------------------------
\subsubsection{Dependence of the results on physics settings}
\label{sec:physics}
%------------------------------------
We first illustrate how the results change when we use the tight coupling approximation for the transfer functions. For the $\mu T$ and $\mu E$ predictions, we do not expect significant differences given that the sourcing occurs at high redshifts where the approximations work well. On the other hand, for the $yT$ and $y E$ we expect noticeable differences (see figure~\ref{fig:monopole_rate}). For the $\mu T$ and $\mu E$ power spectra we confirmed this statement, finding negligible differences in the results (i.e., not visible by eye). In contrast, for the $y T$ and $y E$ power spectra, the tight coupling approximation for the heating rates modifies the predicted signals noticeably, especially at large ($\ell \lesssim 200$) and very small ($\ell \gtrsim 10^3$) angular scales (see figure~\ref{fig:PS_compare_TC}). Overall, we conclude that $\mu T$ and $\mu E$ signals the tight coupling solutions can be used to accurately predict the effective source terms, while for the corresponding $y$-power spectra small departures may be present. This eases the calculations significantly, since the heating rate computations can be carried out more quickly, without the need for interpolating pre-computed transfer functions.
Note, however, that for predictions of the late $y T/E$ contributions, including those arising during reionization \citep[e.g.,][]{Ota2018reionization}, a detailed treatment is required.

Next we changed the sourcing of distortion anisotropies to only include $\ell=0$. For the $\mu$-distortion cross power spectra we do not expect any large differences, since the tight coupling approximation predicts very small contributions from $\ell>0$, as we again confirmed numerically. For the $y$-distortion cross power spectra we find a small difference at the level of a few percent, which we decided to not even illustrate in more detail. Overall, we conclude that for predictions of the distortion power spectra it is sufficient to include sources with $\ell=0$, especially for $\mu T/E$ signals.

%------------------------------------
\begin{figure}
\centering
\includegraphics[width=0.82\columnwidth]{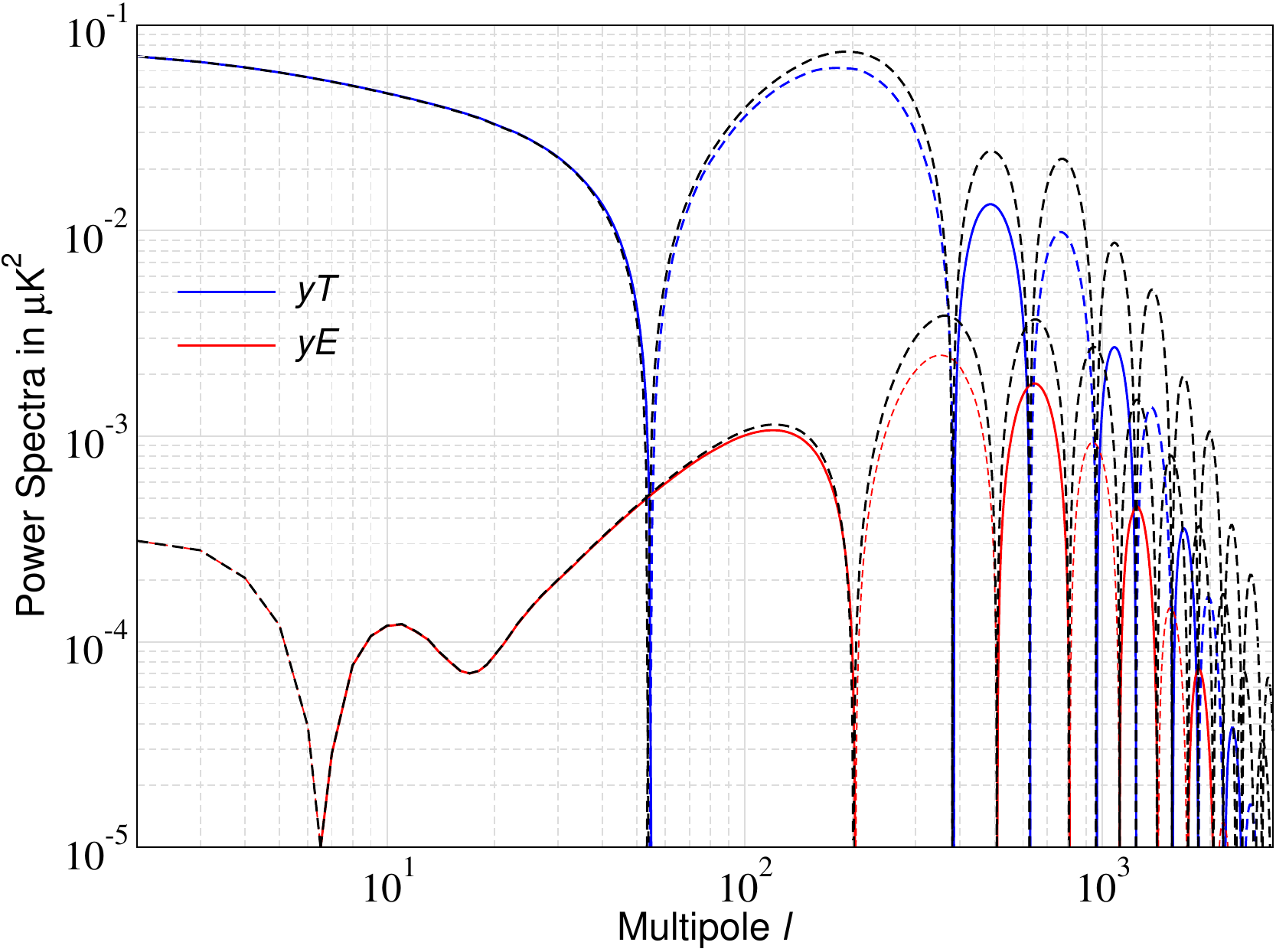}
\\
\caption{Comparison of the detailed prediction for the $y T$ and $y E$ power spectra with those obtained using a $k$-independent heating rate with $k=10^{-3}\,\Mpc^{-1}$ (dashed black lines).}
\label{fig:PS_compare_large}
\end{figure}
%------------------------------------
As a last test, we study how the signal power spectra are affected when we assume that the sourcing is independent of $k$ and can be approximated using $k=10^{-3}\,\Mpc^{-1}$. In this case, we again expect small changes for the $\mu$-distortion predictions while for the $y$-distortion the small scale power spectrum should be enhanced, given that the extra reduction of the sources is not included. As anticipated, for the $\mu T$ and $\mu E$ power spectra we found only small differences at the level of a few percent at $\ell \gtrsim 10^3$. For the $y$-distortion power spectra, we illustrate the results in figure~\ref{fig:PS_compare_large}, which in line with our expectation reveals visible changes at small angular scales. The enhancement also partially explains the difference with respect to \citep{Ravenni2017}.

%------------------------------------
\subsection{Distortion signals for enhanced small-scale power spectra with PNG}
\label{sec:model-II}
%--------------------------------------
In the previous section we focused on the $\Lambda$CDM case with PNG, as was originally motivated by early works \citep{Pajer2012, Ganc2012}. However, one can significantly alter the distortion signal by enhancing the small-scale power spectrum. In particular, larger anisotropies can be created for lower levels of PNG, as stressed in \citep{Chluba2017muT}. In this situation, we will also receive a second contribution to the $\mu T/E$ signals from propagation effects through the perturbed medium, as we illustrate now.  

%------------------------------------
\begin{figure}
\centering
\includegraphics[width=0.82\columnwidth]{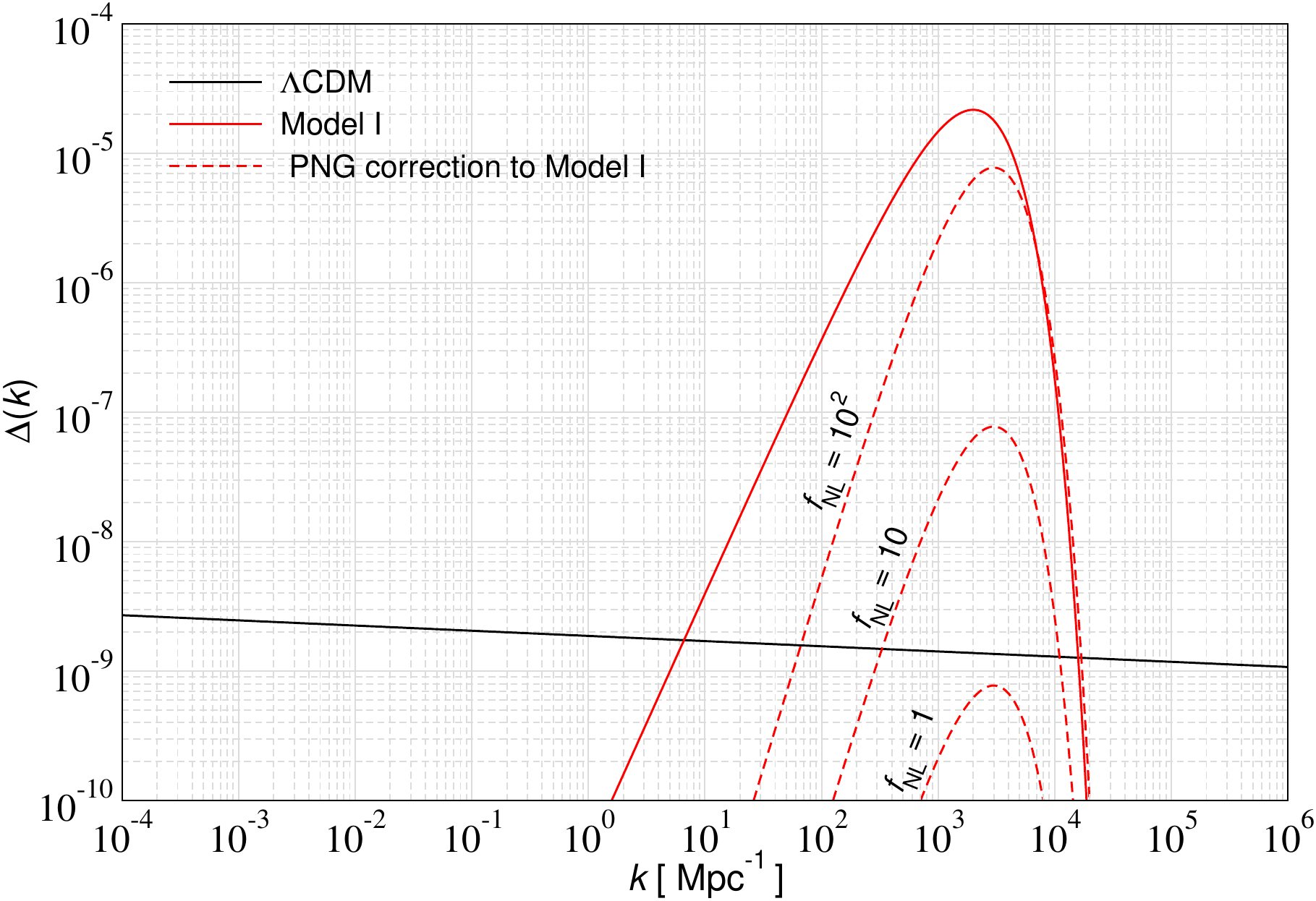}
\\
\caption{Comparison of scalar power spectrum models. The $\Lambda$CDM curve (solid black line) is the simple power-law spectrum with $\nS=0.96$, while for Model I (solid red line) we used $\Delta(k) = A_{\rm p} (k/k_{\rm p})^{n_{\rm p}-1}\,\expf{-k/k_{\rm c}}$ with $A_{\rm p} =\pot{4}{-7}, k_{\rm p}=10^2\,\Mpc^{-1},n_{\rm p}=3$ and $k_{\rm c}=10^3\,\Mpc^{-1}$. We also computed the non-Gaussian contribution to Model I using Eq.~\eqref{eq:local_monopole_dis_source_NG} for varying values of $\fNL$.}
\label{fig:Delta_models}
\end{figure}
%------------------------------------
For the small-scale power spectrum, many models can in principle be considered once working outside of slow-roll inflation \citep[e.g.,][]{Chluba2012inflaton, Clesse2014,
Dimastrogiovanni2016, Putti2024}. For illustration, we assume an extra small-scale curvature power spectrum contribution which is described by $\Delta(k) = A_{\rm p} (k/k_{\rm p})^{n_{\rm p}-1}\,\expf{-k/k_{\rm c}}$ (henceforth 'Model I'). This model could be relevant to scenarios with primordial black hole formation \citep[e.g.,][]{Nakama2018mu}, however, here we merely use this as a {\it simplistic} illustration of the methodology.

In figure~\ref{fig:Delta_models} we give an example in comparison to the standard slow-roll power spectrum. The shown scenario will cause a sizable average $\mu$ distortion comparable to the \COBEF upper limit, while the $y$-signals remain negligible. In this scenario, anisotropic distortions are sourced by propagation effects and the anisotropic sourcing is expected to be detectable by \Planck even for $\fNL\simeq 1$, while corrections from the $\fNL^2$ contribution to the average heating effect can be neglected given that the effective power spectrum contribution is subdominant (see figure~\ref{fig:Delta_models} for illustration).

%------------------------------------
\begin{figure}
\centering
\includegraphics[width=0.82\columnwidth]{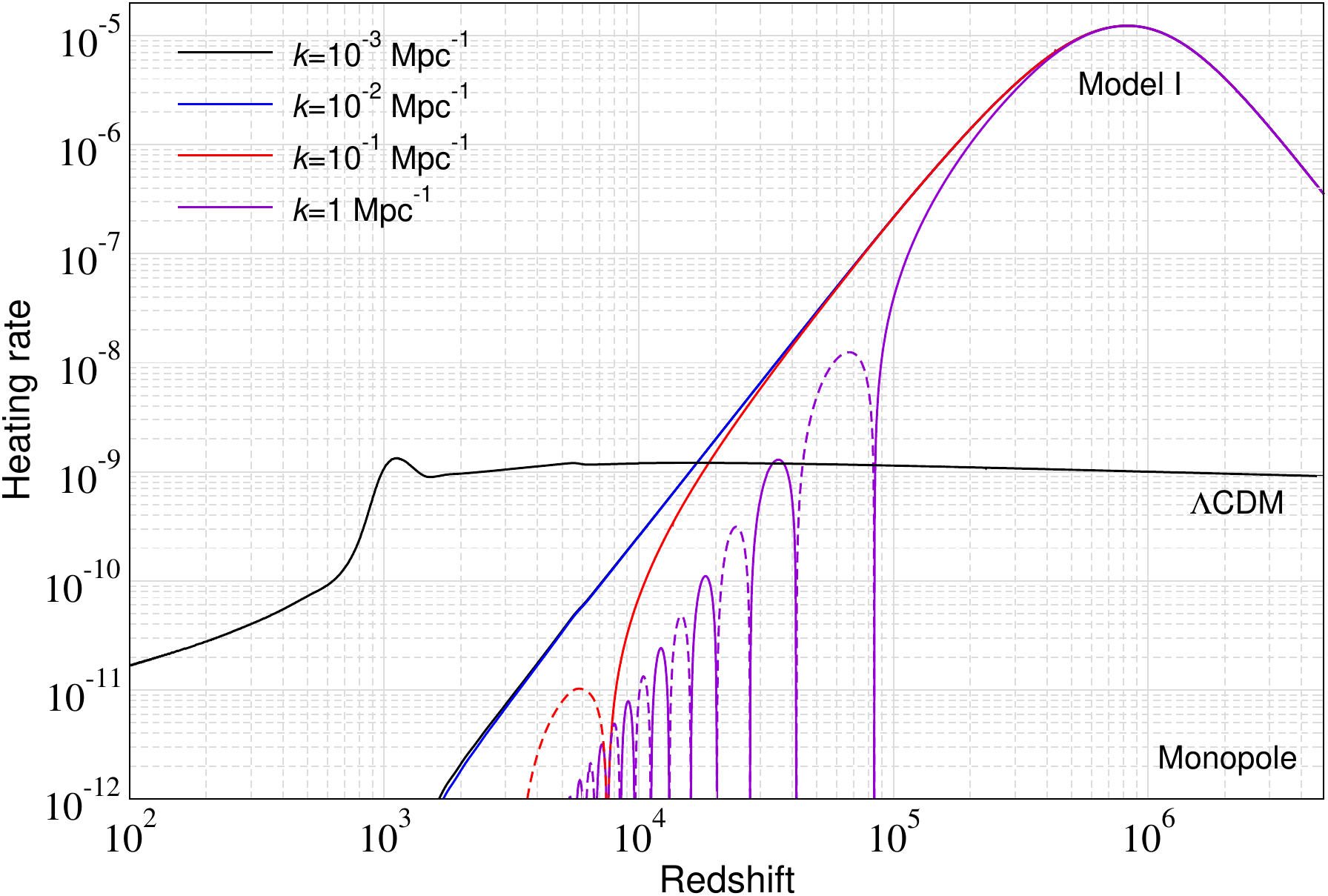}
\\
\caption{Scaled effective heating rates computed from Eq.~\eqref{eq:Q_ell_dot} for Model I as in figure~\ref{fig:Delta_models} for varying values of $k$. For comparison we also show the large scale (uniform) heating rate for the $\Lambda$CDM case. Negative scale-dependent effective heating rates are shown in dashed.
}
\label{fig:Average_heating-TC-M1}
\end{figure}
%------------------------------------
With this in mind, in figure~\ref{fig:Average_heating-TC-M1} we show the scaled effective heating rates for Model I in comparison to the $\Lambda$CDM case for varying values of $k$. Importantly Model I dissipates most of the energy during the $\mu$-era, with minor contributions at $z\lesssim 10^4$, where the standard power spectrum damping dominates. We used the full transfer function database to compute these heating rates, but it is clear that the tight coupling approximation will suffice for Model I. We indeed found no visible differences between the result when switching to the tight coupling treatment which we shall use below. We also did not include $\ell>1$ anisotropic sourcing, given the expected suppression at early times [see discussion surrounding Eq.~\eqref{eq:M_av_approx} and the illustrations in figure~\ref{fig:dipole_rate}].

%------------------------------------
\begin{figure}
\centering
\includegraphics[width=0.80\columnwidth]{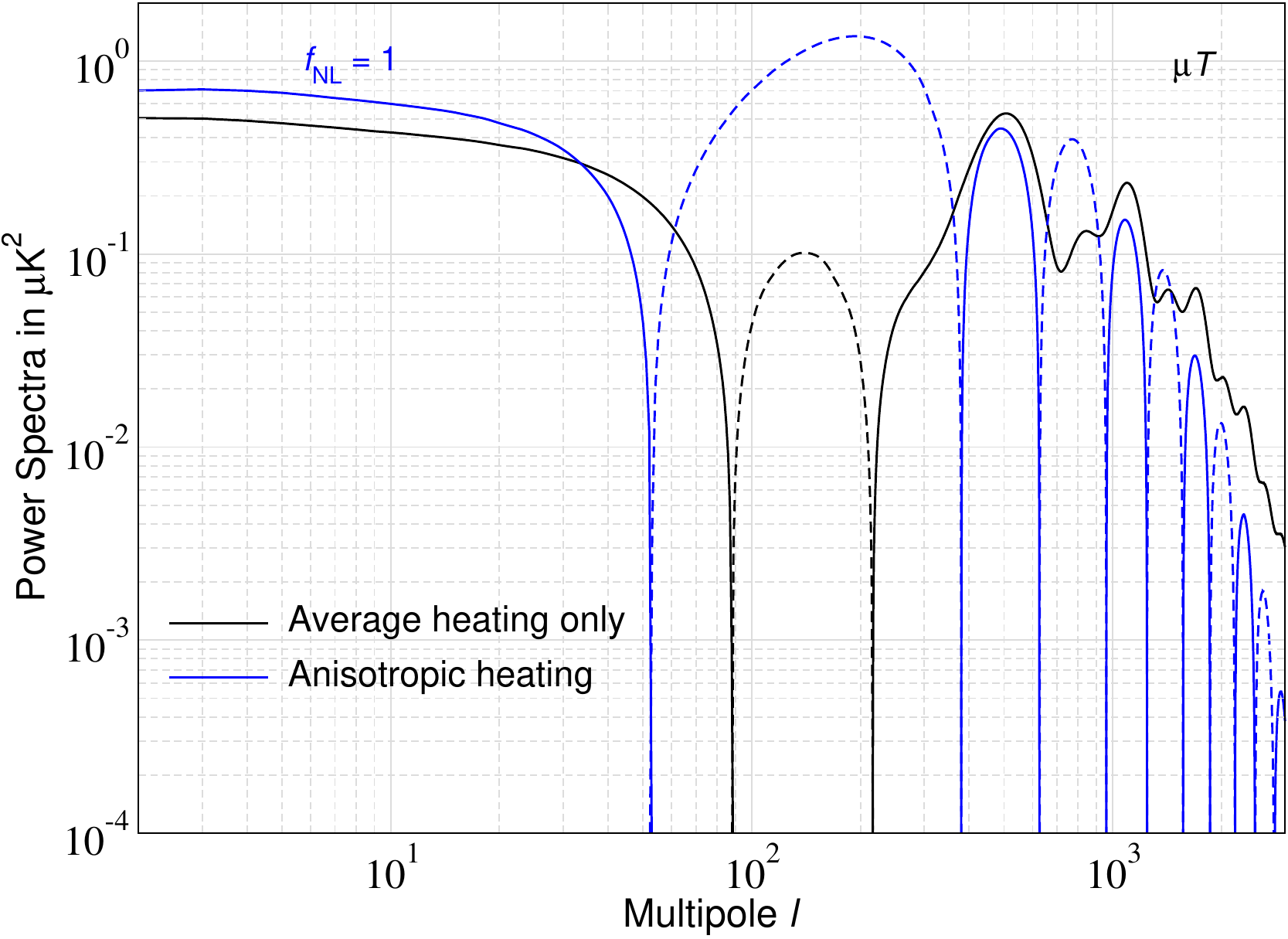}
\\[3mm]
\includegraphics[width=0.80\columnwidth]{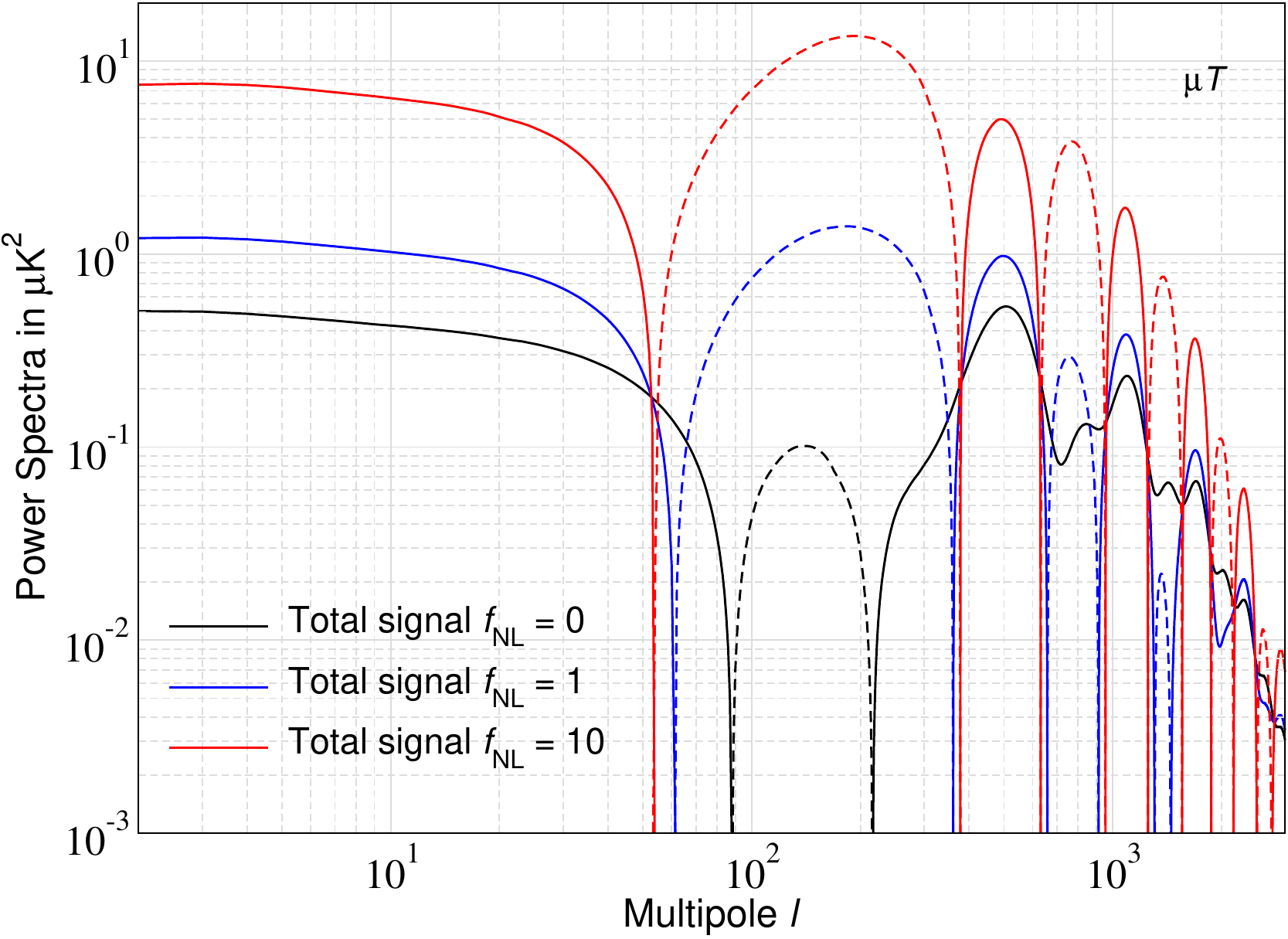}
\\
\caption{Illustration of the $\mu T$ signals for Model I. In the upper panel we show the two contributions to the total signal from i) propagation effects and ii) anisotropic sourcing. In the lower panel we then illustrate the $\ell$ dependence of the total signal for varying values of $\fNL$, demonstrating how the spectro-spatial information can be used to distinguish different models and break degeneracies.}
\label{fig:PS-M1_muT}
\end{figure}
%------------------------------------
To compute the $\mu T$ and $\mu E$ power spectra for this model, we proceed as follows: we first compute the signal contributions using only average (Gaussian) distortion sourcing. This then allows us to capture the effects from propagation through the perturbed universe that will {\it always} be present even in the limit of $\fNL\rightarrow 0$, yielding a signal templates $\propto A_{\rm p}$. We then compute the signals from anisotropic dissipation setting $\fNL=1$. This yields another signal template related to directly-sourced distortion anisotropies $\propto A_{\rm p} \fNL$. Together,
these two templates can then be scaled for varying values of $A_{\rm p}$ and $\fNL$ to describe the total signal. 
In a similar way, we can independently compute the signals from the $\fNL^2$ term, although this is found to remain subdominant for all cases that are not already in strong tension with \Planck and \COBEF, and is thus neglected below.

In this context, it is important to stress that the three contributions scale differently with $A_{\rm p}$ and $\fNL$. This implies that one can in principle distinguish their individual effects and break degeneracies by combining distortion monopole and anisotropy measurements as we will also see below.
As mentioned in the introduction, we stress that from the perspective of perturbation theory, the propagation effects formally appear at $\mathcal O(\mathcal{R}^3)$, and a rigorous treatment in principle requires an expansion to third order in perturbations, which is beyond the scope of this paper. However, we expect our calculation to already capture some of the leading order terms and thus provide a useful illustration.
%
%However, because the {\it average} spectrum becomes distorted at order $\langle \mathcal{R}^2\rangle$, the true perturbation order of this distortion source (i.e., in terms of spatial fluctuations) is only first order $\propto \langle \mathcal{R}^2\rangle \mathcal{R}$, justifying our treatment.

In figure~\ref{fig:PS-M1_muT} we show the results for Model I. From the top panel we can see that the two main contributions to the $\mu T$ signal have a different $\ell$-dependence. The signal caused by propagation effects of the distorted average spectrum through the medium is mostly correlated with the temperature field unless in a window around $\ell \simeq 100-200$. In contrast, the anisotropic dissipation causes an alternating pattern of correlation / anti-correlation across all Doppler peaks. 

In the lower panel of figure~\ref{fig:PS-M1_muT}, we illustrate the total $\mu T$ signal for varying values of $\fNL$. Because the two contributions considered here scale as $\propto A_{\rm p}$ and $\propto A_{\rm p} \fNL$, the spectro-spatial information in particular around the first Doppler peak can in principle alone break the degeneracies and thus allow constraining $A_{\rm p}$ and $\fNL$ independently. A second constraint on $A_{\rm p}$ (and $A_{\rm p}^2 \fNL^2$) can be deduced from measurements of the average distortion signal, providing further leverage on the model space. 

We note that for the chosen parameters, all shown cases should be within reach of \Planck, ACT and SPT. In addition, the spectro-spatial information can in principle be used to directly constrain the scale dependence of the power spectrum. For instance, a reduction of $k_{\rm p}$ in the model increases the distortion anisotropy sourcing during the $y$-era while simultaneously changing the scale dependence of the $\mu T$ signal from propagation effect. A combination of all the available information will thus open a path to constraining details of the model, although a detailed forecast is left to future work.

\section{Conclusion}
\label{sec:conclusions}
%--------------------------------------
CMB spectral distortion anisotropies provide a direct probe of correlations between primordial perturbations on scales that are widely separated in wavenumber. In this paper, we developed a detailed framework for calculating the distortion anisotropies generated by acoustic damping in the presence of PNG. The calculation combines an explicit derivation of the second-order blackbody-mixing sources with the FH of spectral evolution and photon transport. This allows us to describe the production, thermalization and propagation of the distortion perturbations from the early Universe through recombination and reionization using {\tt CosmoTherm}.

We first derived the distortion source associated with the mixing of blackbodies of different temperatures. Combining the relevant Liouville, Thomson-scattering and velocity-dependent terms, we showed that the real-space source can be written compactly [see Eq.~\eqref{eq:dis_source_V_B}] in terms of the gauge-independent photon temperature variable, $\Theta_{\rm g}=\Theta+\Psi - V$. We then projected this expression onto its angular multipoles and transformed it to Fourier space. The resulting source depends not only on the magnitudes of the two dissipating modes, but also on their relative orientation and on their orientation with respect to the observable mode [see Eq.~\eqref{eq:source_ell_m_kz_gen} for the general case]. The full mode-coupling geometry therefore differs from previous treatments, which did not include these aspects consistently. Although these geometric corrections have a modest effect for the local-type bispectrum considered here, they are expected to become more important for general primordial bispectrum shapes \citep[e.g., see][for overview and references]{Planck2013ng}.

Rather than directly evolving the complete second-order photon distribution, we used the structure of the observable cross-correlations to reduce the problem to a set of effective anisotropic heating rates. These source functions are constructed from first-order photon and baryon transfer functions while retaining the primordial bispectrum and the full Fourier-space geometry. To accelerate the calculation we first produce a transfer function database that then can be used in the subsequent computation (see sections~\ref{sec:transfer_database} and \ref{sec:effect_HR_setup} for more details about the setup).
Once tabulated as functions of redshift and observable wavenumber, they can be inserted as external sources into the linear FH. This separation renders the calculation numerically tractable while retaining the information required to predict the distortion–temperature and distortion–polarisation cross-power spectra. However, when the scale-dependence of the considered model changes, the setup has to be rerun, suggesting that in the future emulator-based acceleration methods could be explored.

For a nearly scale-invariant primordial power spectrum with local-type non-Gaussianity, the anisotropies generated by propagation of the average distortion through the perturbed Universe are negligible at currently relevant sensitivities. The observable signal is instead dominated by the direct modulation of acoustic dissipation $\propto \fNL$, as we explain in section~\ref{sec:model-I}. The corresponding effective monopole heating rate approaches the standard average heating rate on large scales, but develops a significant scale dependence at larger wavenumbers (see figure~\ref{fig:monopole_rate}). In particular, heating close to recombination is suppressed as the observable wavenumber increases, so that small-scale distortion anisotropies are sourced predominantly during the $\mu$-era. The dipolar and quadrupolar heating sources are strongly suppressed relative to the monopole, especially at the redshifts relevant for $\mu$-distortion production (see figure~\ref{fig:dipole_rate}).

Using the FH implementation of {\tt CosmoTherm}, we computed the resulting $\mu T/E$ and $yT/E$ cross-power spectra.\footnote{The related auto-power spectra $\mu \mu$ and $y y$ as well as $y \mu$ and residual distortion power spectra are also obtained but were not further discussed here given how sub-leading they are \citep[see some examples in][]{kite_spectro-spatial_2023-III}.} The $\mu T$ spectrum displays alternating regions of correlation and anti-correlation, but is damped slightly faster than the primary temperature spectrum (see figure~\ref{fig:PS_all}). This enhanced damping follows from the absence of acoustic propagation in the distortion variables: in tight coupling, distortion perturbations are overdamped rather than participating in the exchange between photon monopole and dipole perturbations that supports ordinary acoustic waves \citep[e.g.,][]{Chluba2017muT}. The numerical damping scale is consistent with the analytic expectation based on the modified diffusion scale of the distortion perturbations [see Eq.~\eqref{eq:y_ell}].

The $yT$ spectrum has a similar oscillatory structure to the $\mu T$ signal, but is lower in amplitude and more strongly damped on small angular scales. The latter effect arises both from photon diffusion and from the suppression of the anisotropic heating rate during the $y$-era at large wavenumbers. The $\mu E$ and $yE$ spectra are further suppressed and exhibit the expected phase shift relative to the temperature correlations. They also contain large-scale reionisation features, although detecting these features would require sensitivities well beyond those presently available.

Our refined results broadly reproduce the large-scale behavior found in previous approximate calculations (see figure~\ref{fig:PS_compare}). However, visible differences arise from the detailed evolution of the distortion perturbations and from the scale dependence of the source. Previous numerical results provide an accurate approximation to the $\mu T$ spectrum at $\ell\lesssim 400$, but do not capture the additional damping at higher multipoles. For the $yT$ spectrum, earlier approximations underestimate the signal on large angular scales and overestimate it on small angular scales. These differences illustrate why the distortion transfer functions cannot, in general, be replaced by standard temperature transfer functions multiplied by a local heating modulation.

We also tested several approximations that could simplify future calculations (see section~\ref{sec:physics}). Tight-coupling transfer functions reproduce the $\mu T$ and $\mu E$ spectra with negligible error because their sources are generated predominantly at high redshift. The same approximation produces visible, although still moderate, changes in the $yT$ and $yE$ spectra, particularly at very large and very small angular scales. Restricting the anisotropic heating source to its local monopole changes the $\mu$-distortion spectra negligibly and affects the $y$-distortion spectra only at the level of a few percent. Similarly, neglecting the wavenumber dependence of the source is adequate for the $\mu T/E$ spectra except at $\ell\gtrsim {\rm few}\times 10^3$, but produces appreciable changes in the small-scale $yT/E$ signals. These results suggest that accurate $\mu$-distortion predictions can be obtained with substantially simplified source calculations, whereas precision predictions for $y$-distortion anisotropies require greater care.

Finally, for illustration we considered a model with an enhanced small-scale curvature power spectrum (see section~\ref{sec:model-II}). In this case, the average $\mu$-distortion can approach the \COBEF limit, and the propagation of this distorted average spectrum through the perturbed Universe produces an additional anisotropic signal that is no longer negligible. The propagation contribution and the signal generated directly by anisotropic dissipation exhibit differing angular dependences (see figure~\ref{fig:PS-M1_muT}). The former is predominantly correlated with the temperature field, apart from a region around the first acoustic peak, whereas the latter retains an alternating correlation and anti-correlation pattern across the Doppler peaks also found for a quasi-scale independent Gaussian power spectrum.

These contributions also have different dependences on the model parameters. The propagation signal scales with the amplitude $\propto A_{\rm p}$ of the enhanced small-scale power, while the anisotropic-dissipation signal scales as $\propto A_{\rm p}f_{\rm NL}$. Contributions arising from the non-Gaussian correction to the average distortion have a further dependence on $\propto A_{\rm p}^2f_{\rm NL}^2$, although they remain subdominant in the parameter range considered here. The differing angular structures and parameter scalings imply that measurements of the distortion monopole and the distortion anisotropies can jointly constrain the amplitude of the small-scale power spectrum and the primordial non-Gaussianity, rather than only their product. This provides exciting opportunities for further investigations using existing \Planck, ACT and SPT data and might be relevant to discussions about the origin of super-massive primordial black holes \citep[e.g.,][]{Carr2010, Nakama2018mu}.

In summary, the newly developed FH of {\tt CosmoTherm} now makes it possible to exploit the full spectro-spatial information carried by distortion anisotropies in scenarios with PNG. It provides a systematic way to distinguish distortions generated at different epochs, include all relevant photon-transport effects, and connect the observed correlations to the scale dependence and configuration dependence of primordial fluctuations. Extending the calculation to more general bispectrum shapes and small-scale power-spectrum models, possibly by using machine-learning approaches, together with a detailed treatment of experimental sensitivity and astrophysical foregrounds, will allow the constraining power of existing and future multi-frequency CMB observations to be assessed.

{\small
\paragraph{Acknowledgments}
This work was supported by the UKSA grant: LiteBIRD UK ST/Y005945/1. 
AO was supported in part by the National Natural Science Foundation of China under Grant No. 12403001 and 12547101, and New Chongqing YC Project CSTB2024YCJH-KYXM0083.
NB acknowledges support by the INFN InDark Initiative.
NB acknowledges support from the COSMOS network (\url{www.cosmosnet.it}) through the ASI (Italian Space Agency) Grants 2016-24-H.0, 2016-24-H.1-2018 and 2020-9-HH.0.
}

{\small
\bibliographystyle{plain}
\bibliography{Lit-all, Lit-extra}
}

\appendix

\section{Useful relation and functions}
\label{app:Relations}
%--------------------------------------
In this section we summarize a few useful functions that appear repeatedly in our derivations. 

\subsection{Integrals over three spherical Bessel functions}
\label{app:Triple_j_int}
%--------------------------------------
In our derivations we will need integrals over three Bessel functions of the form
%-----------
\beal
\label{eq:I_def}
\mathcal{I}_{\ell}(k, k_1,k_2)&=
\frac{4}{\pi}\int r^2 \id r
\,j_0(k r) \,j_{\ell}(k_1 r)\,j_{\ell}(k_2 r).
\end{align}
%-----------
The more general integral was discussed by \citep{Mehrem1991}, but the special cases we require reduce to \citep{Mehrem2010}
%-----------
\beal
\label{eq:I_2_bessel}
\mathcal{I}_{\ell}(k, k_1,k_2)&=
\frac{\beta_{\rm H}(\Delta)}{k k_1 k_2}\,P_\ell(\Delta), &\Delta &= \frac{k_1^2+k_2^2-k^2}{2k_1 k_2},\nonumber \\
\beta_{\rm H}(y)&=\theta_{\rm H}(1-y)\,\theta_{\rm H}(1+y), &\theta_{\rm H}(y)&=
\begin{cases}
0 & y<0
\\
\frac{1}{2} & y=0
\\
1 & y>0.
\end{cases}
\end{align}
%-----------
These expressions will appear as weight factors for the distortion source terms.

\subsection{Gaunt integrals}
\label{app:Gaunt}
%--------------------------------------
Integrals over products of three spherical harmonics can be nicely written in terms of the Gaunt coefficients
%-----------
\beal
\label{eq:Gaunt}
\mathcal{G}^{\ell,\ell_1,\ell_2}_{m,m_1,m_2}
&=
\int \id \hat{\vek{r}}\,
Y_{\ell m}(\hat{\vek{r}})
\,Y_{\ell_1 m_1}(\hat{\vek{r}})
\,Y_{\ell_2 m_2}(\hat{\vek{r}})
\nonumber\\
&
=\sqrt{
\frac{(2\ell+1)(2\ell_1+1)(2\ell_2+1)}{4\pi}
}
\,\Bigg(
\vspace{-1mm}
\begin{array}{ccc}
\vspace{-4mm}
\ell  & 
\ell_1 & \ell_2\\
0 & 0 & 0
\end{array}
\Bigg)
\,\Bigg(
\vspace{-1mm}
\begin{array}{ccc}
\vspace{-4mm}
\ell  & \ell_1 & \ell_2\\
m & m_1 & m_2
\end{array}
\Bigg),
\end{align}
%-----------
where we used the Wigner-$3J$ symbols. These Wigner symbols enforce $m+m_1+m_2=0$, which means that $\mathcal{G}^{\ell,\ell_1,\ell_2}_{m,m_1,-m_1}$ means $m=0$. Similarly, $(-1)^m \mathcal{G}^{\ell,\ell_1,\ell_2}_{-m,m_1,m_2}$ appears when the first spherical harmonic is conjugated, as often relevant to projection integrals. We also have $[\mathcal{G}^{\ell,\ell_1,\ell_2}_{m,m_1,m_2}]^*=\mathcal{G}^{\ell,\ell_1,\ell_2}_{-m,-m_1,-m_2}$. 

In our calculations, we are interested in the sourcing of distortions involving modes with $\ell=0-2$, since the other modes are only relevant around the time of recombination. Some of the the required Gaunt integrals are
%-----------
\bealf{
\label{eq:Gaunts}
\sqrt{4\pi}(-1)^{m'}\mathcal{G}^{0, \ell',\ell''}_{0,m',-m'}&=\int \id \hat{\vek{r}}\,
P_0(\hat{\vek{z}}\cdot\hat{\vek{r}})
\,Y_{\ell' m'}(\hat{\vek{r}})
\,Y^*_{\ell'' m'}(\hat{\vek{r}})
= \delta_{\ell',\ell''}
\nonumber \\
\sqrt{\frac{4\pi}{3}}(-1)^{m'}\mathcal{G}^{1, \ell',\ell''}_{0,m',-m'}&=\int \id \hat{\vek{r}}\,
P_1(\hat{\vek{z}}\cdot\hat{\vek{r}})
\,Y_{\ell' m'}(\hat{\vek{r}})
\,Y^*_{\ell'' m'}(\hat{\vek{r}})
= C^{m'}_{\ell'+1} \delta_{\ell'+1,\ell''}+C^{m'}_{\ell'}\delta_{\ell'-1,\ell''}
\\ \nonumber
\sqrt{\frac{4\pi}{5}}(-1)^{m'}\mathcal{G}^{2, \ell',\ell''}_{0,m',-m'}&=\int \id \hat{\vek{r}}\,
P_2(\hat{\vek{z}}\cdot\hat{\vek{r}})
\,Y_{\ell' m'}(\hat{\vek{r}})
\,Y^*_{\ell'' m'}(\hat{\vek{r}})
= \frac{3}{2}\,C^{m'}_{\ell'+1}C^{m'}_{\ell'+2} \delta_{\ell'+2,\ell''}+\frac{3}{2}\,C^{m'}_{\ell'}C^{m'}_{\ell'-1}\delta_{\ell'-2,\ell''} 
\nonumber\\ \nonumber
&\qquad\qquad\qquad\qquad\qquad\qquad\qquad\qquad 
+\frac{3(C^{m'}_{\ell'+1})^2+3(C^{m'}_{\ell'})^2-1}{2} \delta_{\ell',\ell''} 
}
%-----------
with $C^m_\ell=\sqrt{(\ell^2-m^2)/(4\ell^2-1)}$.

\section{First order photon temperature perturbations}
\label{app:Theta_conventions}
The first order temperature perturbations have the form $\Theta^{(1)}=\Theta^{(1)}(\eta, \vgh, \vek{r})=\sum_{\ell m} \Theta^{(1)}_{\ell m}(\eta, \vek{r})\,Y_{\ell m}(\vgh)$. In Fourier space we have
%-----------------------
\beal
\Theta^{(1)}_{\ell m}(\eta, \vek{r})&=\int \frac{\id^3 k}{(2\pi)^3}\,\expf{\i \vek{k}\cdot \vek{r}}\,\Theta^{(1)}_{\ell m}(\eta, \vek{k})
=\int \frac{\id^3 k}{(2\pi)^3}\,\expf{\i \vek{k}\cdot \vek{r}}\,\hat{\Theta}^{(1)}_{\ell m}(\eta, \vek{k})\,\mathcal{R}(\vek{k})
\end{align}
%-----------------------
where $\hat{\Theta}^{(1)}_{\ell m}(\eta, \vek{k})$ is the temperature transfer function. When we compute the first order anisotropies, we usually determine the Legendre coefficients, $\hat{\Theta}^{(1)}_{\ell}(\eta, k)$, of the transfer function in Fourier space for an irrotational field sourced by scalars
%-----------------------
\beal
\label{eq:first_order_m_only}
\hat{\Theta}^{(1)}(\eta, k, \vgh\cdot \hat{\vek{k}})=\sum_\ell (-\i)^\ell\, (2\ell +1) \,\hat{\Theta}^{(1)}_{\ell}(\eta, k)\, P_\ell(\vgh\cdot \hat{\vek{k}})
=\sum_{\ell m} 4\pi (-\i)^\ell\, \hat{\Theta}^{(1)}_{\ell}(\eta, k)\, Y_{\ell m}(\vgh)
\,Y^*_{\ell m}(\hat{\vek{k}}).
\end{align}
%-----------------------
This expression implies that 
%-----------------------
\beal
\label{eq:Theta_ellm_irrotational}
\hat{\Theta}^{(1)}_{\ell m}(\eta, \vek{k})
&\equiv 4\pi (-\i)^\ell\, \hat{\Theta}^{(1)}_{\ell}(\eta, k)\,Y^*_{\ell m}(\hat{\vek{k}}).
\end{align}
%-----------------------
For $\hat{\vek{k}}\parallel \vbh \parallel \hat{\vek{z}}$, this further simplifies to $\hat{\Theta}^{(1)}_{\ell m}(\eta, \vek{k})
\equiv \sqrt{4\pi (2\ell+1)} (-\i)^\ell\, \hat{\Theta}^{(1)}_{\ell}(\eta, k)\,\delta_{m 0}$.

\section{Coordinate independent version of the velocity terms}
\label{app:coordinate_indep}
%---------------------
The source terms $S^{\rm mix}_L$ and $S^{\rm mix}_T$ in Eq.~\eqref{eq:Boltzmann_second_order_general_dist_source_LHS} and \eqref{eq:Thomson} are already in coordinate-independent form. However, for $S^{\rm mix}_\beta$ in Eq.~\eqref{eq:velocity} we assumed that $\vbh$ is aligned with the $z$-axis. To make these terms general, we first use the spherical harmonic identities to simplify the terms $\propto \Theta^{(1)}_{2m}$
%-----------
\beal
\label{eq:velocity_coordinate_independent}
\mathcal{S^{\rm mix}_{\beta}}&=
\frac{\vbeta^{(1)}\cdot\vbeta^{(1)}}{3}
-
\beta^{(1)} C^0_1 \,\Theta^{(1)}_{10} \, Y_{00}(\vgh)
-\frac{\beta^{(1)}}{10}
\left\{
\sum_{m=-1}^1 
C^m_2\,\Theta^{(1)}_{1m}
Y_{2m}(\vgh)
+\sum_{m=-2}^2 
C^m_3\,\Theta^{(1)}_{3m}
Y_{2m}(\vgh)
\right\}
\nonumber \\ 
&\qquad +\beta^{(1)} \left[ \Theta^{(1)}_{0}(\vgh)+\frac{1}{10}\,\Theta^{(1)}_{2}(\vgh)\right]\,P_{1}(\vgh\cdot\vbh)
+\frac{11}{30}\,\vbeta^{(1)}\cdot\vbeta^{(1)}\, P_{2}(\vgh\cdot\vbh).
\end{align}
%-----------
This means that now only the dipole and octupole terms are coordinate-dependent.

To get the coordinate-independent version, we went back to the deriving all the scattering terms from scratch. For the velocity-dependent terms we then find the distortion sources
%-----------
\beal
\label{eq:velocity_coordinate_independent_rederived}
\mathcal{S}^{\rm mix}_{\beta}&=
\frac{\vbeta^{(1)}\cdot\vbeta^{(1)}}{3}
-\beta^{(1)}\left[(\vbh\cdot \vgh)\,\Theta^{(1)}(\vgh)\right]_0
-\frac{\beta^{(1)}}{10}\left[(\vbh\cdot \vgh)\,\Theta^{(1)}(\vgh)\right]_2
\nonumber \\ 
&\qquad +\beta^{(1)} \left[ \Theta^{(1)}_{0}(\vgh)+\frac{1}{10}\,\Theta^{(1)}_{2}(\vgh)\right]\,P_{1}(\vgh\cdot\vbh)
+\frac{11}{30}\,\vbeta^{(1)}\cdot\vbeta^{(1)}\, P_{2}(\vgh\cdot\vbh),
\end{align}
%-----------
which already looks very similar to the previous expression.
Using the addition theorem for spherical harmonics, we then have
%-----------
\bsub
\beal
\label{eq:velocity_coordinate_independent_rederived_Wigner}
\left[(\vbh\cdot \vgh)\,\Theta^{(1)}(\vgh)\right]_0
&=\frac{4\pi}{3}\sum_{m'} 
Y_{1m'}(\vbh) 
\int Y^*_{1m'}(\vgh) 
\sum_{\ell m} \Theta^{(1)}_{\ell m} Y_{\ell m}(\vgh) \frac{\id^2\vgh}{4\pi}
=\frac{1}{3}
\sum_{m} 
Y_{1m}(\vbh) \, \Theta^{(1)}_{1 m}
\\
\left[(\vbh\cdot \vgh)\,\Theta^{(1)}(\vgh)\right]_{2m}
&=
\frac{4\pi}{3}\sum_{m'} 
Y^*_{1m'}(\vbh) 
\int Y^*_{2m}(\vgh)\, Y_{1m'}(\vgh) 
\sum_{\ell m''} \Theta^{(1)}_{\ell m''} Y_{\ell m''}(\vgh) \id^2\vgh
\nonumber\\[-2mm]
&=
\sqrt{\frac{4\pi}{3}}
\sum_{m'} 
Y^*_{1m'}(\vbh) 
\,
\sum_{\ell m''} 
\Theta^{(1)}_{\ell m''} 
(-1)^m \sqrt{5 (2\ell+1)}
\,\Bigg(
\vspace{-1mm}
\begin{array}{ccc}
\vspace{-4mm}
2  & 1 & \ell\\
0 & 0 & 0
\end{array}
\Bigg)
\,\Bigg(
\vspace{-1mm}
\begin{array}{ccc}
\vspace{-4mm}
2  & 1 & \ell\\
-m & m' & m''
\end{array}
\Bigg)
\end{align}
\esub
%-----------
The Wigner-$3J$ symbols only allow $\ell=1$ and $\ell=3$.
This is consistent with what follows from Eq.~(C26) of \citep{Chluba2012} although there the octupole term seems to have an extra factor of $\sqrt{3/7}$.
Aligning $\vbh$ parallel to $z$, we have 
%-----------
\bsub
\beal
\label{eq:velocity_coordinate_independent_rederived_Wigner_z}
\left[(\vbh\cdot \vgh)\,\Theta^{(1)}(\vgh)\right]_0
&=C^0_1 \, \Theta^{(1)}_{1 0} \,Y_{00}(\vgh)\equiv \frac{\Theta^{(1)}_{1 0}}{\sqrt{12\pi}}
\\
\left[(\vbh\cdot \vgh)\,\Theta^{(1)}(\vgh)\right]_{2m}
&=
\sum_{\ell} 
\Theta^{(1)}_{\ell m} 
(-1)^m \sqrt{5 (2\ell+1)}
\,\Bigg(
\vspace{-1mm}
\begin{array}{ccc}
\vspace{-4mm}
2  & 1 & \ell\\
0 & 0 & 0
\end{array}
\Bigg)
\,\Bigg(
\vspace{-1mm}
\begin{array}{ccc}
\vspace{-4mm}
2  & 1 & \ell\\
-m & 0 & m
\end{array}
\Bigg)
\equiv C^m_2 \Theta^{(1)}_{1 m} +C^m_3 \Theta^{(1)}_{3 m}.
\end{align}
\esub
%-----------
This confirms the correctness of the general expression in Eq.~\eqref{eq:velocity_coordinate_independent_rederived}.
However, we can further simplify matters by introducing the variable $V^{(1)}=\vbeta^{(1)}\cdot \vgh$. This has the benefit that $V$ now behaves the same way as a CMB temperature variable and thus can be expanded into spherical harmonics \citep{Ota2017}. Since
%-----------
\bsub\beal
\label{eq:V2}
\frac{\vbeta\cdot\vbeta}{3}\equiv [V^2]_0&=\left(\frac{4\pi}{3}\right)^2 \beta^2
\sum_{m m'} 
Y^*_{1m}(\vbh) 
Y_{1m'}(\vbh) 
\int Y^*_{1m}(\vgh) 
Y_{1m'}(\vgh) \,\frac{\id^2 \vgh}{4\pi}
%\\
%[V^2]_{2m}&=\left(\frac{4\pi}{3}\right)^2
%\sum_{m' m''} 
%Y^*_{1m'}(\vbh) 
%Y_{1m''}(\vbh) 
%%
%\int Y^*_{2m}(\vgh)  \,Y^*_{1m'}(\vgh) 
%\,Y_{1m''}(\vgh) \,\id^2 \vgh
%\equiv \frac{\vbeta\cdot\vbeta}{3}
\\
\vbeta\cdot\vbeta\, P_{2}(\vgh\cdot\vbh)
&=\frac{3}{2}\left\{V^2-[V^2]_0\right\}
\end{align}
\esub
%-----------
we then have the final frame-independent mixing term
%-----------
\beal
\label{eq:velocity_coordinate_independent_rederived_V}
\mathcal{S}^{\rm mix}_{\beta}&=
[V^2]_0
+V\left[\Theta_{0}+\frac{1}{10}\Theta_{2}\right]
-\left[V\,\Theta\right]_0
-\frac{1}{10}\left[V\,\Theta\right]_2
+\frac{11}{20}\,V^2-\frac{11}{20}\,[V^2]_0.
\end{align}
%-----------
Because $V^2 \equiv [V^2]_0+[V^2]_2$, we can trade terms as needed. This then agrees with \citep{Ota2017arXiv-v3}. 

\newpage 
\section{Structure of the first order photon evolution equation}
\label{app:evol_eq}
%---------------------
At first order in perturbation theory, we may write the evolution equation for the temperature perturbations as \citep{Ma1995}
%--------------------
\bealf{
\label{eq:evol_I}
\frac{\partial \Theta}{\partial \eta}+\vgh\cdot \nabla \Theta+
\frac{\partial \Phi}{\partial \eta}+ \vgh\cdot \nabla\Psi
&= S+
\tau'\left[\Theta_0+\frac{1}{10}\,\Theta_2 - \Theta + \vbeta \cdot \vgh \right],
}
%--------------------
where we suppressed explicit mentioning of the order. Here, we have $\Theta=\Theta(\eta, \vgh, \vek{r})$, $\Phi=\Phi(\eta, \vek{r})$, $\Psi=\Psi(\eta, \vek{r})$, $\vbeta=\vbeta(\eta, \vek{r})$ and a general temperature source term $S=S(\eta, \vgh, \vek{r})$. The latter can be caused by photon mixing process \citep[e.g.,][]{Chluba2012} and the logic of how to solve the problem in the presence of distortion anisotropies follows similarly. 

Let us first go to Fourier space, which means that the terms $\nabla X\rightarrow \i \vek{k}\,\tilde{X}$, resulting in 
%--------------------
\bealf{
\label{eq:evol_I_F}
\frac{\partial \tilde{\Theta}}{\partial \eta}+\i \vgh\cdot \vek{k} \, \tilde{\Theta}+
\frac{\partial \tilde{\Phi}}{\partial \eta}+ \i \vgh\cdot \vek{k} \, \tilde{\Psi}
&= \tilde{S}+
\tau'\left[\tilde{\Theta}_0+\frac{1}{10}\,\tilde{\Theta}_2 - \tilde{\Theta} + \tilde{\vbeta} \cdot \vgh \right],
}
%--------------------
where $\tilde{X}$ denotes the Fourier coefficients. We will only consider the irrotational component of the velocity field\footnote{At first order in perturbation theory, vortical modes are not sourced and they furthermore damp away very quickly.}, such that $\tilde{\vbeta}(\eta, \vek{k})\approx -\i \tilde{\beta}(\eta, k) \, \hat{\vek{k}}$ \citep{Bartolo2006,Bartolo2007}, such that $\tilde{\vbeta} \cdot \vgh$ becomes $-\i \tilde{\beta}
\,\vgh\cdot\hat{\vek{k}}$.

\subsection{Aligning $\vek{k}$ with the $z$-axis}
\label{app:k_parallel_z}
%--------------------
In the standard treatment of cosmological perturbations at first order, one recognizes that only $\vgh\cdot\hat{\vek{k}}$ appears explicitly in the evolution equation, at least if we ignore the source term, which could have a more general dependence on $\vek{k}$. This means we can simplify the problem by aligning $\vek{k}$ with the $z$-axis, in which case the evolution equation reads:\footnote{In a statistically homogeneous and isotropic Universe this step should always be ok.}
%--------------------
\bealf{
\label{eq:evol_I_F_kz}
\frac{\partial \tilde{\Theta}}{\partial \eta}+\i k \chi\, \tilde{\Theta}+
\frac{\partial \tilde{\Phi}}{\partial \eta}+ \i k \chi\, \tilde{\Psi}
&= \tilde{S}+
\tau'\left[\tilde{\Theta}_0+\frac{1}{10}\,\tilde{\Theta}_2-\tilde{\Theta}  - \i \tilde{\beta} \,\chi \right],
}
%--------------------
with $\chi=\hat{e}_z\cdot\vgh$. We next insert the spherical harmonic expansion, $\tilde{\Theta}(\eta, \vgh, \vek{k})=\sum_{\ell m} \tilde{\Theta}_{\ell m}(\eta, \vek{k})\,Y_{\ell m}(\vgh)$, for the temperature field and use the identity\footnote{This is only applicable because we used $\hat{\vek{k}}\equiv\hat{e}_z$ and therefore the polar angles are given by $\chi=\cos\theta$.} 
%--------------------
\bealf{
\label{eq:chi_Yellm}
\chi Y_{\ell m}(\vgh)&=C^m_{\ell+1}Y_{\ell+1, m}(\vgh)+C^m_{\ell}Y_{\ell-1, m}(\vgh)
}
%--------------------
with $C^m_\ell=\sqrt{(\ell^2-m^2)/(4\ell^2-1)}$, to obtain
%--------------------
\bealf{
\label{eq:evol_I_F_kz_ellm}
&\frac{\partial \tilde{\Theta}_{\ell m}}{\partial \eta}+
\i k \left[C^m_{\ell}\tilde{\Theta}_{\ell-1, m} + C^m_{\ell+1}\tilde{\Theta}_{\ell+1, m}
\right]
+(-\i)^{\ell}\,\left[\sqrt{4\pi}\,\delta_{\ell 0}\frac{\partial \tilde{\Phi}}{\partial \eta}
-   
\sqrt{\frac{4\pi}{3}}\, \delta_{\ell 1}\delta_{m0} \,k\tilde{\Psi}\right]
\nonumber\\
&\qquad\qquad = \tilde{S}_{\ell m}+
\tau'\left[\left(\delta_{\ell 0}+\frac{\delta_{\ell 2}}{10} - 1\right)\,\tilde{\Theta}_{\ell m} + (-\i)^\ell \sqrt{\frac{4\pi}{3}}\,\delta_{\ell 1} \delta_{m0}\,\tilde{\beta}\right].
}
%--------------------
The factors of $\sqrt{4\pi}$ and $\sqrt{4\pi/3}$ and the powers of $\i$ can be absorbed by re-scaling
%--------------------
\bealf{
\label{eq:rescale}
\tilde{\Theta}_{\ell m}\rightarrow (-\i)^{\ell}(2\ell+1)\sqrt{\frac{4\pi}{2\ell+1}}\,\tilde{\theta}_{\ell m}, \qquad
\tilde{S}_{\ell m}\rightarrow (-\i)^{\ell}(2\ell+1)\sqrt{\frac{4\pi}{2\ell+1}}\,\tilde{s}_{\ell m},
}
%--------------------
where we used the normalization convention similar to Eq.~\eqref{eq:first_order_m_only}. 
We then have the photon hierarchy
%--------------------
\bealf{
\label{eq:evol_I_F_kz_ellm_final}
&\frac{\partial \tilde{\theta}_{\ell m}}{\partial \eta}+
k \left[\kappa^m_{\ell+1}\,\frac{\ell+1}{2\ell+1}\tilde{\theta}_{\ell+1, m}-\kappa^m_{\ell}\,\frac{\ell}{2\ell+1}\tilde{\theta}_{\ell-1, m}
\right]
+\delta_{\ell 0}\frac{\partial \tilde{\Phi}}{\partial \eta}
- \delta_{\ell 1} \delta_{m0}\,  
\frac{k}{3}\tilde{\Psi}
\nonumber\\
&\qquad\qquad = \tilde{s}_{\ell m}+
\tau'\left[\left(\delta_{\ell 0}+\frac{\delta_{\ell 2}}{10} - 1\right)\,\tilde{\theta}_{\ell m} + \delta_{\ell 1} \delta_{m0}\,\frac{\tilde{\beta}}{3}\right],
}
%--------------------
with $\kappa^m_{\ell}=\sqrt{(\ell^2-m^2)/\ell^2}$ \citep[see also][]{Hu1997}. This set of equations allows for the full structure in $\vgh$, while only requiring computations for fixed values of $k=|\vek{k}|$ in the frame where $\vek{k}\parallel \hat{e}_z$. To obtain the transfer function $\tilde{\Theta}_{\ell m}(\eta, \vek{k})$ for general direction of $\vek{z}$ we rotate the solutions as
%--------------------
\bealf{
\label{eq:Theta_Rotation}
\tilde{\Theta}_{\ell m}(\eta, \vek{k})&=(-\i)^{\ell}\,\sqrt{4\pi(2\ell+1)}\times \sum_{m'} (-1)^{m'}\sqrt{\frac{4\pi}{2\ell+1}}\,
{_{-m'}}Y^*_{\ell m}(\hat{\vek{k}})\,\expf{-\i m' \psi}\,
\tilde{\theta}_{\ell m'}(\eta, k)
\nonumber\\
&=4\pi(-\i)^{\ell} \sum_{m'} (-1)^{m'}{_{-m'}}Y^*_{\ell m}(\hat{\vek{k}})\,\expf{-\i m' \psi}\,
\tilde{\theta}_{\ell m'}(\eta, k)
\nonumber\\
&\equiv 
4\pi\i^{\ell} \sum_{m'} 
(-1)^{m'} {_{m'}}Y^*_{\ell m}(-\hat{\vek{k}})\,\expf{-\i m' \psi}\,
\tilde{\theta}_{\ell m'}(\eta, k),
}
%--------------------
where we allowed for a general roll angle, $\psi$, in the coordinate rotation. For $m'=0$ this agrees with Eq.~\eqref{eq:Theta_ellm_irrotational}. To write this expression, we used the rotation of the harmonic function and coefficients from the special frame $S'$ to $S$ as described using the Wigner-D matrices \citep{Varshalovich1988}:
%--------------------
\bealf{
\label{eq:Wigner_D-rotation}
Y_{\ell m}(\hat{\vek{k}})&=\sum_{m'}[D^\ell_{m m'}(\phi, \theta, \psi)]^* Y'_{\ell m'}(\hat{\vek{k}'})
\\
X_{\ell m}&=\sum_{m'}D^\ell_{m m'}(\phi, \theta, \psi)\, X'_{\ell m'}.
}
%--------------------
Here, the rotation is defined by the Euler angles, $\phi$, $\theta$ and $\psi$. In our problem, $\phi$ and $\theta$ are directly the angles of $\vek{k}$ in the general frame $S$. The Wigner-D matrices can then be written as 
%--------------------
\bealf{
\label{eq:Wigner_D-rotation_II}
D^\ell_{m' m}(\phi, \theta, \psi)
=\left<\ell m'\big|\hat{R}(\phi, \theta, \psi)\big|\ell m\right>\equiv (-1)^{m}\sqrt{\frac{4\pi}{2\ell+1}}\,
{_{-m}}Y_{\ell m'}(\hat{\vek{k}})\,\expf{\i m \psi}
}
%--------------------
in terms of the spin-harmonics, ${_s}Y_{\ell m}(\hat{\vek{n}})$. Here we also gave the 'Bra-Ket' notation with the rotation operator $\hat{R}(\phi, \theta, \psi)$. Together, this yields the expressions from above.

The final solution for the temperature field is thus given by
%--------------------
\bealf{
\label{eq:Theta_Sol_kz}
\Theta(\eta, \vgh, \vek{r})&=
\int \frac{\id^3 k}{(2\pi)^3}
\,\expf{\i \vek{r}\cdot\vek{k}}
\sum_{\ell m} \, \tilde{\Theta}_{\ell m}(\eta, \vek{k})\,Y_{\ell m}(\vgh).
}
%--------------------
We can then compute the two-point angular correlation function at fixed location $\vek{r}$ and time $\eta$ as
%--------------------
\bealf{
\label{eq:correlation_kz}
\xi(\eta,\vgh, \vgh', \vek{r})&=\left<\Theta(\eta, \vgh, \vek{r})\,
\Theta(\eta, \vgh', \vek{r})\right>
\\ \nonumber
&=
\int \frac{\id^3 k}{(2\pi)^3}\frac{\id^3 k'}{(2\pi)^3}
\,\expf{\i \vek{r}\cdot(\vek{k}+\vek{k}')}
\sum_{\ell m}
\sum_{\ell' m'}
\left<\tilde{\Theta}_{\ell m}(\eta, \vek{k})
\,\tilde{\Theta}_{\ell' m'}(\eta, \vek{k}')\right>
\,
Y_{\ell m}(\vgh)
Y_{\ell' m'}(\vgh').
}
%--------------------
To simplify this result further, we have to specify the form of the correlators $\left<\tilde{\Theta}_{\ell m}(\eta, \vek{k})
\,\tilde{\Theta}_{\ell' m'}(\eta, \vek{k}')\right>$. The result then depends on the statistics of the initial fluctuations and the corresponding statistics of the sources, as we explain now.

\subsubsection{Vanishing source term}
%--------------------
Assuming a vanishing source term, we then have
%--------------------
\bealf{
\label{eq:correlation_kz_II}
\left<\tilde{\Theta}_{\ell m}(\eta, \vek{k})
\,\tilde{\Theta}_{\ell' m'}(\eta, \vek{k}')\right>=\hat{\Theta}_{\ell m}(\eta, \vek{k})
\,\hat{\Theta}_{\ell' m'}(\eta, \vek{k}')
\left<\mathcal{R}(\vek{k})
\,\mathcal{R}(\vek{k}')\right>,
}
%--------------------
using the photon transfer functions $\hat{\Theta}_{\ell m}(\eta,\vek{k})$. For Gaussian, statistically isotropic and homogeneous initial fluctuations, $\langle\mathcal{R}(\vek{k})\,\mathcal{R}(\vek{k}')\rangle=(2\pi)^3\,\delta^{(3)}(\vek{k}+\vek{k}')\,P(k)$, we then obtain
%--------------------
\bealf{
\label{eq:correlation_kz_final}
\xi(\eta,\vgh, \vgh', \vek{r})&=
\int \frac{\id^3 k}{(2\pi)^3}\,P(k)\,
\sum_{\ell m}
\sum_{\ell' m'}
\hat{\Theta}_{\ell m}(\eta, \vek{k})
\,\hat{\Theta}_{\ell' m'}(\eta, -\vek{k})
\,
Y_{\ell m}(\vgh)
Y_{\ell' m'}(\vgh').
}
%--------------------
This expression is close to the final result, but we still need to carry out the integral over the direction of $\vek{k}$. For this we have to include the rotation of the transfer functions as defined in Eq.~\eqref{eq:Theta_Rotation}. For scalar-modes without source term, only $m=0$ would matter and this rotation simply gives $$\tilde{\Theta}_{\ell m}(\eta, \vek{k})=4\pi(-\i)^{\ell} {_{0}}Y^*_{\ell m}(\hat{\vek{k}})\,
\tilde{\theta}_{\ell 0}(\eta, k)\equiv 4\pi(-\i)^{\ell} Y^*_{\ell m}(\hat{\vek{k}})\,
\tilde{\Theta}_{\ell}(\eta, k),$$
consistent with the result in Eq.~\eqref{eq:Theta_ellm_irrotational}. However, we shall not use this simplification here and just keep things general in anticipation of a more general expression when sources are involved.
Inserting $\hat{\Theta}_{\ell m}(\eta, \vek{k})$ from Eq.~\eqref{eq:Theta_Rotation}, we then have
%--------------------
\bealf{
\label{eq:correlation_kz_final_II}
\xi(\eta,\vgh, \vgh', \vek{r})&=
\sum_{\ell m}
\sum_{\ell' m'}
Y_{\ell m}(\vgh)\,
Y_{\ell' m'}(\vgh')
\int \frac{k^2 \id k}{2\pi^2}\,P(k)
\\ \nonumber
&\quad \times
\int \frac{\id \hat{\vek{k}}}{4\pi}
\,
(4\pi)^2(-\i)^{\ell+\ell'} 
\sum_{\mu\,\mu'} (-1)^{\mu+\mu'} {_{-\mu}}Y^*_{\ell m}(\hat{\vek{k}})\,
{_{-\mu'}}Y^*_{\ell' m'}(-\hat{\vek{k}})\,
\expf{-\i(\mu+\mu')\psi}
\,
\hat{\theta}_{\ell \mu}(\eta, k)
\,\hat{\theta}_{\ell' \mu'}(\eta, k).
}
%--------------------
With ${_{-\mu'}}Y^*_{\ell' m'}(-\hat{\vek{k}})=(-1)^{\ell'}{_{\mu'}}Y^*_{\ell' m'}(\hat{\vek{k}})=(-1)^{\ell'+m'+\mu'}{_{-\mu'}}Y_{\ell', -m'}(\hat{\vek{k}})$, from the orthogonality of the spin-weight harmonic functions we then have $\int \id \hat{\vek{k}}\, {_{-\mu}}Y^*_{\ell m}(\hat{\vek{k}})\,
{_{-\mu'}}Y^*_{\ell' m'}(-\hat{\vek{k}})=(-1)^{\ell'+m'+\mu'}\,\delta_{\ell\ell'}\,\delta_{m,-m'}
\,\delta_{\mu,-\mu'}$. Inserting this back we find
%--------------------
\bealf{
\label{eq:correlation_kz_final_B}
\xi(\eta,\vgh, \vgh', \vek{r})&=
\sum_{\ell m}
\sum_{\ell' m'}
4\pi\,Y_{\ell m}(\vgh)\,
Y_{\ell' m'}(\vgh')
\int \frac{k^2 \id k}{2\pi^2}\,P(k)
\nonumber\\
&\qquad \times
(-\i)^{\ell+\ell'} 
\sum_{\mu\,\mu'} 
(-1)^{\ell'+m'+\mu'}
\delta_{\ell\ell'}\,\delta_{m,-m'}
\,\delta_{\mu,-\mu'}
\,
\expf{-\i(\mu+\mu')\psi}
\,
\hat{\theta}_{\ell \mu}(\eta, k)
\,\hat{\theta}_{\ell' \mu'}(\eta, k)
\nonumber\\
&=
\sum_{\ell m}
4\pi\,Y_{\ell m}(\vgh)
\,Y^*_{\ell m}(\vgh')
\int \frac{k^2 \id k}{2\pi^2}\,P(k)
\times
\sum_{\mu} (-1)^{\mu}
\hat{\theta}_{\ell \mu}(\eta, k)
\,\hat{\theta}_{\ell, -\mu}(\eta, k)
\nonumber\\
&=\sum_{\ell}\,\frac{(2\ell+1)}{4\pi}P_\ell(\vgh\cdot\vgh')
\times 
\left[
4\pi\int \frac{k^2 \id k}{2\pi^2}\,P(k)
\times
\sum_{\mu}|\hat{\theta}_{\ell \mu}(\eta, k)|^2
\right]
\nonumber\\
&=\sum_{\ell}\,\frac{(2\ell+1)}{4\pi}P_\ell(\vgh\cdot\vgh')\times C_\ell^{\Theta\Theta}(\eta).}
%--------------------
This is the final result with all intermediate rotations carried out consistently. The temperature power spectrum is then given by
%--------------------
\bealf{
\label{eq:correlation_kz_final_C}
C_\ell^{\Theta\Theta}(\eta) 
&=4\pi\int \frac{k^2 \id k}{2\pi^2}\,P(k)
\times
\sum_{\mu}|\hat{\theta}_{\ell \mu}(\eta, k)|^2.
}
%--------------------
Starting only with scalar initial sources, we then only have to consider the modes for $\mu=0$.

\subsubsection{Non-vanishing source term}
%--------------------
For a non-vanishing source term, the final solutions have two contributions:
%--------------------
\bealf{
\label{eq:Transfer_kz_source}
\tilde{\Theta}_{\ell m}(\eta, \vek{k})&=\hat{\Theta}_{\ell m}(\eta, \vek{k})\,\mathcal{R}(\vek{k})
+\int_0^\eta \hat{\Theta}_{\ell m}(\eta', \eta, \vek{k})\,S_{\ell m}(\eta', \vek{k}) \id \eta',
}
%--------------------
where $\hat{\Theta}_{\ell m}(\eta', \eta, \vek{k})$ is the transfer function for a mode with $\vek{k}$ from $\eta'\rightarrow \eta$. Here we used that the equations we are solving are linear such that a Green's function approach is justified.
For the correlators, we then have three contributions, one from the initial value problem, one with initial value and source, and one with two source terms:
%--------------------
\bealf{
\label{eq:correlation_kz_III}
\left<\tilde{\Theta}_{\ell m}(\eta, \vek{k})
\,\tilde{\Theta}_{\ell' m'}(\eta, \vek{k}')\right>&=\hat{\Theta}_{\ell m}(\eta, \vek{k})
\,\hat{\Theta}_{\ell' m'}(\eta, \vek{k}')
\left<\mathcal{R}(\vek{k})
\,\mathcal{R}(\vek{k}')\right>
\nonumber\\
&\;\;+\int_0^\eta \hat{\Theta}_{\ell m}(\eta, \vek{k})\,\hat{\Theta}_{\ell' m'}(\eta', \eta, \vek{k}')\,\left<\mathcal{R}(\vek{k})\,S_{\ell' m'}(\eta', \vek{k}')\right> \id \eta'
\nonumber\\
&\;\;\;+\int_0^\eta \hat{\Theta}_{\ell m}(\eta', \eta, \vek{k})\,\hat{\Theta}_{\ell' m'}(\eta, \vek{k}')\,\left<\mathcal{R}(\vek{k}')\,S_{\ell m}(\eta', \vek{k})\right> \id \eta'
\\ \nonumber
&\;\;\;\;+\int_0^\eta \int_0^\eta \hat{\Theta}_{\ell m}(\eta', \eta, \vek{k})\,\hat{\Theta}_{\ell' m'}(\eta'', \eta, \vek{k}')\,\left<S_{\ell m}(\eta', \vek{k})\,S_{\ell' m'}(\eta'', \vek{k}')\right> \id \eta'\id \eta''.
}
%--------------------
If we assume scalar perturbations, then the transfer functions $\hat{\Theta}_{\ell m}(\eta, \vek{k})$ can be obtained from Eq.~\eqref{eq:Theta_ellm_irrotational} from the solutions $\hat{\Theta}_{\ell}(\eta, k)$. For the other two, a general rotations from the frame with $\vek{k}\parallel \vek{e}_z$ is required, however, this can only be understood by considering the details of the source term.

\section{Required distortion sources for monopole}
\label{app:distortion_sources}
%--------------------------------------
To compute the temperature and distortion cross power spectra, we need to consider the correlators of $\Theta_{\ell m}(\eta, \vek{K})$ and $\mu_{\ell m}(\eta, \vek{k})$. For the temperature perturbations, we know that the transfer function $\hat{\Theta}^{(1)}_{\ell m}(\eta, \vek{K})
\equiv 4\pi (-\i)^\ell\, \hat{\Theta}^{(1)}_{\ell}(\eta, K)\,Y^*_{\ell m}(\hat{\vek{K}})$ is applicable, while for the distortions we keep things general for now, given that they are sourced at second order. The source of the distortion anisotropies is related to $\mathcal{S}^{\rm mix}(\eta, \vek{k})$ which in contrast to the temperature perturbations is not simply solved as an initial condition problem. Assuming no initial distortion anisotropies, we can write the formal solution 
%-----------
\beal
\label{eq:mu_sourced}
\mu_{\ell m}(\eta, \vek{k})
&=\int \id \eta' \hat{\mu}_{\ell m}(\eta', \eta, \vek{k}) \, \mathcal{S}_0^{\rm mix}(\eta', \vek{k})
\end{align}
%-----------
using the distortion transfer function from $\eta'$ to $\eta$ and for now assuming that distortions are {\it purely sourced by monopole terms}, which we generalize in the main text (Sect.~\ref{sec:source_ellm}). We therefore need to consider unequal-time correlators of the form $\left<\Theta^*_{\ell m}(\eta, \vek{K})\,\mathcal{S}^{\rm mix}_0(\eta', \vek{k})\right>$. The results can be seen in the easiest way by symmetrizing the source term integral. Instead of Eq.~\eqref{eq:local_monopole_dis_source_k}, we shall use
%-----------
\beal
\label{eq:local_monopole_dis_source_k_sym}
\mathcal{S}^{\rm mix}_0(\eta, \vek{k})
&=
\int \frac{\id^3 k_1}{(2\pi)^3}\frac{\id^3 k_2}{(2\pi)^3}\,
(2\pi)^3\delta^{(3)}(\vek{k}_2-\vek{k}_1+\vek{k})\,
\,\mathcal{R}(\vek{k}_1)\,\mathcal{R}^*(\vek{k}_2)\,
\hat{\mathcal{M}}_0(\eta, k_1, k_2, \chi_{12})
\end{align}
%-----------
with $\chi_{12}=\hat{\vek{k}}_1\cdot\hat{\vek{k}}_2$. The relevant correlators are then
%-----------
\beal
%\label{eq:Source_correlators}
\left<\mathcal{R}^*(\vek{K})\,\mathcal{S}^{\rm mix}_0(\eta', \vek{k})\right>
&=
\int \!\!\id^3 k_1\frac{\!\!\id^3 k_2}{(2\pi)^3}\,
\delta^{(3)}(\vek{k}_2-\vek{k}_1+\vek{k})
\left<\mathcal{R}^*(\vek{K})\,\mathcal{R}(\vek{k}_1)\,\mathcal{R}^*(\vek{k}_2)\right>
\hat{\mathcal{M}}_0(\eta', k_1, k_2, \chi_{12})
\nonumber \\ \nonumber
&=
\int 
\!\!\id^3 k_1
\!\id^3 k_2\,
\delta^{(3)}(\vek{k}_2-\vek{k}_1+\vek{k})\,
\delta^{(3)}(\vek{k}_2-\vek{k}_1+\vek{K})\,
B(k_1,k_2,K)\,
\hat{\mathcal{M}}_0(\eta', k_1, k_2, \chi_{12}),
\end{align}
%-----------
where we introduced the bispectrum, $\left<\mathcal{R}^*(\vek{K})\,\mathcal{R}(\vek{k}_1)\,\mathcal{R}^*(\vek{k}_2)\right>=(2\pi)^3\,\delta^{(3)}(\vek{k}_2-\vek{k}_1+\vek{K})\,
B(K, k_1,k_2)$.
The two Dirac $\delta$ functions enforce $\vek{K}\equiv \vek{k}$, which means we can alternatively write
%-----------
\beal
\label{eq:Source_correlators_II}
\left<\mathcal{R}^*(\vek{K})\,\mathcal{S}^{\rm mix}_0(\eta',\vek{k})\right>
%&=
%\delta^{(3)}(\vek{K}-\vek{k})\int 
%\!\!\id^3 k_1
%\!\id^3 k_2\,
%\delta^{(3)}(\vek{k}_2-\vek{k}_1+\vek{k})\,
%B(k, k_1,k_2)\,
%\hat{\mathcal{M}}_0(\eta', k_1, k_2, \chi_{12})
%\nonumber\\
&=
(2\pi)^3\,\delta^{(3)}(\vek{K}-\vek{k})
\,\Sigma_0(\eta', \vek{k}) 
\nonumber\\[1mm] 
\Sigma_0(\eta', \vek{k}) 
&=
\int 
\frac{\!\id^3 k_1
\!\id^3 k_2}{(2\pi)^3}\,
\delta^{(3)}(\vek{k}_2-\vek{k}_1+\vek{k})\,
B(k, k_1,k_2)\,
\hat{\mathcal{M}}_0(\eta', k_1, k_2, \chi_{12}).
\end{align}
%-----------
The easiest way forward is to carry out the integral over $\id^3 k_2$ using the $\delta$-function. We can then go to the special system where $\vek{k}\parallel \vek{e}_z$, which simplifies matters to $m=0$. This then yields
%-----------
\beal
\Sigma_0(\eta', \vek{k}) 
&=
\int 
\frac{k_1^2\!\id k_1}{2\pi^2}\,
\int_{-1}^1\frac{\id \chi_1}{2}\,
B(k, k_1,|\vek{k}_1-\vek{k}|)\,
\hat{\mathcal{M}}_0\left(\eta', k_1, |\vek{k}_1-\vek{k}|, \frac{k_1^2-\vek{k}\cdot \vek{k}_1}{k_1 |\vek{k}_1-\vek{k}|}\right),
\end{align}
%-----------
where $\chi_1=\hat{\vek{k}}\cdot \hat{\vek{k}}_1$. Defining $k_2=|\vek{k}_1-\vek{k}|=\sqrt{k^2+k_1^2-2 k k_1 \chi_1}$ and trading variables one then alternatively has
%-----------
\beal
\Sigma_0(\eta', \vek{k}) 
&=
\int 
\frac{k_1^2\!\id k_1}{2\pi^2}\,
\int_{|k_1-k|}^{k_1+k}\,\frac{k^2_2 \id k_2}{2 \pi^2}
\,\frac{\pi^2 B(k, k_1, k_2)}{k k_1 k_2}\,
\hat{\mathcal{M}}_0\left(\eta', k_1, k_2, \frac{k_1^2+k_2^2-k^2}{2 k_1 k_2}\right).
\end{align}
%-----------
This expression can be evaluated given the bispectrum and first order transfer function solutions.
For local-type non-Gaussianity, we can use 
$B(k, k_1,k_2)=\frac{6}{5}\,f_{\rm NL}[P(k_1)P(k_2)+P(k_1)P(k)+P(k_2)P(k)]$ to tabulate $\Sigma_{0}(\eta', \vek{k})$. The required source term for computing the transfer solution is then 
%-----------
\beal
\label{eq:source_monopole}
s_{\ell m}=\delta_{\ell 0}\delta_{m0} \frac{\Sigma_{0}(\eta', \vek{k})}{P(k)},
\end{align}
%-----------
which then gives the final power-spectrum normalized solution $\hat{\mu}_{\ell m}^{\rm source}(\eta, k)$ that can be used in
%--------------------
\bealf{
\label{eq:correlation_kz_final_B_source}
C_\ell^{\Theta\mu}(\eta) 
&=4\pi\int \frac{k^2 \id k}{2\pi^2}\,P(k)
\times \hat{\theta}_{\ell 0}(\eta, k)\,\hat{\mu}^{\rm source}_{\ell 0}(\eta, k)
}
%--------------------
to compute the related distortion cross power spectrum.

\subsection{Alternative evaluation of source term}
%-----------
Instead of carrying out the integral over the $\delta$-function in Eq.~\eqref{eq:Source_correlators_II}, we can use the identity
%-----------
\beal
\label{eq:delta_identity}
\delta^{(3)}(\vek{k}_2-\vek{k}_1+\vek{k})
&=
\int \frac{\id^3 r}{(2\pi)^3}\,
\expf{\i \vek{r}\cdot\left(\vek{k}_2-\vek{k}_1+\vek{k}\right)}\,
\nonumber
\\[1mm]
&\equiv 
\int 8 r^2\id r \sum_{\ell,m}\sum_{\ell_1,m_1}\sum_{\ell_2,m_2}\,\i^{\ell+\ell_1+\ell_2}
j_\ell(k r)\,j_{\ell_1}(k_1 r)\,j_{\ell_2}(k_2 r)
\nonumber\\
&\qquad
\times \int \id \hat{\vek{r}}\,
Y^*_{\ell m}(\hat{\vek{r}})Y_{\ell m}(\hat{\vek{k}})
\,
Y_{\ell_1 m_1}(\hat{\vek{r}})Y^*_{\ell_1 m_1}(-\hat{\vek{k}}_1)
\,
Y_{\ell_2 m_2}(\hat{\vek{r}})Y^*_{\ell_2 m_2}(\hat{\vek{k}}_2)
\nonumber \\
&=
\int 8 r^2\id r \sum_{\ell,m}\sum_{\ell_1,m_1}\sum_{\ell_2,m_2}\,\i^{\ell+\ell_1+\ell_2}
j_\ell(k r)\,j_{\ell_1}(k_1 r)\,j_{\ell_2}(k_2 r)
\nonumber\\
&\qquad
\times (-1)^m \,\mathcal{G}^{\ell,\ell_1,\ell_2}_{-m,m_1,m_2}
\,
Y_{\ell m}(\hat{\vek{k}})
\,
Y^*_{\ell_1 m_1}(-\hat{\vek{k}}_1)
\,
Y^*_{\ell_2 m_2}(\hat{\vek{k}}_2),
\end{align}
%-----------
where $\mathcal{G}^{\ell,\ell_1,\ell_2}_{m,m_1,m_2}$ is related to the Gaunt coefficients (Appendix~\ref{app:Gaunt}). Aligning $\vek{k}\parallel \vek{e}_z$, we then have
%-----------
\beal
\label{eq:delta_identity_kz}
\delta^{(3)}(\vek{k}_2-\vek{k}_1+\vek{k})
&\equiv 
\int 8 r^2\id r \sum_{\ell, \ell_1, \ell_2}
\sqrt{\frac{2\ell+1}{4\pi}} 
\,\i^{\ell+\ell_1+\ell_2}
\,j_\ell(k r)\,j_{\ell_1}(k_1 r)\,j_{\ell_2}(k_2 r)
\nonumber\\
&\qquad
\times \sum_{m_1} (-1)^{\ell_1+m_1}\mathcal{G}^{\ell,\ell_1,\ell_2}_{0,m_1,-m_1}
\,
Y^*_{\ell_1 m_1}(\hat{\vek{k}}_1)
\,
Y_{\ell_2 m_1}(\hat{\vek{k}}_2).
\end{align}
%-----------
With this, we then find
%-----------
\beal
\label{eq:Sigma_ellm_Source_correlators_II}
\Sigma_{0}(\eta', \vek{k})
&=\int \frac{8 r^2 \id r}{(2\pi)^3}\, \sum_{\ell} \sqrt{\frac{2\ell+1}{4\pi}} j_\ell(k r)\,
\sum_{\ell_1, \ell_2}
\i^{\ell+\ell_1+\ell_2}
\int k_1^2\id k_1 \, k_2^2\id k_2\,j_{\ell_1}(k_1 r)\,j_{\ell_2}(k_2 r)\,B(k, k_1,k_2)
\nonumber
\\
&\qquad
\times 
\sum_{m_1}(-1)^{\ell_1+m_1}
\mathcal{G}^{\ell,\ell_1,\ell_2}_{0,m_1,-m_1}\,
\int \id \hat{\vek{k}}_1 \id \hat{\vek{k}}_2\,
Y^*_{\ell_1 m_1}(\hat{\vek{k}}_1)
\,
Y_{\ell_2 m_1}(\hat{\vek{k}}_2)
\,\hat{\mathcal{M}}_0(\eta', k_1, k_2, \chi_{12}).
\end{align}
%-----------
Because we can write $\hat{\mathcal{M}}_0(\eta', k_1, k_2, \hat{\vek{k}}_1\cdot \hat{\vek{k}}_2)=\sum_\ell (2\ell +1)\,\hat{\mathcal{M}}_{0,\ell}(\eta', k_1, k_2) \,P_\ell(\hat{\vek{k}}_1\cdot \hat{\vek{k}}_2)$ we then find
%-----------
\beal
\label{eq:Sigma_ellm_Source_correlators_IIb}
\int \id \hat{\vek{k}}_1 \id \hat{\vek{k}}_2\,
Y^*_{\ell_1 m_1}(\hat{\vek{k}}_1)
\,
Y_{\ell_2 m_1}(\hat{\vek{k}}_2)
\,\hat{\mathcal{M}}_0(\eta', k_1, k_2, \chi_{12})
&=
4\pi \, \delta_{\ell_1 \ell_2}
\hat{\mathcal{M}}_{0,\ell_1}(\eta', k_1, k_2).
\end{align}
%-----------
Inserting this back into Eq.~\eqref{eq:Sigma_ellm_Source_correlators_II} yields
%-----------
\beal
\label{eq:Sigma_ellm_Source_correlators_III}
\Sigma_{0}(\eta', \vek{k})
&=\int \frac{4 r^2 \id r}{\pi^2}\, \sum_{\ell} \sqrt{\frac{2\ell+1}{4\pi}}\,\i^{\ell}
j_\ell(k r)\,
\sum_{\ell_1}
\sum_{m_1}(-1)^{m_1}
\mathcal{G}^{\ell,\ell_1,\ell_1}_{0,m_1,-m_1}\,
\nonumber
\\
&\qquad
\times 
\int k_1^2\id k_1 \, k_2^2\id k_2\,j_{\ell_1}(k_1 r)\,j_{\ell_1}(k_2 r)\,B(k, k_1,k_2)
\hat{\mathcal{M}}_{0,\ell_1}(\eta', k_1, k_2).
\end{align}
%-----------
For the sum over $m_1$, we can furthermore write
$$
\sum_{m_1}
(-1)^{m_1}\mathcal{G}^{\ell,\ell_1,\ell_1}_{0,m_1,-m_1}
=\delta_{\ell 0}\,\frac{(2\ell_1+1)}{\sqrt{4\pi}}
$$
giving the final source term
%-----------
\beal
\label{eq:Sigma_ellm_Source_correlators_IV}
\Sigma_{0}(\eta', \vek{k})
&=\int \frac{k_1^2\id k_1}{2\pi^2} \, \frac{k_2^2\id k_2}{2\pi^2}\,\pi^2 B(k, k_1,k_2)
\times 
\sum_{\ell_1}
(2\ell_1+1) \,\mathcal{I}_{\ell_1}(k, k_1,k_2)\,\hat{\mathcal{M}}_{0,\ell_1}(\eta', k_1, k_2),
\end{align}
%-----------
where $\mathcal{I}_{\ell}(k, k_1,k_2)$ is an integral over three spherical Bessel functions as defined in Eq.~\eqref{eq:I_2_bessel}.

\section{Simplified computation of the distortion sources}
\label{app:old_sources}
%-----------
Here we briefly compute the source terms for $\ell=1-3$ when neglecting the effects of angles between the Fourier modes. This does not capture all the effects but is good for comparison. 

\subsection{Dipole and quadrupole distortion sources}
%-----------
Like for the monopolar distortion source we first compute the corresponding Legendre coefficient of the dipole distortion source term\footnote{The temperature field at a given location has azimutal symmetry, $\Theta^{(1)}=\Theta^{(1)}(\eta, \vek{x}, \chi)$ so we do not need the full spherical harmonic coefficients.}:
%-----------
\beal
\tilde{\mathcal{S}}^{\rm mix}_{1}(\eta, \vek{x})
%&=\i \int \left[S^{\rm mix}_{\rm L}+S^{\rm mix}_{\rm T}+S^{\rm mix}_{\rm \beta}\right] \chi \frac{\id \Omega}{4\pi}
%\nonumber\\
&=
\int 
\i \,\Theta^{(1)}\left[\Theta^{(1)}-\Theta^{(1)}_0-\frac{1}{10}\Theta^{(1)}_2-\beta^{(1)}\chi \right]\chi\frac{\id \chi}{2}
\nonumber\\
&\qquad
+
\int 
\frac{\i}{2}\left[
 \left[(\Theta^{(1)})^2\right]_0+\frac{1}{10}\left[(\Theta^{(1)})^2\right]_2-(\Theta^{(1)})^2
 \right]\chi\frac{\id \chi}{2}
+
\frac{\tilde{\beta}^{(1)}}{3}\,
\left[\tilde{\Theta}^{(1)}_{0}
-\frac{\tilde{\Theta}^{(1)}_{2}}{5}
\right]
\nonumber\\
&=
\int 
\left[\frac{\i(\Theta^{(1)})^2}{2}
-\frac{\i}{10}\Theta^{(1)}\,\Theta^{(1)}_2
-\tilde{\beta}^{(1)}\Theta^{(1)}\chi \right]\chi\frac{\id \chi}{2}
-\tilde{\Theta}^{(1)}_0\,\tilde{\Theta}^{(1)}_1
+
\frac{\tilde{\beta}^{(1)}}{3}\,
\left[\tilde{\Theta}^{(1)}_{0}
-\frac{\tilde{\Theta}^{(1)}_{2}}{5}
\right].
\nonumber
\end{align}
%-----------
To simplify matters we can now use $\chi^2=[2P_2(\chi)+1]/3$, $P_2(\chi)\,\chi=[2P_1(\chi)+3P_3(\chi)]/5$ and the recurrence relation $\chi\,P_\ell(\chi)=[(\ell+1)P_{\ell+1}(\chi)+\ell P_{\ell-1}(\chi)]/(2\ell+1)$, which yields
%-----------
\beal
\label{eq:local_dipole_dis_source}
\tilde{\mathcal{S}}^{\rm mix}_{1}(\eta, \vek{x})
&=
\int 
\frac{\i}{2}\sum_{\ell\ell'} (-\i)^{\ell+\ell'}(2\ell+1)\tilde{\Theta}^{(1)}_\ell\tilde{\Theta}^{(1)}_{\ell'}P_\ell(\chi)
\left[(\ell'+1)P_{\ell'+1}(\chi)
+\ell'P_{\ell'-1}(\chi)\right] \frac{\id \chi}{2}
\nonumber\\
&\qquad
-\frac{\tilde{\beta}^{(1)}\tilde{\Theta}^{(1)}_0}{3}
+\frac{2\tilde{\beta}^{(1)}\tilde{\Theta}^{(1)}_2}{3}
+\frac{\tilde{\Theta}^{(1)}_1\,\tilde{\Theta}^{(1)}_2}{5}
-\frac{3\tilde{\Theta}^{(1)}_2\,\tilde{\Theta}^{(1)}_3}{10}
-\tilde{\Theta}^{(1)}_0\,\tilde{\Theta}^{(1),\rm g}_1
-\frac{\tilde{\beta}^{(1)}\tilde{\Theta}^{(1)}_{2}}{15}
\nonumber\\
&=
-\frac{1}{2}\sum_{\ell\ell'} (-\i)^{\ell+\ell'+1}\tilde{\Theta}^{(1)}_\ell\tilde{\Theta}^{(1)}_{\ell'}
  \left[(\ell'+1)\delta_{\ell,\ell'+1}
+\ell'\delta_{\ell,\ell'-1}\right]
\nonumber\\
&\qquad
-\frac{\tilde{\beta}^{(1)}\tilde{\Theta}^{(1)}_0}{3}
+\frac{2\tilde{\beta}^{(1)}\tilde{\Theta}^{(1)}_2}{3}
-\frac{3\tilde{\Theta}^{(1)}_2\,\tilde{\Theta}^{(1)}_3}{10}
-\tilde{\Theta}^{(1)}_0\,\tilde{\Theta}^{(1),\rm g}_1
+\frac{\tilde{\Theta}^{(1),\rm g}_1\tilde{\Theta}^{(1)}_{2}}{5}
\end{align}
%-----------
The remaining sum can be rewritten as $\sum_{\ell=1}^\infty (-1)^{\ell+1} \ell\,\tilde{\Theta}^{(1)}_{\ell-1}\tilde{\Theta}^{(1)}_\ell$. The leading terms $\tilde{\Theta}^{(1)}_0\tilde{\Theta}^{(1)}_1-2\tilde{\Theta}^{(1)}_1\tilde{\Theta}^{(1)}_2+3\tilde{\Theta}^{(1)}_2\tilde{\Theta}^{(1)}_3$ can be grouped with the remaining velocity terms, yielding the gauge-invariant expression
%-----------
\beal
\label{eq:local_dipole_dis_source_I}
\tilde{\mathcal{S}}^{\rm mix}_{1}(\eta, \vek{x})
&=
-\frac{9}{5}\,\tilde{\Theta}^{(1),\rm g}_1\,\tilde{\Theta}^{(1)}_{2}
+\frac{27}{10}\,\tilde{\Theta}^{(1)}_2\,\tilde{\Theta}^{(1)}_3
-
\sum_{\ell=4}^\infty (-1)^{\ell} \ell\,\tilde{\Theta}^{(1)}_{\ell-1}\tilde{\Theta}^{(1)}_\ell
\end{align}
%-----------
This result is in agreement with the $y$-source given by \cite{Ota2017}, but here including all higher order terms. 

For the quadrupole, the distortion source is readily obtained with similar steps. Before the Fourier transformation we have
%-----------
\beal
\label{eq:local_quadrupole_dis_source_II}
\tilde{\mathcal{S}}^{\rm mix}_{2}(\eta, \vek{x})
%&=\i^2 \int \left[S^{\rm mix}_{\rm L}+S^{\rm mix}_{\rm T}+S^{\rm mix}_{\rm \beta}\right] P_2(\chi) \frac{\id \Omega}{4\pi}
%\nonumber\\
&=
\int 
\i^2 \left[\frac{(\Theta^{(1)})^2}{2}
-\Theta^{(1)}\Theta^{(1)}_0+\frac{\left[(\Theta^{(1)})^2\right]_2}{20}-\frac{\Theta^{(1)}\Theta^{(1)}_2}{10}-\beta^{(1)}\Theta^{(1)}\chi \right]P_2(\chi)\frac{\id \chi}{2}
\nonumber\\
&\qquad\qquad
+
\frac{\tilde{\beta}^{(1)}}{5}\,
\left(\frac{11 \tilde{\beta}^{(1)}}{30}
-\frac{\tilde{\Theta}^{(1)}_{1}}{5} 
+\frac{3\,\tilde{\Theta}^{(1)}_{3}}{10} 
\right)
\nonumber\\
&=
\int 
\i^2 \left[\frac{(\Theta^{(1)})^2}{2}
+\frac{\left[(\Theta^{(1)})^2\right]_2}{20} \right]P_2(\chi)\frac{\id \chi}{2}
-\frac{\tilde{\Theta}^{(1)}_0\tilde{\Theta}^{(1)}_2}{10}
+\frac{(\tilde{\Theta}^{(1)}_2)^2}{7}
-\frac{9\tilde{\Theta}^{(1)}_2\tilde{\Theta}^{(1)}_4}{35}
\nonumber\\
&\qquad
-\tilde{\Theta}^{(1)}_0\tilde{\Theta}^{(1)}_2
-\frac{2}{5}\tilde{\beta}^{(1)}\tilde{\Theta}^{(1)}_1
+\frac{3}{5}\tilde{\beta}^{(1)}\tilde{\Theta}^{(1)}_3
+
\frac{\tilde{\beta}^{(1)}}{5}\,
\left(\frac{11 \tilde{\beta}^{(1)}}{30}
-\frac{\tilde{\Theta}^{(1)}_{1}}{5} 
+\frac{3\,\tilde{\Theta}^{(1)}_{3}}{10} 
\right)
\end{align}
%-----------
where we used $P^2_2(\chi)=[7P_0(\chi)+10P_2(\chi)+18P_4(\chi)]/35$. For the remaining integral we have
%-----------
\beal
\label{eq:local_quadrupole_dis_source}
\int 
\i^2 &\left[\frac{(\Theta^{(1)})^2}{2}
+\frac{\left[(\Theta^{(1)})^2\right]_2}{20} \right]P_2(\chi)\frac{\id \chi}{2}
=\int
\i^2 \left[\frac{(\Theta^{(1)})^2}{2}
+\frac{(\Theta^{(1)})^2}{20}\right] P_2(\chi)\frac{\id \chi}{2}
\nonumber\\
&=\frac{11}{20} \int
\sum_{\ell\ell'} (-\i)^{\ell+\ell'+2}
(2\ell+1)(2\ell'+1)
\tilde{\Theta}^{(1)}_\ell\tilde{\Theta}^{(1)}_{\ell'}
\nonumber\\
&\qquad\qquad
\Bigg[
\frac{3}{2}\frac{(\ell+1)P_{\ell+1}(\chi)
+\ell P_{\ell-1}(\chi)}{2\ell+1}
\frac{(\ell'+1)P_{\ell'+1}(\chi)
+\ell'P_{\ell'-1}(\chi)}{2\ell'+1}
-\frac{P_{\ell}P_{\ell'}}{2}
\Bigg] \frac{\id \chi}{2}
\nonumber\\
&=\frac{11}{20}
\sum_{\ell\ell'} (-\i)^{\ell+\ell'+2}
\tilde{\Theta}^{(1)}_\ell\tilde{\Theta}^{(1)}_{\ell'}
\Bigg[
\frac{3}{2}
\Bigg(
\frac{(\ell+1)(\ell'+1)}{(2\ell+3)}\delta_{\ell+1, \ell'+1}
+\frac{(\ell+1)\ell'}{(2\ell+3)}\delta_{\ell+1, \ell'-1}
\nonumber\\
&\qquad\qquad\qquad
+\frac{\ell(\ell'+1)}{(2\ell-1)}\delta_{\ell-1, \ell'+1}
+\frac{\ell\ell'}{(2\ell-1)}\delta_{\ell-1, \ell'-1}
\Bigg)
-\frac{(2\ell+1)}{2}\delta_{\ell,\ell'}
\Bigg]
\nonumber\\
&=\frac{11}{20}
\sum_{\ell} 
(-1)^{\ell} \Bigg[
-\frac{(2\ell+1)(\ell+1)\ell}{4\ell(\ell+1)-3}\,
(\tilde{\Theta}^{(1)}_\ell)^2
+\frac{(\ell+1)(\ell+2)}{2\ell+3}\,
\tilde{\Theta}^{(1)}_{\ell}\tilde{\Theta}^{(1)}_{\ell+2}
+\frac{\ell(\ell-1)}{2\ell-1}\,
\tilde{\Theta}^{(1)}_{\ell}\tilde{\Theta}^{(1)}_{\ell-2}
\Bigg]
\nonumber\\
&=
\frac{11}{10}\tilde{\Theta}^{(1)}_0\tilde{\Theta}^{(1)}_2
+
\frac{33}{50}(\tilde{\Theta}^{(1)}_1)^2
-
\frac{11}{14}(\tilde{\Theta}^{(1)}_2)^2
-
\frac{99}{50}\tilde{\Theta}^{(1)}_1\tilde{\Theta}^{(1)}_3
+
\frac{99}{35}\tilde{\Theta}^{(1)}_2\tilde{\Theta}^{(1)}_4
\nonumber\\
&\qquad\qquad+
\frac{11}{20}
\sum_{\ell=3} 
(-1)^{\ell} \Bigg[
-\frac{(2\ell+1)(\ell+1)\ell}{4\ell(\ell+1)-3}\,
(\tilde{\Theta}^{(1)}_\ell)^2
+\frac{3(\ell+1)(\ell+2)}{2\ell+3}\,
\tilde{\Theta}^{(1)}_{\ell}\tilde{\Theta}^{(1)}_{\ell+2}
\Bigg].
\end{align}
%-----------
Inserting this back into Eq.~\eqref{eq:local_quadrupole_dis_source_II} and collecting terms we finally find
%-----------
\beal
\label{eq:local_dipole_dis_source_IIII}
\tilde{\mathcal{S}}^{\rm mix}_{2}(\eta, \vek{x})
&=
\frac{33}{50}\,(\tilde{\Theta}^{(1),\rm g}_1)^2
-\frac{9}{14}\,(\tilde{\Theta}^{(1)}_2)^2
-\frac{99}{50}\tilde{\Theta}^{(1),\rm g}_1\tilde{\Theta}^{(1)}_3
+\frac{18}{7}\tilde{\Theta}^{(1)}_2\tilde{\Theta}^{(1)}_4
\\ \nonumber
&\qquad\qquad+
\frac{11}{20}
\sum_{\ell=3} 
(-1)^{\ell} \Bigg[
-\frac{(2\ell+1)(\ell+1)\ell}{4\ell(\ell+1)-3}\,
(\tilde{\Theta}^{(1)}_\ell)^2
+\frac{3(\ell+1)(\ell+2)}{2\ell+3}\,
\tilde{\Theta}^{(1)}_{\ell}\tilde{\Theta}^{(1)}_{\ell+2}
\Bigg],
\end{align}
%-----------
which again is gauge-independent. 
This result again is in agreement with the $y$-source given by \cite{Ota2017}, but here including all higher order terms.
In the tight-coupling regime, the second term will provide the dominant source of distortion anisotropies. Indeed, it sources {\it negative} $y$-distortion anisotropies. The remaining terms are suppressed by at least one small factor.

\subsection{Octupolar distortion source}
%-----------
Looking at Eq.~\eqref{eq:velocity}, we see that at least one additional distortion source term involving only multipoles with $\ell\leq 2$ is expected for the octupole part of the spectrum. Neglecting all multipoles with $\ell>2$ we find
%-----------
\beal
\label{eq:local_octupole_dis_source}
\tilde{\mathcal{S}}^{\rm mix}_{3}(\eta, \vek{x})
&\approx
\frac{81}{70}\,\tilde{\Theta}^{(1),\rm g}_1 \tilde{\Theta}^{(1)}_2
\end{align}
%-----------
as the leading contribution. This term was omitted by \cite{Ota2017}, although it should be of similar order as many of the other terms. Nevertheless, this contribution is again suppressed in the tight-coupling era such that we can neglect it for now.
The required source matrix element is then obtained as before.

\end{document}